\documentclass[aps,prl,reprint,superscriptaddress,amsmath,amssymb,floatfix]{revtex4-2}

\usepackage{graphicx}
\usepackage{dcolumn}
\usepackage{bm}
\usepackage{soul}
\usepackage{subfigure}
\usepackage{amsthm}
\usepackage{amsfonts}
\usepackage{braket}
\usepackage{comment}
\usepackage{enumitem}
\usepackage{float}
\usepackage{dsfont}
\usepackage{xcolor}
\usepackage{mathtools}
\usepackage{soul}
\usepackage{hyperref}
\usepackage[capitalize]{cleveref} 
\hypersetup{
    breaklinks=true,
    colorlinks=true,
    linkcolor=blue,
    urlcolor=blue,
    citecolor=blue
}
\usepackage{bibunits}
\defaultbibliography{references,outputNotes}
\defaultbibliographystyle{apsrev4-2}

\def\be{\begin{equation}}
\def\ee{\end{equation}}
\def\bs{\boldsymbol}

\def\blambda{{\boldsymbol{\lambda}}}
\def\bmu{{\boldsymbol{\mu}}}
\newcommand{\fab}[1]{{\color{red}\ifmmode\text{\footnotesize(FE) #1}\else\footnotesize{(FE) #1}\fi}}
\newcommand{\sups}[1]{^{\mkern-2mu(\mkern-2mu #1 \mkern-2mu)}}
\newcommand{\curlb}[1]{\{\mkern-2mu \bs \lambda \sups #1 \mkern-2mu \}}
\newcommand{\curlbI}[1]{\{\mkern-2mu \bs I \sups #1 \mkern-2mu \}}

\providecommand{\selectlanguage}[1]{\relax}

\begin{document}

\preprint{APS/123-QED}

\begin{bibunit}

\title{Non-equilibrium quantum dynamics of interacting integrable models \\by Monte Carlo sampling Lehmann representations}
\author{Riccardo Senese}
\affiliation{SISSA and INFN, Via Bonomea 265, 34136 Trieste, Italy}
\affiliation{Rudolf Peierls Centre for Theoretical Physics, University of Oxford, Oxford OX1 3PU, United Kingdom}
\author{Fabian H.L. Essler}%
\affiliation{Rudolf Peierls Centre for Theoretical Physics, University of Oxford, Oxford OX1 3PU, United Kingdom}

\date{\today}

\begin{abstract}
Determining the dynamics of interacting integrable many-particle quantum systems at finite times after homogeneous quantum quenches is a long-standing challenge. We present a Monte Carlo sampling scheme that numerically evaluates the Lehmann representation for time-dependent expectation values of local operators, allowing us to access system sizes and times significantly beyond the reach of existing methods. The approach accommodates both the full Lehmann sum and the Quench Action formalism. We benchmark against exact results for non-interacting lattice and continuum models and short-time results at weak interactions, finding excellent agreement. We apply the method to quantum quenches from a Bose-Einstein condensate in the repulsive Lieb-Liniger model and determine the time evolution of the order parameter for a wide range of interaction strengths. We discuss the emergence of a “sign problem” for more general dynamical correlators and setups.
\end{abstract}

\maketitle
Non-equilibrium dynamics in isolated many-particle quantum systems has attracted a huge amount of attention over the last two decades. Experimental advances in ultra-cold atomic systems and quantum computing platforms have made it possible to access new regimes in quantum physics and address a wide range of fundamental questions \cite{bloch2012quantum,polkovnikov_colloquium_2011, calabrese2016introduction,dalessio_quantum_2016,gogolin_equilibration_2016,ueda2020quantum,bouchoule2022generalized,doyon2025generalized}. In spite of remarkable progress, an efficient description of quantum dynamics in spatially extended systems has remained elusive. The key difficulty is the growth of quantum entanglement under time evolution \cite{calabrese_evolution_2005,calabrese2016quantum,nahum2017quantum,vonkeyserlingk_operator_2018}: even if we start in a product state, under time evolution it quickly becomes highly entangled and therefore difficult to describe analytically or by classical computational methods. In practice this imposes severe limits on the time window accessible by numerical techniques \cite{paeckel2019time,wu2025real}. Quantum field theories — relevant for many experiments — pose additional obstacles to numerical approaches.
As a result the dynamics in the intermediate and late time regimes in presence of strong interactions has remained largely out of reach, even in the paradigmatic class of integrable models \footnote{A notable exception is the large-scale structure after quantum quenches from inhomogeneous initial states, which can be determined from generalized hydrodynamics \cite{fagotti2016transport,castroalvaredo2016emergent,bouchoule2022generalized,doyon2025generalized}}. These are characterized by having extensive numbers of conservation laws with local densities \cite{korepin_quantum_1993} and have played an important role in shaping our understanding of non-equilibrium dynamics in extended many-particle quantum systems \cite{calabrese2016introduction,essler_quench_2016,bouchoule2022generalized,doyon2025generalized}. 
They provide valuable benchmarks for numerical approaches and quantum simulators and at the same time provide a mechanism for ergodicity breaking that can generate exotic non-thermal steady states \cite{astrakharchik2005beyond,batchelor2005evidence,piroli2016multiparticle}. While integrability provides exact expressions for energy eigenstates, using them to obtain the non-equilibrium dynamics of local observables for large systems with hundreds of particles has been a long-standing problem. The related but simpler case of zero temperature equilibrium dynamics have been solved by using Lehmann representations, which express correlation functions in terms of sums involving matrix elements of operators in energy eigenstates \cite{caux2005computation,caux2009correlation}. In this letter we show how to Monte Carlo sample Lehmann representations in order to determine non-equilibrium correlators in interacting integrable models. In particular we consider the Lieb-Liniger model \cite{lieb_exact_1963,lieb_exact_1963a}, a key paradigm for integrable interacting quantum field theories \cite{korepin_quantum_1993} with important applications in cold atom experiments \cite{kinoshita2004observation,paredes2004tonks,fabbri2015dynamical,meinert2015probing,fang2016momentum,kinoshita2006quantum,hofferberth2007non,cheneau2012light,bouchoule2022generalized}.

\emph{Correlation functions and MC scheme.} \ 
Energy eigenstates in quantum integrable models can be parametrized in terms of rapidity variables $\blambda=\{\lambda_1, \ldots,\lambda_N\}$ \cite{korepin_quantum_1993,takahashi_thermodynamics_1999} 
\begin{equation}
    \hat{H} \ket \blambda = E_\blambda \ket \blambda \ , \qquad \hat{P} \ket \blambda = P_\blambda \ket \blambda \ ,
\end{equation}
where $\hat H$ and $\hat P$ are respectively the Hamiltonian and momentum operator. Using the completeness relation $\mathds{1}=\sum_{\bs \lambda} \ket{\bs \lambda}\bra{\bs \lambda}$ one can derive Lehmann representations for generic correlation functions in any density matrix $\hat \rho$
\begin{align}
\label{eq:kpf}
    &C(\boldsymbol{s})\equiv
    {\rm Tr}\big[ \hat \rho \prod_{\ell=1}^{r}O_\ell(x_\ell,t_\ell) \big]=\sum_{ \curlb j}
\alpha_{ \curlb j} (\bs s) 
    \, F_{ \curlb j} \ , \nonumber \\
    &F_{ \curlb j}=\braket{\blambda \sups{r+1}|\hat \rho|\blambda \sups 1}\prod_{\ell=1}^{r}\braket{\blambda \sups \ell|O_\ell(0,0)|\blambda \sups{\ell+1}} \ .
\end{align}
Here $\curlb j=\{\bs \lambda\sups 1,\ldots,\blambda \sups{r+1}\}$, $O_\ell(x,t)$ denotes the Heisenberg-picture evolution of a local operator with support around position $x$, ${\bs s}=(x_1,t_1;x_2,t_2;\dots;x_{r},t_{r})$ and
\begin{align}
\label{eq:alphas}
\alpha_{ \curlb j} (\bs s) &= \prod_{\ell=1}^{r}
e^{i[E_{\blambda \sups \ell}-E_{\blambda \sups{\ell+1}}]t_\ell-
i[P_{\blambda \sups \ell}-P_{\blambda \sups{\ell+1}}]x_\ell} .
\end{align}
The correlators in \cref{eq:kpf} capture both equilibrium dynamics in (generalized) Gibbs ensembles \cite{rigol_relaxation_2007, ilievski_complete_2015,essler_quench_2016} and non-equilibrium protocols like global quantum quenches \cite{greiner_quantum_2002, greiner_collapse_2002, kinoshita_quantum_2006, calabrese_evolution_2005, calabrese_time_2006, polkovnikov_colloquium_2011, essler_quench_2016}. A key feature of integrable models is that explicit expressions for the \emph{form factors} $\braket{\bs \lambda |O |\bs \mu}$ and \emph{overlaps} $\braket{\blambda|\hat \rho|\bmu}$ are available in a number of cases. Yet, how to perform the sums over eigenstates that appear in \cref{eq:kpf} by either analytical or numerical means has generally remained an open problem. On the numerical side, the main difficulty is that in interacting theories the number of eigenstates that contributes meaningfully to the spectral sum scales exponentially with the system size $L$ \cite{essler_statistics_2024, senese_finite_2026, rottoli_eigenstate_2026, orlov_multiscale_2026} \footnote{Furthermore, identifying the eigenstates that contribute non-negligibly to the correlators is non-trivial given the rich structure in the statistics of matrix elements of integrable models \cite{essler_statistics_2024, rottoli_eigenstate_2026, orlov_multiscale_2026}.}.
In a recent work \cite{senese_finite_2026} we showed that in the case of $2$-point correlation functions -- \cref{eq:kpf} for $r = 2$ and identical operators -- in equilibrium states this difficulty can be efficiently overcome by employing a Monte Carlo (MC) scheme in which the eigenstates are sampled according to their spectral weight, \emph{cf.}~Refs \cite{gritsev_exact_2010, buccheri_structure_2011, faribault_integrabilitybased_2013, faribault_spin_2013, burovski_momentum_2014, alba_simulating_2015, alba_quench_2016, gamayun_impact_2018, bouchoule_effect_2020, zhang_monte_2024} for related earlier approaches and \cite{bastianello2026exotic,zeng2026realization} for a recent application of the method. 
The MC scheme can be generalized to the generic correlators in \eqref{eq:kpf} by expressing them as
\begin{align}
\label{eq:Csampling}
    C(\bs s)&= Z \sum_{\curlb j} \alpha_{ \curlb j}(\bs s) \, e^{i \theta_{\curlb j}} \, \frac{|F_{\curlb j}|}{Z} \ ,\\
\label{eq:thetaAndZ}
    \theta_{\curlb j}&=\arg (F_{\curlb j}) \ ,  \quad Z = \sum_{\curlb j} |F_{\curlb j}| \ ,
\end{align}
and then carrying out the sums by MC sampling sets of rapidities $\curlb j$ according to the normalized probability distribution $\mathcal P(\curlb j) = |F_{\curlb j}|/Z$. In integrable models this can be achieved very efficiently \cite{senese_finite_2026, alba_simulating_2015} via Markov chain MC sampling \cite{metropolis_equation_1953, hastings_monte_1970,rubinstein_simulation_2016, gilks_markov_1995, feller_introduction_2009} (see Appendix \hyperlink{target:appendixA}{A}). Denoting the sets of rapidities sampled at MC step $\ell$ by $\curlb j_\ell$, we obtain an estimate
\begin{equation}
\label{eq:sampscheme}
    C(\bs s) \approx  \ Z \left[ \frac{1}{\ell_{\rm max}}\sum_{\ell=1}^{\ell_{\rm max}} \alpha_{\curlb j_\ell}(\bs s) \, \exp\big(i \theta_{\curlb j_\ell }\big)\right] \ ,
\end{equation}
where the term in square brackets is the output of the MC algorithm. In practice MC sampling can be used to determine $C(\bs s)$ only if the normalization $Z$ is not too large. In situations where $Z$ grows exponentially with system size $L$, a “sign problem” arises due to strong fluctuations in the phases $\exp(i \theta_{\curlb j})$. In the absence of a sign problem, MC sampling allows the full reconstruction of correlators in all space-time regions in which $|C(\bs s)|/Z$ is not vanishingly small, with a resolution that increases as $\sqrt{\ell_{\rm max}}$ \footnote{The normalization $Z$ can be extracted from the knowledge of $C(\bs 0)$ and the output of MC at $x_i = t_i = 0 \ \forall \ i$. Indeed, the value of $Z$ is not needed in the MC sampling, see Appendix A.}. 
In the following we demonstrate that the MC scheme makes it possible to determine the finite-time evolution of local observables after global quantum quenches in spatially extended interacting integrable theories. Specifically, we show that quenches from the paradigmatic class of “integrable” initial states \cite{ghoshal_boundary_1994, fioretto_quantum_2010, sotiriadis_zamolodchikov_2012, bertini_quantum_2014, piroli_what_2017} do not suffer from a sign problem. 

\emph{Non-equilibrium dynamics. \ } 
Specifying \cref{eq:kpf} to the case of the expectation value of a local operator $O$ after a quantum quench with initial state $|\psi(0)\rangle\equiv \ket{\psi}$ gives
\begin{equation}
\label{eq:quench}
\begin{aligned}
    \braket{O(t)}\equiv\braket{\psi(t)|O|\psi(t)} &= \sum_{\bs \lambda, \bs \mu} \alpha_{\bs \lambda, \bs \mu}(t) \, F_{\bs \lambda, \bs \mu}  \ , \\
    F_{\bs \lambda, \bs \mu} &= \braket{\psi|\bs \lambda}\braket{\bs \lambda |O|\bs \mu}\braket{\bs \mu |\psi} \ ,
\end{aligned}
\end{equation}
where $\alpha_{\bs \lambda, \bs \mu}(t)=e^{i(E_{\bs \lambda}-E_{\bs \mu})t}$. We refer to (\ref{eq:quench}) as the direct sum (DS) representation.
The “Quench Action” (QA) method \cite{caux_time_2013, caux_quench_2016} provides a dramatic simplification of 
the spectral representation \eqref{eq:quench} in the limit of large system sizes/particle numbers
\begin{align}
\label{eq:quenchQA}
     \lim_{L \to \infty}\braket{O(t)} &=  \lim_{L \to \infty}\left(\frac{\braket{\psi|O(t)|\bs \lambda_{\rm sp}}}{2 \braket{\psi|\bs \lambda_{\rm sp}}} +
     \frac{\braket{\bs \lambda_{\rm sp}|O(t)|\psi}}{2 \braket{\bs \lambda_{\rm sp}|\psi}}     \right)\nonumber\\
    & = \lim_{L \to \infty} \sum_{\bs \mu} \left(\alpha_{\bs \mu}(t)F_{\bs \mu}^{O} +\alpha_{\bs \mu}^*(t)(F_{\bs \mu}^{O^\dagger})^*\right) \ , \nonumber\\
 &\hskip -50pt   F_{\bs \mu}^{O} = \frac{\braket{\psi|\bs \mu }\braket{\bs \mu |O|\bs \lambda_{\rm sp}}}{2 \braket{\psi|\bs \lambda_{\rm sp}}}\ ,\ \alpha_{\bs \mu}(t)=e^{i(E_{\bs \mu}-E_{\bs \lambda_{\rm sp}})t} \ .
\end{align}
Here $\ket{\bs \lambda_{\rm sp}}$ is a single representative “saddle-point” eigenstate that is fixed by the expectation values of the conserved charges in the initial state \cite{caux_time_2013, caux_quench_2016}, see Supplemental Material (SM). 
The key simplification in \eqref{eq:quenchQA} compared to \eqref{eq:quench} is that only a single sum over eigenstates appears, which has been very useful in analytical approaches to quench dynamics \cite{bertini_quantum_2014, nardis_relaxation_2015, granet_finite_2020, granet_systematic_2021, salvo_quantum_2023, salvo_relaxation_2025}. Until now the validity of \eqref{eq:quenchQA} at finite times after a quantum quench has been established only in integrable models that can be mapped to free theories \cite{caux_time_2013}. 
Using that the MC scheme \eqref{eq:sampscheme} can be applied to both \eqref{eq:quench} and \eqref{eq:quenchQA}, we have verified that the DS and QA representations indeed give identical results in the Lieb-Liniger model within the accuracy limits of the MC method. This highly non-trivial check represents the first finite-time benchmark of the QA method in an interacting integrable theory.

\begin{figure}[!b]
     \centering
     \begin{minipage}[b]{0.239\textwidth}
         \centering
         \includegraphics[width=\textwidth]{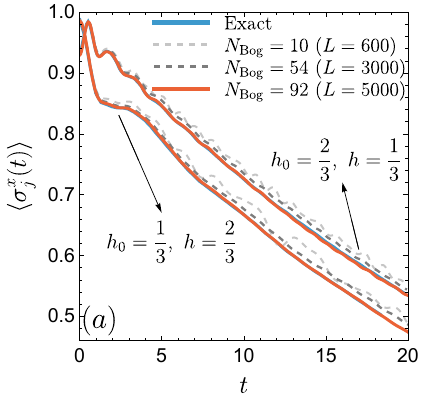}
     \end{minipage}
     \hspace{-7pt}
     \begin{minipage}[b]{0.239\textwidth}
         \centering
         \includegraphics[width=\textwidth]{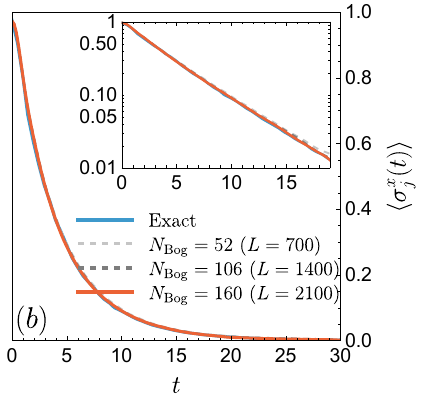}
     \end{minipage}
     \vspace{-10pt}
     \caption{Comparison of $\braket{\sigma_j^x(t)}$ from MC (orange and dashed lines) and exact results from Ref.~\cite{calabrese_quantum_2012} (blue lines), starting from the GS of $H_{\rm TFIC}(h_0)$ and quenching $h_0 \to h$.  \eqref{eq:quenchQA} (QA) is sampled by running $10$-$50$ Markov chains in parallel up to $\ell_{\rm max}=10^6$. (a) Quenches for $L=600,3000,5000$ where $(h_0,h)=(1/3,2/3)$ for the lower curves and $(h_0,h)=(2/3,1/3)$ for the upper curves. $N_{\rm Bog}$ denotes the number of quasiparticles excited by the quench. (b) Quench with $(h_0,h)=(0.5,0.99)$ (inset: same data on a log scale).}
     \label{fig:TFIC}
\end{figure}
\emph{Benchmarks in TFIC.} \ We first consider the transverse-field Ising chain (TFIC) \cite{lieb_two_1961} 
\vspace{-7pt}
\begin{equation}
\label{eq:TFICh}
    \hat{H}_{\rm TFIC}(h) = -J \sum_{j=1}^L(\sigma^x_j \sigma_{j+1}^x + h\, \sigma_j^z) \qquad J > 0 \ .
\end{equation}
This model is a equivalent to a free theory of spinless fermions via the Jordan-Wigner mapping \cite{jordan_ber_1928,lieb_two_1961}. We consider the quench dynamics of the order parameter $O=\sigma_j^x$ from the ferromagnetic ground state (GS) $\ket \psi$ of $H_{\rm TFIC}(h_0)$ for $h_0 < 1$, which spontaneously breaks the $\mathbb Z_2$ symmetry $\sigma^x \to -\sigma^x$ for $L \to \infty$. This is a classic example of an “integrable” initial state.
Even though the model is non-interacting, the Lehmann representation for $\braket{\sigma^x_j(t)}$ is very similar to that of interacting integrable theories, in that the number of relevant eigenstates to be summed over in \eqref{eq:quench} or \eqref{eq:quenchQA} is exponentially large in $L$ 
\cite{calabrese_quantum_2011, calabrese_quantum_2012}. Analytical and numerically exact results for $\braket{\sigma_j^x(t)}$ have been obtained by free fermion techniques in Refs.~\cite{calabrese_quantum_2011, calabrese_quantum_2012}, and provide us with a direct benchmark of the MC method for the dynamics at finite times after a quantum quench.
\cref{fig:TFIC} shows MC results for the QA representation (\ref{eq:quenchQA}) for various quenches within the ordered phase $h_0,h<1$. The order parameter $\braket{\sigma_j^x(t)}$ decays to zero as a consequence of dynamical symmetry restoration \cite{essler_quench_2016,collura2020order}. The agreement between the MC estimate and the exact results from Ref.~\cite{calabrese_quantum_2012} is seen to be excellent \footnote{To obtain convergence with $L$ within the QA formalism of \eqref{eq:quenchQA} requires $L = \mathcal{O}(10^3)$, which corresponds to a number of free Bogoliubov excitations of order $N_{\rm Bog}=\mathcal{O}(10^1$-$10^2)$ (see SM). Large $L$  and $N_{\rm Bog}$ values are expected given the saddle-point argument underpinning the QA approach. Convergence in DS \eqref{eq:quench} occurs at significantly smaller $L$, at the cost of introducing a second spectral sum.}. For details on the MC algorithm see Appendix \hyperlink{target:appendixA}{A}. In the SM we prove the absence of a sign problem for $h_0 < h$ \footnote{For $h_0 > h$ there is no sign problem even though $F_{\curlb j}$ is \emph{not} positive definite. Indeed, this property of the $F_{\curlb j}$ enables the growth of the order parameter seen in \cref{fig:TFIC}(a) at early times.}.

\emph{Repulsive Lieb-Liniger (LL) model.} \ We now turn to the repulsive ($c > 0$) LL model of bosons \cite{lieb_exact_1963, lieb_exact_1963a} on a ring of length $L$ (setting $\hbar = 2m = 1$)
\begin{equation}
\label{eq:H_LL}
    \hat{H}_{\rm LL} = \int_0^L dx  \Big[-\phi^\dag(x)\partial_x^2 \phi(x) + c  \big(\phi^\dag(x)\big)^2 \big(\phi(x)\big)^2\Big]
\end{equation}
where $\phi(x)$ is a complex Bose field with commutation relations $[\phi(x),\phi^\dag(y)]=\delta(x-y)$. In the limits $c = 0$ and $c = \infty$ \cite{girardeau_relationship_1960, creamer_quantum_1980} the model \eqref{eq:H_LL} reduces respectively to non-interacting bosons and fermions. 
\begin{figure}[!b]
     \centering
     \begin{minipage}[b]{0.243\textwidth}
         \centering
         \includegraphics[width=\textwidth]{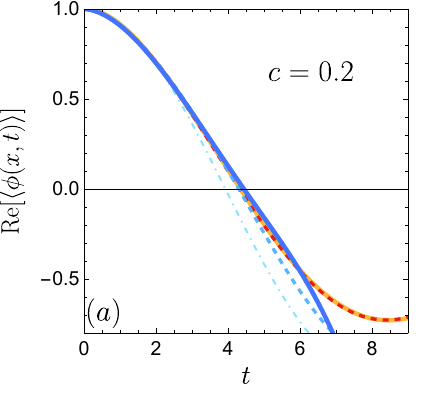}
     \end{minipage}
     \hspace{-9pt}
     \begin{minipage}[b]{0.243\textwidth}
         \centering
         \includegraphics[width=\textwidth]{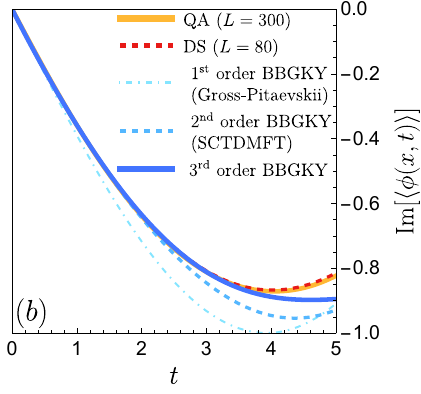}
     \end{minipage}
     \vspace{-20pt}
     \caption{MC results from QA \eqref{eq:quenchQA} and the full double sum (DS) \eqref{eq:quench} for $\braket{\phi(x,t)}$ in the shallow quench $c = 0 \to c =0.2$ (with density $n=\overline{N}/L = 1$). The MC results are benchmarked against the prediction of the BBGKY hierarchy truncated at $1^{\rm st}$,$2^{\rm nd}$ and $3^{\rm rd}$ order (see Appendix B). The MC data is obtained by running 100 Markov chains in parallel with $\ell_{\rm max}=10^6$-$10^7$. (a) Real part. (b) Imaginary part.}
     \label{fig:BBGKY_c_p2}
\end{figure}
We focus on quenches from the BEC state (GS of the $c = 0$ theory) in the symmetry broken description \cite{bogoliubov_theory_1947}
\begin{equation}
\label{eq:GSfreetheory}
    \ket{\psi}=\ket{{\rm GS}}_\eta=e^{- |\eta|^2/2}e^{\eta  \tilde\phi_0^\dag}\ket 0 \quad \quad |\eta|^2 = \overline N \ ,
\end{equation}
and track the dynamics of the order parameter $\braket{\phi(x,t)} = \braket{\psi(t)|\phi(x)|\psi(t)}$ under evolution with $\hat H_{\rm LL}(c>0)$. In \eqref{eq:GSfreetheory} we choose $\eta$ real, $\ket{0}$ denotes the boson vacuum, $\tilde \phi_k = \int_0^L dx \, e^{ikx}\phi(x)/\sqrt{L}$,  and $\overline{N}=\braket{\psi|\hat N|\psi}$ denotes the average number of particles. 
Closed expressions for the form factors $\braket{\bs \lambda |\phi(0)|\bs \mu}$ ~\cite{kojima_determinant_1997,caux_oneparticle_2007,piroli_exact_2015} and overlaps $\braket{\psi|\bs \lambda}$ \cite{denardis_solution_2014,brockmann_overlaps_2014,brockmann_neelxxz_2014} needed in \cref{eq:quench,eq:quenchQA} are known (see SM). 
Using their explicit expressions and the fact that $\ket{\psi}$ is an integrable initial state, i.e.~$\braket{\psi|\bs \lambda}\neq0$ only if $\ket{\bs \lambda}=\ket{-\bs \lambda}$, one can show that $F_{\bs \lambda , \bs \mu}$ from \eqref{eq:quench} and $F_{\bs \mu}^{O}$ from \eqref{eq:quenchQA} are positive definite and hence there is no sign problem (see SM).
Starting from the initial value $\braket{\phi(x,0)}=(\overline N /L)^{1/2}$, we expect $\braket{\phi(x,t)}$ to decay to zero over time as a consequence of the finite energy density injected by the quench, which leads to restoration of the initially broken U(1) symmetry \cite{essler_quench_2016,collura2020order}. This is the same mechanism behind the decay of $\braket{\sigma_j^x(t)}$ in TFIC. Details of the MC scheme for sampling the DS representation \eqref{eq:quench} and the QA representation \eqref{eq:quenchQA} in the LL quench are given in Appendix \hyperlink{target:appendixA}{A}. 

\begin{figure*}[!t]
     \centering
     \begin{minipage}[b]{0.29\textwidth}
         \centering
         \includegraphics[width=\textwidth]{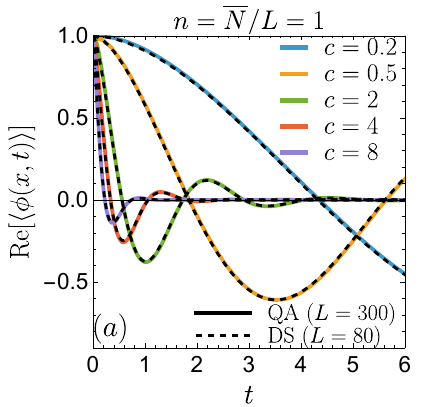}
     \end{minipage}
     \begin{minipage}[b]{0.29\textwidth}
         \centering
         \includegraphics[width=\textwidth]{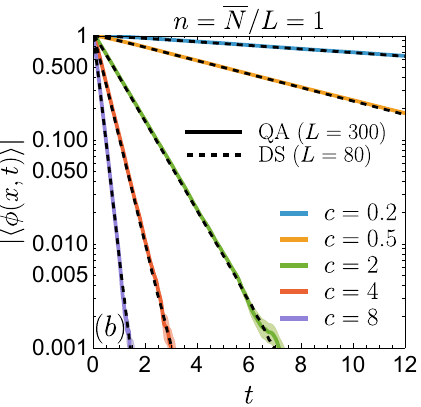}
     \end{minipage}
     \begin{minipage}[b]{0.29\textwidth}
         \centering
         \includegraphics[width=\textwidth]{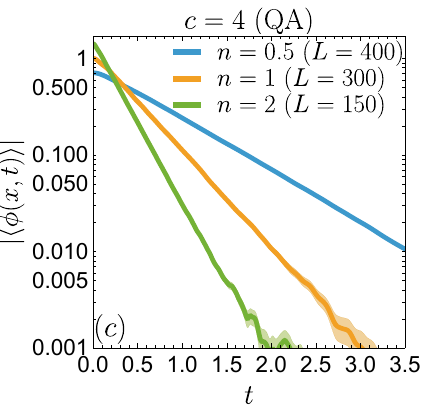}
     \end{minipage}
     \vspace{-10pt}
     \caption{MC predictions from QA \eqref{eq:quenchQA} and DS \eqref{eq:quench} for the dynamics of $\braket{\phi(x,t)}$ in quenches from the BEC initial state (\ref{eq:GSfreetheory}) for a range of interaction strengths $c$ and average densities $n = \overline N / L$. The MC data are obtained by running 100-200 parallel Markov chains with $\ell_{\rm max}=10^6$-$10^7$. The shaded bands around the lines indicate uncertainty, as given by the standard error on the mean of all parallel Markov chains run. The curves shown are converged with $L$ (see Appendix C). (a) Linear scale plot for ${\rm Re}[\braket{\phi(x,t)}]$ at fixed density $n = 1$ and varying $c$. (b) Log scale plot for $|\braket{\phi(x,t)}|$ at fixed density $n = 1$ and varying $c$. (c) Log scale plot for $|\braket{\phi(x,t)}|$ from QA, at fixed final interaction $c$ and varying density $n$.}
     \label{fig:n_c_plots}
\end{figure*}
\emph{Benchmarks at small $c$ and short times. \ }To obtain a benchmark for our MC results we note that in the initial state  \eqref{eq:GSfreetheory} all $k$-point cumulants of $\phi(x)$ and $\phi^\dag(x)$ vanish for $k\ge2$. This makes it possible to quantitatively describe the short-time dynamics for weak interactions by truncations of the Bogoliubov–Born–Green–Kirkwood–Yvon (BBGKY) hierarchy \cite{martin_theory_1959, bonitz_quantum_2016}. We have derived a truncation scheme that retains cumulants with $k\leq 3$, see Appendix \hyperlink{target:appendixB}{B}. Given the singular nature of the momentum distribution function in our initial state, truncations of the BBGKY hierarchy are expected to capture early-time dynamics better than classical-field approaches \cite{blakie_dynamics_2008, polkovnikov_phase_2010}.
In \cref{fig:BBGKY_c_p2} we compare the predictions of MC sampling DS \eqref{eq:quench} and QA \eqref{eq:quenchQA} with those of BBGKY truncations up to order 3, for a shallow quench with $\hat H_{\rm LL}(c=0.2)$ and average density $n = \overline{N}/L=1$ (results for $c = 0.5$ are given in the SM). The first and second order truncations correspond respectively to the Gross-Pitaevskii equation \cite{pitaevskii_vortex_1961, gross_hydrodynamics_1963} and to self-consistent time-dependent mean-field theory (SCTDMFT) \cite{chang_quantum_1975, boyanovsky_evolution_1998, sotiriadis_quantum_2010,van_Nieuwkerk_2019,vannieuwkerk2021Josephson,robertson2023simple,senese_outofequilibrium_2024}. To our knowledge, third-order BBGKY truncations remain largely unexplored in the literature, \emph{cf.}~\cite{kronke_bornbogoliubovgreenkirkwoodyvon_2018}. Our first observation is that the MC results for DS and QA are in perfect agreement. This provides a non-trivial check for the validity of the QA approach in interacting theories at finite times. Our second observation is that BBGKY truncations are in excellent agreement with MC data at early times, and that the time window over which the result agree well grows with the order of the truncation. This
provides a non-trivial benchmark for the MC approach. The failure of the $r=2,3$ truncations to capture the late-time relaxation to zero of $\braket{\phi(x,t)}$ (see SM) showcases the intrinsic difficulty in describing quench dynamics from a BEC in LL. As a final benchmark we have considered the dynamics of $\braket{\psi(t)|\phi(x)\phi(0)|\psi(t)}$ in the limit $c=\infty$, where we can compare our MC method to exact results from Ref.~\cite{denardis_analytical_2014}. We again find excellent agreement (see SM).

\emph{Results for repulsive LL. \ }In \cref{fig:n_c_plots} we report MC results for various post-quench interactions $c$ and initial densities $n = \overline{N}/L$. The curves are converged with respect to the system size $L$, see Appendix \hyperlink{target:appendixC}{C}. In \cref{fig:n_c_plots}(a) and (b) the density is fixed at $n = 1$ and the interaction $c$ is varied from weak ($c = 0.2$) to strong ($c = 8$) coupling. We again observe that QA \eqref{eq:quenchQA} and DS \eqref{eq:quench} yield identical results, providing a meaningful check of the validity of the MC results for both representations. Indeed, while QA is based on knowledge of the saddle-point state $\ket{\bs \lambda_{\rm sp}}$, DS has a priori no direct information on $\ket{\bs \lambda_{\rm sp}}$ because we initialize the DS Markov chain run in a state $\ket{\bs \lambda}$ that can be significantly different from $\ket{\bs \lambda_{\rm sp}}$ \footnote{Furthermore, while the MC sampling of QA operates at a fixed number of particles $N_{\rm sp}$ set by $\ket{\bs \lambda_{\rm sp}}$, 
$N$ varies significantly over the course of the Markov chain run for DS due to the variations in $N$ present in the initial state \eqref{eq:GSfreetheory} (see Appendix A).}. The necessity of reaching $L$ of the order of $200$-$300$ to remove finite-size effects in QA (see Appendix \hyperlink{target:appendixC}{C}), compared to $L$ of the order of $50$-$80$ for DS, simply reflects the large $L$ values required for the saddle-point arguments behind QA \cite{caux_time_2013, caux_quench_2016} to apply. 
In \cref{fig:n_c_plots}(c) we fix the post-quench interaction $c$ and vary the initial density $n$, with larger $n$ indicating an initially higher degree of symmetry breaking. We find that the initial states with higher symmetry breaking restore the symmetry more quickly. This is an instance of the quantum Mpemba effect \cite{ares_entanglement_2023, ares_quantum_2025}, which was predicted to occur in the present quench setup in Ref.~\cite{rylands_microscopic_2024}. Beyond a short initial time all the curves in \cref{fig:n_c_plots} are well fitted by (see SM)
\begin{equation}
    \braket{\phi(x,t)}\propto  e^{-i \omega(c,n) t}e^{-t/\tau(c,n)} \ .
\end{equation}
The frequency $\omega(c,n)>0$ and decay time $\tau(c,n)>0$ are expected to have a non-trivial analytical dependence on $c$ and $n$, \emph{cf.}~the case of $\braket{\sigma^x_j(t)}$ in TFIC from Ref.~\cite{calabrese_quantum_2012}. 

\begin{figure}[!b]
     \centering
     \begin{minipage}[b]{0.242\textwidth}
         \centering
         \includegraphics[width=\textwidth]{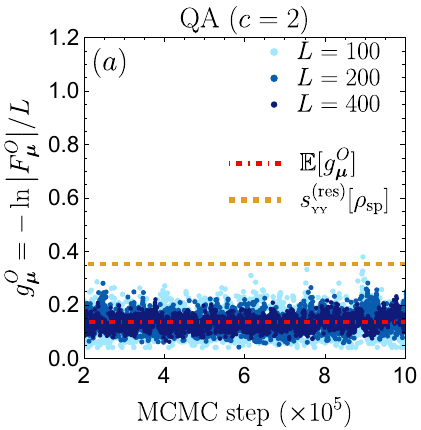}
     \end{minipage}
     \hspace{-7pt}
     \begin{minipage}[b]{0.242\textwidth}
         \centering
         \includegraphics[width=\textwidth]{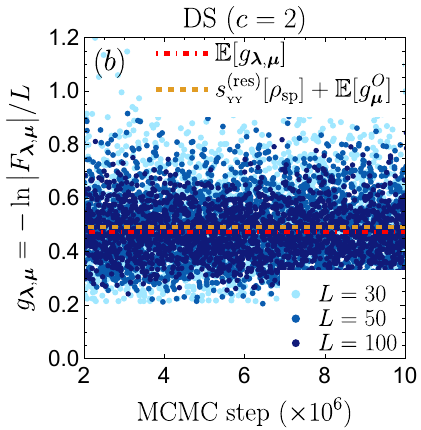}
     \end{minipage}
     \vspace{-20pt}
     \caption{Dots show QA and DS spectral weights \eqref{eq:g} sampled in a single Markov chain run in the LL quench for $c =2$, $n = 1$. $\mathbb E[\ldots]$ denotes average over the largest-$L$ dataset. (a) Results for QA. The average $\mathbb E[g_{\bs \mu}^{O}]$ is compared with the restricted Yang-Yang entropy density $s_\text{\tiny YY}^{\rm (res)}[\rho_{\rm sp}]$ of the saddle-point QA macrostate (see SM). (b) Results for DS, where the average $\mathbb E[g_{\bs \lambda,\bs \mu}]$ is compared with $s_\text{\tiny YY}^{\rm (res)}[\rho_{\rm sp}]+\mathbb E[g_{\bs \mu}^{O}]$.}
     \label{fig:weights_c_2}
\end{figure}

\noindent
The MC method provides very useful information on the structure of the spectral sum \cite{senese_finite_2026}. As in interacting theories the relevant $|F_{\curlb j}|$ are exponentially small in $L$ \cite{essler_statistics_2024, senese_finite_2026}, we define
\begin{equation}
\label{eq:g}
    g_{\curlb j}=-\ln|F_{\curlb j}| / L \ .
\end{equation}
\cref{fig:weights_c_2} shows the stationary MC distributions of $g_{\curlb j}$ sampled in a single Markov chain run for both QA and DS, in the LL quench with $c = 2, n =1$ (see SM for similar plots at $c = 8$). We observe that the variance of the sampled $g_{\curlb j}$ decreases (as a power law, see Appendix \hyperlink{target:appendixD}{D}) with increasing $L$. This is compatible with our expectation (Appendix \hyperlink{target:appendixD}{D}) that for very large $L$ all sampled $g_{\curlb j}$ are narrowly peaked around a single value $g^*$. \cref{fig:weights_c_2}(a) for QA shows that $g^*_{\rm QA}<s_\text{\tiny YY}^{(\rm res)}[\rho_{\rm sp}]$, where $s_\text{\tiny YY}^{(\rm res)}[\rho_{\rm sp}]$ is the \emph{restricted} Yang-Yang entropy density of the QA saddle-point macrostate (see SM). This was expected based on earlier results of \cite{essler_statistics_2024, senese_finite_2026}. Under the assumptions underlying QA, one also expects $g^*_{\rm DS} = g^*_{\rm QA}+s_\text{\tiny YY}^{(\rm res)}[\rho_{\rm sp}] $. Crucially, we recover this relation exceptionally well in \cref{fig:weights_c_2}(b), despite the use of data from a single Markov chain run and the moderate values of $L$ used for DS.

\emph{Sign problem. \ } When the normalization \eqref{eq:thetaAndZ} grows as $Z\sim e^{\nu L}$ ($\nu >0$), it becomes impossible to determine correlators $C(\bs s)$ via \eqref{eq:sampscheme}, even at moderate values of $L$. This is because the MC output has to estimate quantities that scale as $\sim e^{-\nu L}$ (given that $C(\bs s)= \mathcal{O}(L^0)$). Given the statistical uncertainty of the sampling scheme \eqref{eq:sampscheme} \cite{rubinstein_simulation_2016, gilks_markov_1995}, this leads to an exponentially decaying signal-to-noise ratio, requiring an exponentially large number of samples to obtain a statistically significant result \footnote{This difficulty is similar (although here it has a different origin) to the one arising from the sign problem that plagues fermionic quantum Monte Carlo simulations, which has been proved to be NP hard \cite{troyer_computational_2005}.}. 
A sign problem arises whenever the spectral phases $e^{i \theta_{\curlb j}}$ in \eqref{eq:Csampling} are highly fluctuating, resulting in strong cancellations in the sum over $\curlb j$. 
We stress that the phases $\alpha_{\curlb j}(\bs s)$ from the space-time evolution do not cause a sign problem, because for finite $\bs s$ they vary slowly when  $\curlb j_\ell \to \curlb j_{\ell + 1}$. 
Our present understanding of the sign problem is as follows:

(i) A sign problem \emph{never} arises for $2$-point functions (of the same operator) in equilibrium \cite{senese_finite_2026}, because $F_{\curlb j}=|F_{\curlb j}|$ and hence $Z = C(\bs 0)=\mathcal{O}(L^0)$; 
(ii) A sign problem emerges for higher-order correlators \eqref{eq:kpf} in equilibrium for $r> 2$ (see SM for supporting evidence);
(iii) A sign problem can emerge in global quenches (see SM), but is absent for quenches from integrable initial states.

\emph{Discussion.} \ We have introduced a MC method for computing Lehmann representations of correlation functions in interacting quantum integrable models and applied it to the dynamics after global quantum quenches.
Our results demonstrate that reconstructing interacting quench dynamics from the paradigmatic class of integrable initial states \cite{piroli_what_2017} is a low-complexity task, provided exact form factors and overlaps are known from integrability. In more general situations the method faces an intrinsic “sign problem”, and it would be interesting to identify other physical scenarios that might evade it. Crucial next steps would be to extend our method to a broader class of observables, initial states, and integrable models, particularly those hosting bound states \cite{korepin_quantum_1993, takahashi_thermodynamics_1999, essler_onedimensional_2005}, and to explore whether progress is possible even when analytical expressions for the overlaps are not available \cite{robinson_computing_2021}.

\begingroup
\renewcommand{\addcontentsline}[3]{}

\begin{acknowledgments}
\textit{Acknowledgments.---}%
This work was supported by the EPSRC under grant EP/X030881/1 (FHLE) and European Research Council (ERC) Advanced Grant MOSE (No.~101199196) (RS). R.S. thanks Pasquale Calabrese and Alvise Bastianello for useful discussions.
\end{acknowledgments}

\clearpage

\onecolumngrid
\begin{center}
    \vspace*{\fill}
    {\Large \textbf{End Matter}}
    \vspace*{\fill}
\end{center}
\twocolumngrid

\noindent
\hypertarget{target:appendixA}{\textbf{Appendix A: MC sampling Lehmann representations in integrable models}}

In LL each eigenstate $\ket{\bs \lambda}$ is in one-to-one correspondence with a set $\bs I=\{I_1, I_2, \ldots\}$ of \emph{distinct} (half-odd) integers. While the rapidities $\bs \lambda$ satisfy a non-trivial set of quantization conditions (the Bethe equations, see SM), the sets $\bs I$ can be chosen freely within a given range. Hence sampling over sets of eigenstates in \eqref{eq:sampscheme} corresponds to sampling sets of (half-odd) integers $\curlbI j$, where $j$ runs over the independent sums over eigenstates in the Lehmann representation. 
At each MC step $\ell$ we pass from integers 
$\{ {\bs I}^{(j)}\}_\ell$ to rapidities $\curlb j_\ell$
by numerically solving the Bethe equations by a Newton-Raphson method. In TFIC passing from (half-odd) integers $\bs I$ to momenta $\bs \lambda$ simply amounts to $\bs \lambda = 2\pi \bs I/L$. The MC scheme follows from a standard application of the Metropolis-Hastings algorithm \cite{metropolis_equation_1953, hastings_monte_1970} (see Appendix B of Ref.~\cite{senese_finite_2026} for a brief review of MC sampling relevant to the present context). (i) Starting from $\curlbI j_\ell\equiv y$ we propose a move to new sets $\curlbI j'\equiv y'$
with probability $P_{y \to y'}$. (ii) We accept the move with Metropolis-Hastings probability
\begin{align}
\label{eq:alphaMHdef}
\alpha_{\rm MH}=\min \left[\frac{\mathcal P(\curlb j')}{\mathcal P(\curlb j_\ell)}\frac{P_{y' \to y}}{P_{y \to y'}},1\right] \ ,
\end{align}
and reject it with probability $1 -\alpha_{\rm MH}$. Here $\mathcal P(\curlb j)$ is defined below \cref{eq:Csampling} and $P_{y' \to y}$ denotes the probability of proposing the inverse move $\curlbI j'\to\curlbI j_\ell$. (iii) In case of acceptance we set $\curlbI j_{\ell +1}=\curlbI j'$, otherwise $\curlbI j_{\ell +1}=\curlbI j_\ell$. 
We stress that in \eqref{eq:alphaMHdef} the normalization $Z$ from \eqref{eq:thetaAndZ} cancels, hence its knowledge is not required. By iterating the MC step many times one obtains a collection of $\curlb j_\ell$ (approximately) sampled according to $\mathcal P(\curlb j)$ \cite{rubinstein_simulation_2016, gilks_markov_1995}, as needed in \eqref{eq:sampscheme}. 

We always perform the following checks on the MC algorithms implemented: (i) the proposal scheme must be chosen such that $\alpha_{\rm MH}$ is sufficiently high (e.g. $\gtrsim10\%$), to ensure efficient sampling; (ii) it must be verified that the distribution of sampled $\curlb j_\ell$ appears stationary at large $\ell$ and it does not depend on the initial choice $\curlb j_1$; 
(iii) in the case of a field theory (like LL) convergence with respect to the UV cutoff (the maximal allowed (half-odd) integers in the sampling) must be ensured; (iv) the statistical uncertainties on the final values of the dynamical observables are estimated as standard errors on the mean of several Markov chains run in parallel up to large $\ell$. For details on the previous points, see SM.

We have specialized the general scheme to sampling in TFIC and LL by means of single-integer updates. Given the parity-invariance of all eigenstates relevant for quenches from integrable initial states, the sets $\curlbI j$ can be restricted to include only positive (half-odd) integers. 

\emph{QA for TFIC}. \ According to \cref{eq:quenchQA} the sampling is performed over a single set $\bs I$. At each MC step $\ell$ we propose to update a single integer in $\bs I_\ell$ according to three possibilities: adding an integer to $\bs I_\ell$; removing an integer from $\bs I_\ell$; replacing an occupied integer in $\bs I_\ell$ with an unoccupied one (a particle-hole move). Full details are given in the SM. The presence of moves that change the number of elements in $\bs I$ reflects the fact that the order parameter $\sigma^x_j$ can connect eigenstates with different numbers of Bogoliubov excitations. 

\emph{QA for LL.} \ The order parameter $\phi(x)$ can only connect eigenstates whose numbers of particles differ by 1, which implies that all the sampled $\bs I_\ell$ possess the same number of elements, see \eqref{eq:quenchQA}. The only moves we need to propose are therefore particle-hole moves (replacing an occupied integer with an unoccupied one). The proposal scheme is identical to the one of Ref.~\cite{senese_finite_2026}, where by construction $P_{y \to y'}=P_{y' \to y}$ and therefore \eqref{eq:alphaMHdef} simplifies. For details on the computation of $\mathcal P(\bs \lambda)$ see SM. 

\emph{DS for LL.} \ According to \eqref{eq:quench} we have to sample over two sets of (half-odd) integers $\bs J$ and $\bs I$, where the number of elements in $\bs I$ are always one less than in $\bs J$. Due to the particle number fluctuations in the initial BEC state \eqref{eq:GSfreetheory}, the possible move proposals are: adding one unoccupied integer to $\bs J$ and one to $\bs I$; removing one occupied integer from $\bs J$ and one from $\bs I$; performing a particle-hole move either in $\bs J$ or in $\bs I$. Further details are given in the SM. \\

\begin{figure*}[!t]
     \centering
     \begin{minipage}[b]{0.35\textwidth}
         \centering
         \includegraphics[width=\textwidth]{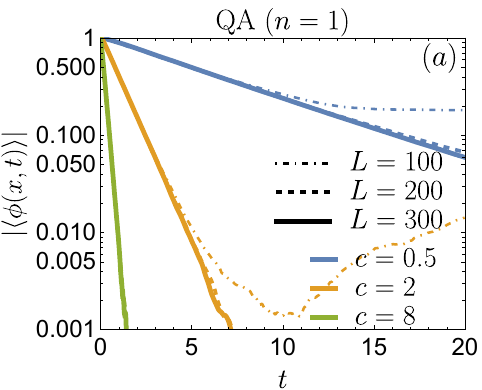}
     \end{minipage}
     \hspace{10pt}
     \begin{minipage}[b]{0.35\textwidth}
         \centering
         \includegraphics[width=\textwidth]{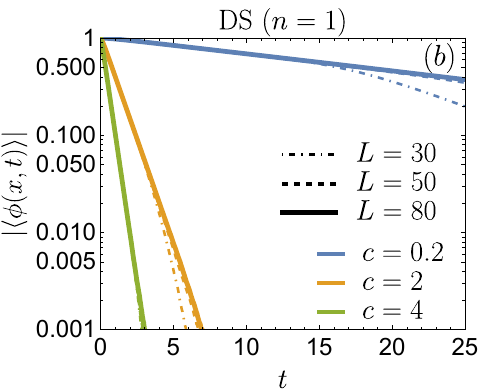}
     \end{minipage}
     \vspace{-10pt}
     \caption{Log-scale plots showing convergence with $L$ of the MC results for $|\braket{\phi(x,t)}|$ in the LL quench described in the main text for several interaction strengths $c$ and $n = \overline N / L = 1$. Uncertainties in the MC data are not reported here, only in \cref{fig:n_c_plots}. (a) QA results for $L = 100,200,300$. (b) DS results for $L = 30,50,80$.}
     \label{fig:finite_size}
\end{figure*}
\noindent
\hypertarget{target:appendixB}{\textbf{Appendix B: Truncations of the BBGKY hierarchy}}

We start from the exact Heisenberg equations of motions relevant for the description of $\braket{\phi(x,t)}$. The first three levels of the BBGKY hierarchy read schematically
\begin{align}
\label{eq:BBGKYeoms}
    &\frac{d}{dt}\braket{\tilde\phi_k(t)}= i\braket{e^{i\hat H_{\rm LL}t} [\hat H_{\rm LL},\tilde \phi_k]e^{-i\hat H_{\rm LL}t}}=\gamma_{k,\phi}^{(3)}(t) \ ,\nonumber\\
&\frac{d}{dt}\braket{\tilde\phi^\dag_k(t)\tilde \phi_k(t)} = \gamma_{k,\phi^\dag \phi}^{(4)}(t) \ , \nonumber\\
&\frac{d}{dt}\braket{\tilde\phi_k(t)\tilde \phi_{-k}(t)} = \gamma_{k,\phi \phi}^{(4)}(t) \ ,\nonumber\\
&\frac{d}{dt}\braket{\tilde\phi^\dagger_{k+p}(t)\tilde\phi_k(t)\tilde \phi_{p}(t)} = \gamma_{k,p,\phi^\dagger\phi \phi}^{(5)}(t) \ ,\nonumber\\
&\frac{d}{dt}\braket{\tilde\phi_{k+p}(t)\tilde\phi_{-k}(t)\tilde \phi_{-p}(t)} = \gamma_{k,p,\phi\phi \phi}^{(5)}(t) \ .
\end{align}
Here $\gamma_{\bs k,O}^{(n)}$ denotes a linear combination of normal-ordered equal-time $q$-point functions of $\tilde \phi_k$, $\tilde \phi_k^\dag$ with $q \le n$, and $\tilde \phi_k = \int_0^L dx \, e^{ikx}\phi(x)/\sqrt{L}$.  
To close the system (\ref{eq:BBGKYeoms}) one considers order-$r$ truncations that neglect all $q$-point \emph{cumulants} with $q > r$ in the terms $\gamma_{\bs k,O}^{(n\geq 3)}$. Given that all cumulants higher than $r = 1$ identically vanish in the initial state \eqref{eq:GSfreetheory}, such approximations are expected to yield accurate predictions for dynamics at early times (see, e.g., Refs.~\cite{bertini_prethermalization_2015, bertini_thermalization_2016, robertson_decay_2024, robertson2023simple, senese_outofequilibrium_2024}). In the SM we report the truncated hierarchy at order $r = 1$ (Gross-Pitaevskii), $r = 2$ (SCTDMFT) and $r = 3$. The curves in \cref{fig:BBGKY_c_p2} are obtained by numerical integration of these (now closed) sets of ODEs. \\

\noindent
\hypertarget{target:appendixC}{\textbf{Appendix C: Finite-size effects}}

As we are ultimately interested in thermodynamic limit results it is important to check finite-size effects in our MC scheme. In the case of QA, the limit of large $L$ is also a crucial assumption behind the validity of the method. \cref{fig:finite_size} shows MC results for the LL quench at various interaction strengths $c$ and for increasing system sizes $L$ (at $n=\overline N / L = 1$).  The convergence of the curves both for QA and DS demonstrates that the results of \cref{fig:n_c_plots} are representative of the thermodynamic limit. The comparison between QA and DS also showcases that finite-size effects are more pronounced in QA than in DS (as expected), given the larger $L$ values needed to obtain convergence.\\

\noindent
\hypertarget{target:appendixD}{\textbf{Appendix D: Saddle-point for $F$-resolved entropy}}\\
It is useful to define the $F$-resolved entropy density $\sigma(g)$
\begin{equation}
\begin{aligned}
    \exp[L \sigma(g)] &\equiv \sum_{\curlb j} \delta(g - g_{\curlb j}) \ ,\\
    |F_{\curlb j}| &= \exp\big[- L g_{\curlb j}\big] \ .
\end{aligned}
\end{equation}
In the limit of asymptotically large $L$, $\sigma(g)$ becomes a smooth function of $g$. 
If we consider the general representation \eqref{eq:Csampling}, and focus on $\bs s = \bs 0$ and the simplest case $F_{\curlb j}=|F_{\curlb j}|$ (which implies absence of a sign problem, as in LL), we get
\begin{equation}
\label{eq:integralsw}
    \mathcal{O}(L^0) = C(\bs 0) = \int dg \, \exp\big[L\big(\sigma(g) - g\big)\big]  \ .
\end{equation}
The previous equation can hold for $L \to \infty$ only if $\lim_{L \to \infty}\big(\sigma(g) - g\big)\le 0$ and there is (at least) one saddle point $g^*$ such that
\begin{equation}
\label{eq:saddlepointg}
    \sigma(g^*) - g^* = \mathcal{O}(1/L) \xrightarrow[]{L \ \to \ \infty} 0 \ .
\end{equation}
This then tells us that in the large-$L$ limit only $\exp(g^* L)$ terms in the Lehmann representation make non-negligible contribution to $C(\bs 0)$, and by extension to $C(\bs s)$ for finite $\bs s$. \cref{fig:weights_c_2} shows that in the LL quench the distribution sampled by MC settles around the saddle point value $g^*$, which is hard to determine analytically. 
\begin{figure}[!b]
     \centering
     \begin{minipage}[b]{0.238\textwidth}
         \centering
         \includegraphics[width=\textwidth]{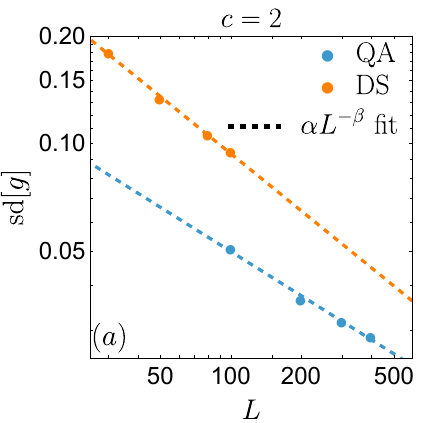}
     \end{minipage}
     \hspace{-3pt}
     \begin{minipage}[b]{0.238\textwidth}
         \centering
         \includegraphics[width=\textwidth]{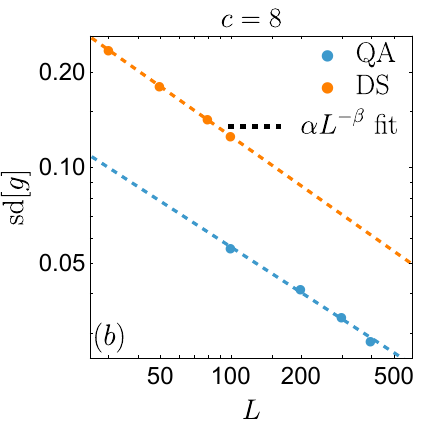}
     \end{minipage}
     \vspace{-20pt}
     \caption{Log-log plots showing the decay with $L$ of the standard deviations $\text{sd}[g]$ for the $g$ values (\cref{eq:g}) sampled in the LL quench with $c = 2,8$. Data are shown for both QA and DS. $\text{sd}[g]$ is obtained by including points from 10 Markov chain runs. The dashed lines indicate power-law fits $\alpha L^{-\beta}$. (a) c = 2, with fitting parameters $\beta \approx 0.41$ (QA) and $\beta \approx0.53$ (DS). (b) c = 8, with $\beta \approx 0.48$ (QA) and $\beta \approx0.52$ (DS).}
     \label{fig:spectral_sd}
\end{figure}
The saddle point argument suggests that the variance of the sampled $g$ values scales as $L^{-1}$. This is already visible in \cref{fig:weights_c_2} (see also $c = 8$ plot in SM). In \cref{fig:spectral_sd} we show that the MC standard deviations $\text{sd}[g]$ for the LL quench in fact decay as power laws $\alpha L^{-\beta}$, with $\beta \approx 0.5 \pm 0.1$. 

\newpage
\putbib[references,outputNotes]
\endgroup
\end{bibunit}

\clearpage
\onecolumngrid

\begin{center}
  \textbf{\large Supplemental Material}\\
    Riccardo Senese and Fabian H.L. Essler
\end{center}

\addtocontents{toc}{\protect\setcounter{tocdepth}{3}}
\tableofcontents
\vspace{2ex}

\begin{bibunit}

\makeatletter
\renewcommand{\hyper@natanchorstart}[1]{\Hy@raisedlink{\hyper@anchorstart{cite.SM.#1}}}
\renewcommand{\hyper@natlinkstart}[1]{\hyper@linkstart{cite}{cite.SM.#1}}
\makeatother

\setlength{\parindent}{0pt}

\setcounter{secnumdepth}{3}

\setcounter{section}{0}
\setcounter{equation}{0}
\setcounter{figure}{0}
\setcounter{table}{0}
\setcounter{page}{1}

\renewcommand{\theequation}{S\arabic{equation}}
\renewcommand{\thefigure}{S\arabic{figure}}
\renewcommand{\thetable}{S\arabic{table}}
\renewcommand{\thesubsection}{\thesection.\Alph{subsection}}

\renewcommand{\theHequation}{S\arabic{equation}}
\renewcommand{\theHfigure}{S\arabic{figure}}
\renewcommand{\theHtable}{S\arabic{table}}
\renewcommand{\theHsection}{\Alph{section}}
\renewcommand{\theHsubsection}{\theHsection.\Alph{subsection}}

\makeatletter
\renewcommand{\p@subsection}{}
\makeatother


\thispagestyle{empty}

\section{Brief review of basic concepts: macrostates, Yang-Yang entropy and QA saddle-point}
\label{sec:briefreview}

Eigenstates $\ket{\bs \lambda}$ of integrable Hamiltonians are labelled by sets of rapidities $\bs \lambda = \{\lambda_1, \lambda_2, \ldots\}$, with $\lambda_j \in \mathbb{C}$ in general \cite{korepin_quantum_1993, takahashi_thermodynamics_1999}. In free theories like the TFIC of \cref{eq:TFICh} (main text) one has $\lambda_j = 2 \pi I_j / L$, where $\{I_j\}$ are a set of (half-odd) integers (see \cref{sec:diagonTFIC} below). In the LL model with repulsive interactions $c > 0$ of \cref{eq:H_LL} (main text) the rapidities $\bs \lambda$ are all real \cite{korepin_quantum_1993} and in the sector with $N$ bosons, each $\bs \lambda=\{\lambda_1, \ldots, \lambda_N\}$ is uniquely determined by a set of $N$ distinct (half-odd integers) integers $\{I_j\}$ for $N$ (even) odd, as the solution of the following Bethe equations \cite{korepin_quantum_1993}
\begin{equation}
\label{eq:logBetheEq}
    L \lambda_j + \sum_{k=1}^N \theta(\lambda_j - \lambda_k) = 2 \pi I_j \qquad j = 1, \ldots, N \  .
\end{equation}
Here $\theta(x)=2 \arctan(x/c)$. The important difference compared to free models is that each $\lambda_j$ now depends on the full set $\bs \lambda$ via \eqref{eq:logBetheEq}.

In the limit of $L \to \infty$ it is convenient to consider entire classes of eigenstates, known as \emph{macrostates}, that share the same distribution $\rho(\lambda)$ of rapidities. In TFIC and LL $\rho(\lambda)$ is defined as
\begin{equation}
\big\lfloor L \rho(\lambda) d\lambda\big\rfloor = \text{number of } \lambda_j \text{\ in} \ [\lambda, \lambda + d \lambda] \ .
\end{equation}
Importantly, all eigenstates belonging to a same macrostate $\rho(\lambda)$ have (up to finite-size corrections) identical local properties, e.g.~expectation values of local observables \footnote{This represents the integrable analogue of the diagonal ETH.}\cite{korepin_quantum_1993}. Given a macrostate $\rho(\lambda)$ one can define the Yang-Yang entropy density \cite{yang_thermodynamics_1969, korepin_quantum_1993} as 
\begin{equation}
s_\text{\tiny YY}[\rho] = \lim_{L \to \infty}\frac{\ln\big(\text{number of eigenstates $\in$ }\rho(\lambda)\big)}{L} \, .
\end{equation}
For example, in LL
{\allowdisplaybreaks
\begin{align}
\label{eq:yangyangLL}
    s_{\text{\tiny YY}}[\rho]&=\int_{-\infty}^{\infty}d\lambda \, \bigg[\rho(\lambda)\log \frac{\rho_t(\lambda)}{\rho(\lambda)}+\rho_h(\lambda)\log \frac{\rho_t(\lambda)}{\rho_h(\lambda)} \bigg] \ , \nonumber \\
    \rho_t(\lambda)&=\frac{1}{2\pi}\left[1+\int_{-\infty}^\infty d \mu \, K(\lambda-\mu)\rho(\mu)\right]  \ ,
\end{align}
}
where $\rho_h(\lambda) = \rho_t(\lambda) - \rho(\lambda)$ (hole density) and $K(\lambda)= \frac{2c}{c^2+\lambda^2}$. 
By maximizing $s_\text{\tiny YY}[\rho]$ under the constraints of fixed energy and particle densities, one obtains the saddle-point thermal macrostate $\rho_{\rm th}(\lambda)$ that determines Gibbs averages at the corresponding inverse temperatures $\beta$ and chemical potential $h$. Generalizations to GGEs are conceptually straightforward \cite{mossel_generalized_2012}. 
QA generalizes such saddle-point techniques to quench dynamics from an initial state $\ket{\psi}$ \cite{caux_time_2013, caux_quench_2016}. In particular, the QA method states that the representative state $\ket{\bs \lambda_{\rm sp}}$ in \eqref{eq:quenchQA} (main text) belongs to the macrostate $\rho_{\rm sp}(\lambda)$ that dominates in
\begin{equation}
\label{eq:QAoverlapeq}
\begin{aligned}
\braket{\psi|\psi}&=\sum_{\bs \lambda} |\braket{\bs \lambda|\psi}|^2 \quad \xrightarrow[]{\ L \gg 1 \ } \quad \sim \int D[\rho]\exp \Big[L \big(s^{\rm (res)}_\text{\tiny YY}[\rho] - s_{\psi}[\rho]\big) \Big] \ .
\end{aligned}
\end{equation} 
Here $s^{(\rm res)}_\text{\tiny YY}[\rho]$ is the Yang-Yang entropy density \emph{restricted} to count only eigenstates that have a nonzero overlap with $\ket \psi$, and we have defined $s_\psi(\bs \lambda) = -  \ln(|\braket{\bs \lambda|\psi}|^2)/L$ and assumed that to leading order in $L$ it becomes only a function of macrostate information $\rho(\lambda)$ \cite{caux_time_2013, brockmann_gaudinlike_2014, caux_quench_2016, denardis_solution_2014, pozsgay_correlations_2014,wouters_quenching_2014, bertini_quantum_2017, bertini_entanglement_2018}. Then $\rho_{\rm sp}(\lambda)$ is the solution of the maximization $\delta(s^{(\rm res)}_\text{\tiny YY}[\rho] - s_{\psi}[\rho])/\delta \rho = 0$. The saddle-point distribution $\rho_{\rm sp}(\lambda)$ dominates the functional integral \eqref{eq:QAoverlapeq}, and satisfies $s^{(\rm res)}_\text{\tiny YY}[\rho_{\rm sp}] - s_{\psi}[\rho_{\rm sp}]=0$ because $\ket \psi$ is normalized. At any finite $L$, $\ket{\blambda_{\rm sp}}$ from \cref{eq:quenchQA} (main text) can be constructed by requiring the finite-$N$ distribution of its rapidities to smoothly overlap with the profile of $\rho_{\rm sp}(\lambda)$ \cite{essler_statistics_2024}. \\

\section{Sign problem for higher-point correlation functions at finite temperature}

In this section we argue analytically and numerically that higher-point functions ($r > 2$) in equilibrium
\begin{equation}
\label{eq:nptdyn}
    C_r(\bs x, \bs t)={\rm Tr}[\hat \rho \, O(x_1,t_1)O(x_2,t_2) \ldots O(x_r,t_r)] 
\end{equation}
cannot be accessed via the MC scheme discussed in the main text, because of the emergence of a “sign problem”. Here $\hat \rho$ indicates, e.g., a Gibbs ensemble at inverse finite temperature $\beta$ or a GGE with nonzero entropy density, and we have assumed that aside for space and time translations all operators in \eqref{eq:nptdyn} are the same (our conclusions are not expected to change for different operators). The correlations \eqref{eq:nptdyn} are the central objects entering non-linear response theory \cite{kubo_statisticalmechanical_1957}, and have been used as probes of operator spreading and information scrambling \cite{maldacena_bound_2016, nahum_operator_2018, vonkeyserlingk_operator_2018, khemani_operator_2018, rakovszky_diffusive_2018, chan_solution_2018, parker_universal_2019, murthy_bounds_2019, foini_eigenstate_2019}.

To showcase the origin of a sign problem we focus on $r = 3$ in \eqref{eq:nptdyn}, the case $r>3$ is analogous. Our considerations apply to both lattice and continuum models with homogeneous $\hat\rho$ and $H$ (Hamiltonian). We discuss both quantum chaotic and integrable systems. In the integrable case we focus on interacting models, or free models that have local operators $O$ with exponentially decaying (in system size) form factors, which in this sense mimic the interacting case for generic observables, see e.g.~Ref.~\cite{essler_statistics_2024} or the case of TFIC in \cref{sec:diagonTFIC}. 

In the thermodynamic limit $L \to \infty$, the product of any finite number of $O(x_i,t_i)$ for fixed, finite $x_i$ and $t_i$ represents a local observable \footnote{Despite the absence of a strict, state-independent Lieb-Robinson bound \cite{lieb_finite_1972} for operator spreading in theories with unbounded spectra and momenta, the restriction to a finite energy density—imposed by the state $\hat\rho$ in \eqref{eq:nptdyn}—acts as an effective ultraviolet regulator. Consequently, the high-energy modes are suppressed, and contributions to expectation values of products of $O(x_i,t_i)$ from arbitrarily far tails are exponentially suppressed (see also \cite{lemm_liebrobinson_2026} and references therein).} \cite{lieb_finite_1972, bravyi_liebrobinson_2006}. Given that eigenstates of quantum chaotic (integrable) systems at the same energy density (in the same macrostate) have identical expectation values for local operators it follows that
\begin{equation}
\label{eq:4pultamicro}
    \lim_{L \to \infty} C_3(\bs x,\bs t) = \lim_{L \to \infty} \braket{n|O(x_1,t_1)O(x_2,t_2)O(x_3,t_3)|n} \ ,
\end{equation}
where $\ket{n}$ can be chosen to be \emph{any} representative eigenstate at the energy density (in the macrostate) fixed by $\hat \rho$ \cite{korepin_quantum_1993, dalessio_quantum_2016}. Expanding \eqref{eq:4pultamicro} in a Lehmann sum, and dropping the “$\lim_{L \to \infty}$” from hereon (it remains implicit throughout), we obtain
\begin{equation}
\label{eq:C3expandedLehm}
    C_3(\bs x, \bs t) = \sum_{n_1,n_2} \alpha_{n,n_1,n_2}(\bs x, \bs t) \braket{n|O|n_1}\braket{n_1|O|n_2}\braket{n_2|O|n} \ ,
\end{equation}
where $\alpha_{n,n_1,n_2}(\bs x, \bs t)$ is the phase defined in  \cref{eq:alphas} (main text). 
We now restrict the sum $\sum_{n_1,n_2}$ to contain only terms for which $n_i \neq n \ \forall \ i$, $n_1\neq n_2$, and denote the corresponding correlator as $\widetilde{C}_3(\bs x, \bs t)$. This is the most interesting part of the 3-point function \eqref{eq:C3expandedLehm}, as all the terms we dropped represent lower-order cumulants \cite{foini_eigenstate_2019, pappalardi_eigenstate_2022}. 
Our MC strategy relies on the following estimate (cf.~\cref{eq:sampscheme} of main text)
\begin{equation}
\label{eq:MCMCestimateC3}
    \widetilde{C}_3(\bs x, \bs t) \ \to \ \widetilde{Z}_3 \,  \bigg(\frac{1}{\ell_\text{max}}\sum_{\ell = 1}^{\ell_\text{max}} \alpha_{n, \, n_1^{(\ell)},n_2^{(\ell)}}(\bs x, \bs t)\, e^{i \theta\big(n, \, n_1^{(\ell)},n_2^{(\ell)}\big)} \bigg)\ ,
\end{equation}
\begin{equation}
    \theta\big(n, \, n_1,n_2\big)=\arg\big[\braket{n|O|n_1}\braket{n_1|O|n_2}\braket{n_2|O|n} \big]  \ ,\qquad
\label{eq:pfZtilde}
   \widetilde{Z}_3 = \sum_{\substack{n_1,n_2 \\ (n_1 \neq n_2 \neq n)}}\big|\braket{n|O|n_1}\braket{n_1|O|n_2}\braket{n_2|O|n}\big| \ .
\end{equation}
The normalization $\widetilde{Z}_3$ is unphysical due the presence of the absolute value within the sum. We will now argue that $\widetilde{Z}_3$ grows exponentially with system size $L$, and hence gives rise to a sign problem. In the chaotic case, assuming validity of the ETH ansatz, our conclusion on the presence of a sign problem is definitive. In the integrable case the situation is more complex due to a richer structure in the statistics of matrix elements \cite{essler_statistics_2024, rottoli_eigenstate_2026, orlov_multiscale_2026}, but our arguments shed significant light on the origin of the sign problem, which we then verify numerically.
In both cases, the strongly oscillating phases that cause a sign problem are the ones associated with $e^{i \theta(n, \, n_1^{(\ell)},n_2^{(\ell)})}$ in \cref{eq:MCMCestimateC3}. \\

To understand the scaling of $\widetilde{Z}_3$ we start by focusing on the variance of $O$ in a generic eigenstate $\ket r$, which is a finite quantity given the assumption of locality of $O$
\begin{equation}
\label{eq:Ol0varr}
    \mathcal{O}(L^0) = {\rm Var}_{\ket r}[O] = \braket{r|O^2|r} - \braket{r|O|r}^2 = \sum_{s \, (\neq r)}|\braket{r|O|s}|^2 \ .
\end{equation}
We can now apply the same reasoning of Appendix D from the main text. From the fact that relevant off-diagonal matrix elements in interacting theories are exponentially suppressed in $L$ we define $g_{r,s}$ as
\begin{equation}
\label{eq:grsdef}
    |\braket{r|O|s}|^2 = e^{- g_{r,s} L} \ ,
\end{equation}
as well as the matrix-element-resolved entropy density $\sigma^{(r)}(g)$, cf.~Appendix D of main text
\begin{equation}
\label{eq:defsigmaentropy}
    e^{L \sigma^{(r)}(g)} = \sum_{s \, (\neq r)} \delta(g - g_{r,s}) \ .
\end{equation}
This counts the number of off-diagonal matrix-elements of magnitude $e^{-gL}$. In the thermodynamic limit $\sigma^{(r)}(g)$ becomes a smooth function of $g$, so that for large $L$
\begin{equation}
\label{eq:varianceOint}
    \mathcal{O}(L^0) = {\rm Var}_{\ket r}[O] = \int dg \, e^{L\big(\sigma^{(r)}(g) - g\big)}  \ .
\end{equation}
For this to remain finite in the $L\to\infty$ limit we must have
\begin{equation}
    \lim_{L \to \infty}\big(\sigma^{(r)}(g)-g\big) \le 0 \ \ \ \ \forall \ g \ .
\end{equation}
Furthermore, to converge to a finite nonzero $L$-independent value there must be (at least) one saddle point $g^*_r$ where
\begin{equation}
\label{eq:saddlepoint2point}
    \sigma^{(r)}(g^*_r) - g^*_r = \mathcal{O}(1/L) \xrightarrow[]{L \ \to \ \infty} 0 \ ,
\end{equation}
and which totally dominates the integral \eqref{eq:varianceOint}. Therefore the only (but still exponentially many) eigenstates $\ket s$ that matter for ${\rm Var}_{\ket r}[O]$ are those for which $-\log( |\braket{r|O|s}|^2)/L$ tends to $g^*_r$ for $L \to \infty$. In quantum chaotic models the ETH ansatz implies that $g^*_r = s_{\rm th}(e_r)$, where $s_{\rm th}(e)$ is the thermodynamic entropy density at energy density $e$, and $e_r$ is the energy density of the reference eigenstate $\ket r$ \eqref{eq:Ol0varr}. In other words, all eigenstates at the appropriate energy density contribute to \eqref{eq:varianceOint} for $L\to\infty$. In quantum integrable models the statistics of matrix elements has a richer structure \cite{essler_statistics_2024, rottoli_eigenstate_2026, orlov_multiscale_2026}. In \cite{senese_finite_2026} it was shown how $g^*_r$ can be estimated using MC sampling in the context of dynamical 2-point correlation functions in integrable models, and it was found that $0<g^*_r < s_\text{\tiny YY}[\rho_r]$ and $\rho_r(\lambda)$ denotes the macrostate $\ket r$ belongs to (cf.~\cref{sec:briefreview}). This means that in interacting quantum integrable models only a sub-entropic (yet exponentially large) number of matrix elements contributes to 2-point dynamical correlation functions, as expected from earlier results \cite{essler_statistics_2024}.

With the knowledge of the saddle point $g^*$ that governs 2-point functions, we now turn to the normalization $\widetilde{Z}_3$ of 3-point functions \eqref{eq:pfZtilde}. We have
\begin{equation}
    \widetilde{Z}_3 > \sum_{\substack{n_1 \in {\rm sp_2(n)}\\ (n_1 \neq n)}}|\braket{n|O|n_1}|  \sum_{\substack{n_2 \in {\rm sp_2(n_1)}\\ (n_2 \neq n_1 \neq n)}}|\braket{n_1|O|n_2}| |\braket{n_2|O|n}| \ ,
\end{equation}
where ${\rm sp}_2(r)$ denotes the $W_r =e^{L g^*_r}$ eigenstates that form the 2-point function saddle point \eqref{eq:saddlepoint2point} for ${\rm Var}_{\ket r}[O]$. Using \cref{eq:grsdef,eq:defsigmaentropy,eq:saddlepoint2point} and denoting  sub-exponential contributions as $(\ldots)$, we obtain
\begin{equation}
\label{eq:finalineqZ}
    \widetilde{Z}_3 > (\ldots)\,  e^{Lg^*_n/2} \bigg(\frac{1}{W_n}\sum_{\substack{n_1 \in {\rm sp_2(n)}\\ (n_1 \neq n)}} e^{L g_{n_1}^*/2} \bigg(\frac{1}{W_{n_1}}\sum_{\substack{n_2 \in {\rm sp_2(n_1)}\\ (n_2 \neq n_1 \neq n)}} |\braket{n_2|O|n}| \bigg) \bigg) \ .
\end{equation}
We now consider the implications of this lower bound.
\begin{itemize}
    \item Applying \eqref{eq:finalineqZ} to the case of quantum chaotic models (where ETH applies), we have in the limit of large $L$
        \begin{equation}
        \label{eq:keyeqQC}
            g_n^*=g_{n_1}^*=-2 \log\big(|\braket{n_2|O|n_1}|\big)/L =  s_{\rm th}(e_n)\ ,\qquad
            n_1 \in {\rm sp}_2(n), n_2 \in {\rm sp}_2(n_1).
        \end{equation}
This holds for almost all eigenstates at energy density $e_n$. We conclude that
        \begin{equation}
            \widetilde{Z}_3 > (\ldots)  \,  e^{L s_{\rm th}(e_n)/2} \ .
        \end{equation}
        This exponential divergence of $\widetilde{Z}_3$ is indeed the reason why in the $q$-point (full) ETH ansatz \cite{foini_eigenstate_2019} the entropic factor is $e^{-S(q-1)}$ rather than the naive generalization $e^{-q S/2}$ of 2-point ETH. This reduction arises from the phases $\theta\big(n_1, \ldots, n_q\big)$, which lead to massive cancellations when considering averages of “neighbouring” matrix elements. 
    \item In integrable models, the eigenstates $\ket s$ belonging to ${\rm sp}_2(r)$ not only are part of the same macrostate as $\ket r$, but represent a sub-entropic set of eigenstates anomalously close (in rapidity space) to $\ket r$ \cite{essler_statistics_2024, senese_finite_2026}. For this reason one expects for large $L$ that 
        \begin{equation}
            g_n^*=g_{n_1}^* \qquad \quad  \ n_1 \in {\rm sp}_2(n) \ .
        \end{equation}
        In fact, it was conjectured in Ref.~\cite{senese_finite_2026} that $g_n^*$ depends only on macrostate information, i.e.~$g^*=g^*[\rho(\lambda)]$ (as well as on the operator $O$). 
        From this we obtain
\begin{equation}
\label{eq:mathbbEn2n}
\widetilde{Z}_3 > (\ldots) \, e^{L g_n^*} \, \mathbb E[\, |\braket{n_2 | O|n}|\, ] \ ,\qquad
 \mathbb E[\, |\braket{n_2 | O|n}|\, ] =\frac{1}{W_n}\sum_{\substack{n_1 \in {\rm sp_2(n)}\\ (n_1 \neq n)}} \bigg(\frac{1}{W_{n_1}}\sum_{\substack{n_2 \in {\rm sp_2(n_1)}\\ (n_2 \neq n_1 \neq n)}} |\braket{n_2|O|n}| \bigg) \ .
\end{equation}
The key difference to \eqref{eq:keyeqQC} is that in integrable models typical off-diagonal matrix elements between eigenstates in the same macrostate are negligible in the large-$L$ limit, e.g.~$|\braket{r|O|s}|\sim e^{- c' \,L\ln L}$ \cite{essler_statistics_2024, rottoli_eigenstate_2026, orlov_multiscale_2026}, unlike in ETH where all matrix elements are merely exponentially suppressed ($\sim e^{-\alpha L}$). 
By construction all $\ket {n_1} \in {\rm sp}_2(n)$ are atypically close (in terms of their respective distribution of (half) integers) to $\ket n$, and all $\ket {n_2} \in {\rm sp}_2(n_1)$ are similarly close to their $\ket{n_1}$.
This suggests that matrix elements entering the average $\mathbb E[\, |\braket{n_2 | O|n}|\, ]$ are
exponentially small $\sim e^{-cL}$ \cite{essler_statistics_2024,senese_finite_2026} rather than proportional to $\sim e^{- c' \,L\ln L}$. However, the coefficient $c$ in the exponent could  be larger than $g_n^*/2$ for a subset of states $|n_2\rangle$ \footnote{This is because many of the eigenstates $\ket{n_2}$ are expected to be further (in rapidity space) from $\ket{n}$ than the eigenstates $\ket{n_1}$ are.}. This leaves open the possibility that $\mathbb E[\, |\braket{n_2 | O|n}|\, ]$ might decay faster than $e^{-L g_n^*}$, resulting in the absence of a sign problem. Even so, our expectation is that $g_{E}=-\lim_{L \to \infty}\ln \mathbb E[\, |\braket{n_2 | O|n}|\, ]/L < g_n^*$, because a large fraction of matrix elements in \eqref{eq:mathbbEn2n} will have $c=g_n^*/2$ (or even smaller).
 This, in turn, would imply a sign problem in the sense that $\widetilde{Z}> \exp(\nu L)$ with $\nu =g_n^*-g_E>0$.  
\end{itemize}
\begin{figure}[ht]
     \centering
     \begin{minipage}[b]{0.48\textwidth}
         \centering
         \includegraphics[width=\textwidth]{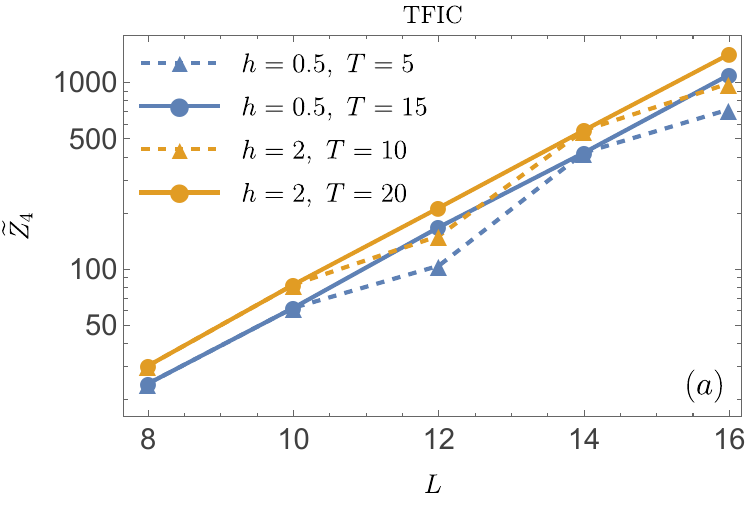}
     \end{minipage}
     \hfill
     \begin{minipage}[b]{0.48\textwidth}
         \centering
         \includegraphics[width=\textwidth]{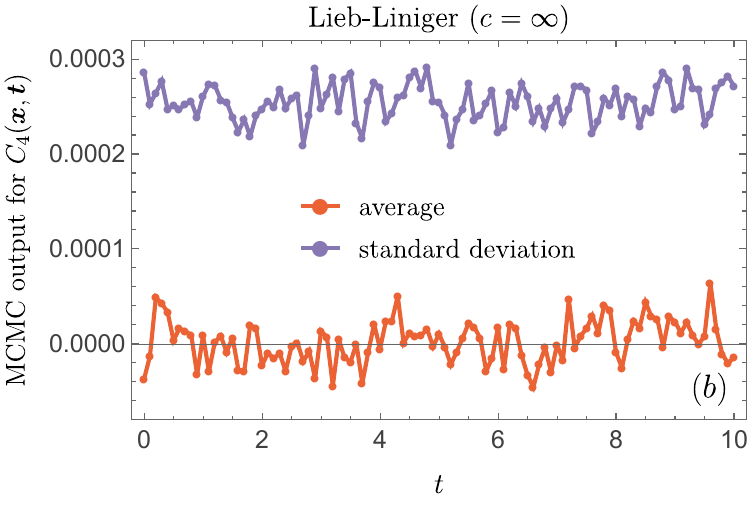}
     \end{minipage}
     \vspace{-10pt}
     \caption{(a) Log-scale plot for the growth of $\widetilde{Z}_4$ with $L$ in the TFIC. The exact values for $\widetilde{Z}_4$ are obtained by a full sum over all intermediate eigenstates in the Hilbert space (see \eqref{eq:pfZtilde} for the $k=3$ case). (b) Output of MC sampling for estimating the 4-point function $C_4(x,t)$ of \cref{eq:C4xtLLT5} in the impenetrable limit ($c = \infty$) of the LL model (for $L = N =50$, $x = 10$ and temperature $T = 10$). The value we are trying to estimate is too small to be resolved via Markov chain Monte Carlo (MCMC) sampling, due to the strong cancellations that underlie the sign problem.}
     \label{fig:signproblem}
\end{figure}
We have tested these considerations numerically both in TFIC and in LL. In \cref{fig:signproblem}(a) we plot the normalization $\widetilde{Z}_4$ (trivial generalization of \eqref{eq:pfZtilde} to the case of 4-point functions) for small values of $L$ in TFIC, where $O = \sigma^x_i$.  The exponential growth of $\widetilde{Z}_4$ with $L$ is evident even at such small system sizes. The value of $\widetilde{Z}_4$ is obtained by a full triple sum (cf.~\eqref{eq:pfZtilde} for the case $r=3$) over all eigenstates in the Hilbert space. The reference eigenstate $\ket{n}$ at each $L$ is obtained as a smooth representative microstate \cite{essler_statistics_2024} of a Gibbs ensemble at temperature $T$ for the Hamiltonian $H_{\rm TFIC}(h)$. The choice of $r = 4$ instead of $r=3$ is due to the fact that $3$-point functions of $\sigma^x$ vanish identically by symmetry in TFIC, while 4-point functions do not. We stress that the spectrum of the TFIC possesses massive degeneracies as a consequence of the single-particle dispersion $\varepsilon_h(k)=\varepsilon_h(-k)$ being symmetric, see \cref{sec:diagonTFIC}.
Therefore, there exists a large amount of freedom in the choice of an orthonormal eigenbasis, and $\widetilde{Z}_4$ depends slightly on this choice because of its unphysical nature. However, the presence or absence of a sign problem is not expected to be related to the choice of the basis, so we worked with the basis discussed in \cref{sec:diagonTFIC}. 

Turning our attention to LL in the impenetrable limit $c = \infty$, in \cref{fig:signproblem}(b) we plot the MC output (cf.~term in brackets in \eqref{eq:MCMCestimateC3} for the $r=3$ case) for the following normal-ordered $4$-point function of the Bose field $\phi(x)$
\begin{equation}
\label{eq:C4xtLLT5}
    C_4(x,t) = \braket{n |\phi^\dag(0,0)\phi^\dag(x,t)\phi(x,t)\phi(0,0)|n} \ ,
\end{equation}
where $\ket n$ is a representative eigenstate of the thermal Gibbs ensemble at temperature $T = 5$ for $L = 50$ and $N/L = 1$. For $x$ values significantly larger than the thermal correlation length $\xi$ and at small times $t$, $C_4(x,t)$ is approximately equal to $\braket{n|\phi^\dag(0,0) \phi(0,0)|n}^2=(N/L)^2=1$ by clustering properties. However, the numerical output of the MC for $x = 10$ (from running 100 parallel Markov chains and considering their average and standard deviation at each $t$ point) is seen to noisily fluctuate around zero, with a standard deviation significantly larger (in absolute value) than the average value. Similar results are obtained for any other value of $x$ (ranging from zero to very large distances) and again for all times. This is a clear signature of the sign problem: the value that the MC is trying to estimate is so small that it remains hidden behind the statistical uncertainty of the sampling. Indirectly, this confirms that also in this case the normalization $\widetilde{Z}_4$ is very large, see \eqref{eq:MCMCestimateC3}.
We have verified that the issue persists in the interacting case $c < \infty$. 

\section{Sign problem for global quantum quenches from arbitrary initial states}

We now focus on the possible origin of a sign problem in global quantum quenches from \emph{arbitrary} initial states $\ket \psi$, i.e.~states for which the overlaps $\braket{n|\psi}$ might not possess any particular structure similar to the highly restrictive one of integrable initial states \cite{piroli_what_2017} (see \cref{sec:diagonTFIC,sec:inistateLL} below). Also in this case it is instructive to consider first the quantum chaotic case.
For quantum quenches the unphysical normalization $\widetilde{Z}$ reads (see Eq.~(5) of main text)
\begin{equation}
\label{eq:tildeZquench}
    \widetilde{Z} = \sum_{\substack{m,n\\(m \neq n)}} \big|\braket{\psi|m}\braket{n|\psi}\braket{m|O|n}\big| \ ,
\end{equation}
for some local operator $O$. As before, we consider only off-diagonal terms, which are those that drive the time evolution. The initial state $\ket \psi$ is usually a lowly entangled state with good clustering properties. These types of states naturally possess a nonzero overlap with exponentially many (in $L$) eigenstates $\ket n$ of the Hamiltonian $H$. Under the assumption that $H$ is quantum chaotic, the general expectation is that the leading-order decay of the overlaps at large $L$ is dictated by a smooth function $s_\psi(e)$ of the energy density \cite{hartmann_gaussian_2004, santos_entropy_2011}
\begin{equation}
    |\braket{n|\psi}|^2 = (\ldots) \, e^{-L s_{\psi}(e_n)} \qquad \qquad e_n = \frac{E_n}{L} \ ,
\end{equation}
where $(\ldots)$ indicate subexponential contributions (which include the pseudorandom eigenstate-to-eigenstate fluctuations). For example, this implies at large $L$
\begin{equation}
    e \equiv \frac{\braket{\psi|H|\psi}}{L} = \sum_n \frac{E_n}{L} |\braket{n|\psi}|^2 = \int d e'(\ldots)\, \exp\big[{L\big(s_{\rm th}(e')-s_\psi(e')\big)}\big] \ ,
\end{equation}
where $s_{\rm th}(e')$ is the thermodynamic entropy density at energy density $e'$. The previous equation is generally satisfied because of the existence of a unique saddle point at energy density $e$, i.e.~$s_{\rm th}(e)=s_{\psi}(e)$ in the limit $L \to \infty$. Indeed, $s_{\psi}(e')$ is expected in general to take the form $s_\psi(e')=(e'-e)^2/(2\sigma^2) + s_{\rm th}(e')$ \cite{hartmann_gaussian_2004} in the vicinity of the saddle point $e$. The existence of the saddle point implies that $\ket{\psi}$ possesses subextensive energy fluctuations, matching the behaviour expected from the clustering properties of $\ket \psi$ and the locality of $H$ (see, e.g., \cite{dalessio_quantum_2016}). Denoting with ${\rm sp}(e)$ a (exponentially large) set of eigenstates that live in a subextensive window around $E=eL$, we can lower-bound $\widetilde{Z}$ as
\begin{equation}
\label{eq:ZquenchQC}
    \widetilde{Z}>\sum_{\substack{m,n \in {\rm sp}(e)\\(m \neq n)}} \big|\braket{\psi|m}\braket{n|\psi}\braket{m|O|n}\big| = (\ldots)\,e^{L s_{\rm th}(e)/2} \ ,
\end{equation}
where we used ETH $|\braket{m|O|n}| \sim (\ldots)\,e^{-Ls_{\rm th}(e)/2}$ for $m,n \in {\rm sp}(e)$ \cite{srednicki_approach_1999, dalessio_quantum_2016, patil_eigenstate_2026} and we are denoting again as $(\ldots)$ generic subexponential terms. Due to the exponential growth on the right-hand side, \cref{eq:ZquenchQC} proves the presence of a sign problem under very natural assumptions for quantum chaotic systems. 

We now turn to the integrable case. The sign problem, if present, emerges equally well in the full direct sum (DS) (see \cref{eq:quench} of main text) and in the Quench Action (QA) (see \cref{eq:quenchQA} of main text) representations. Indeed, the saddle-point strategy of QA cannot solve an intrinsic issue of strongly fluctuating phases. We therefore work directly with QA, both because it is easier and because we would need to use some of the assumptions underlying QA anyway. The relevant normalization $\widetilde{Z}$ for the QA representation of the quench is
\begin{equation}
\label{eq:ZtildeQAf}
    \widetilde{Z}=\sum_{\substack{m\\(m \neq n_{\rm sp})}} \left| \frac{\braket{\psi|m}}{\braket{\psi|n_{\rm sp}}} \braket{m|O|n_{\rm sp}}\right| \ ,
\end{equation}
where again we restrict to the off-diagonal terms that drive the evolution, and we indicate with $\ket{n_{\rm sp}}$ the representative eigenstates of the QA saddle-point macrostate $\rho_{\rm sp}(\lambda)$. A crucial assumption behind the QA method is that the exponential suppression of the nonzero overlaps depends to leading order only on macrostate information $\rho(\lambda)$, i.e. for asymptotically large $L$
\begin{equation}
    - \ln(|\braket{\psi|m}|^2)/L = s_{\psi}[\rho] + o(L^0) \ ,
\end{equation}
where $\rho(\lambda)$ denotes the macrostate $\ket m$ belongs to. This assumption has been verified for several integrable initial states $\ket \psi$ in free and interacting integrable models \cite{caux_time_2013, brockmann_gaudinlike_2014, denardis_solution_2014, pozsgay_correlations_2014,wouters_quenching_2014, caux_quench_2016} (cf.~also \cref{sec:diagonTFIC,sec:inistateLL}), and in free models also for more generic (i.e.~non-integrable) initial states that do \emph{not} possess the opposite-momentum-pair structure \cite{bertini_quantum_2017, bertini_entanglement_2018}. Given that the only relevant eigenstates $\ket m$ in \eqref{eq:ZtildeQAf} are those belonging to the same saddle-point macrostate $\rho_{\rm sp}(\lambda)$ of $\ket{n_{\rm sp}}$ (those belonging to a different macrostate have form factors $\braket{m|O|n_{\rm sp}}$ too strongly suppressed \cite{essler_statistics_2024, rottoli_eigenstate_2026, senese_finite_2026, orlov_multiscale_2026}), we see that the ratio of overlaps in \eqref{eq:ZtildeQAf} only contributes subextensively to the sum, i.e.
\begin{equation}
  \lim_{L \to \infty} \frac{1}{L} \ln\left( \left|\frac{\braket{\psi|m}}{\braket{\psi|n_{\rm sp}}} \right|\right)= 0 \ .
\end{equation}
Denoting as $(\ldots)$ subexponential (with $L$) terms, the previous equation implies
\begin{equation}
\label{eq:ZtildeQAf2}
    \widetilde{Z}=\sum_{\substack{m\\(m \neq n_{\rm sp})}} (\ldots)  \left| \braket{m|O|n_{\rm sp}}\right| \ .
\end{equation}
We can now use the results of Eqs.~\eqref{eq:Ol0varr} to \eqref{eq:saddlepoint2point}, which prove that there exist $\exp[\sigma^{(n_{\rm sp})}(g^*)L]\sim\exp[g^*L]$ many eigenstates $\ket m$ such that $|\braket{m|O|n_{\rm sp}}|^2 \sim \exp[- g^* L]$, for some saddle point $g^*$ \cite{senese_finite_2026}. Denoting as ${\rm sp}(n_{\rm sp})$ the set of such eigenstates, we can lower bound $\widetilde{Z}$ as
\begin{equation}
\label{eq:ZtildeQAf3}
    \widetilde{Z}>\sum_{\substack{m\in {\rm sp}(n_{\rm sp})\\(m \neq n_{\rm sp})}} (\ldots)  \left| \braket{m|O|n_{\rm sp}}\right| = (\ldots) \exp\big[(\tilde \sigma^{(n_{\rm sp})}(g^*)-g^*/2)L\big]\ ,
\end{equation}
where $\tilde \sigma^{(n_{\rm sp})}(g)$ denotes the \emph{restriction} of  $\sigma^{(n_{\rm sp})}(g)$ that counts only eigenstates with nonvanishing overlap with $\ket \psi$. Therefore, in general, $\tilde \sigma^{(n_{\rm sp})}(g^*) \le \sigma^{(n_{\rm sp})}(g^*)$ (as obvious, e.g., in the case of integrable initial states). 
We can distinguish three different cases:
\begin{enumerate}[label=(\roman*)]
    \item $\ket \psi$ is an integrable initial state, therefore $\braket{\psi|m}\neq0$ only if $\ket m$ is parity invariant, i.e.~it is composed entirely of pairs opposite rapidities (e.g.~$\ket m=\ket{\lambda_1,-\lambda_1, \ldots,\lambda_{N/2},-\lambda_{N/2}}$). It is evident that the Yang-Yang density $ s^{(\rm res)}_\text{\tiny YY}[\rho]$ restricted to this class of states (assuming $\rho(\lambda)$ is symmetric around $\lambda = 0$) is exactly \emph{one half} of the full Yang-Yang density $s_\text{\tiny YY}[\rho]$. Although nontrivial to prove rigorously, this suggests a similar relation between the form-factor-resolved entropy densities: $\tilde \sigma^{(n_{\rm sp})}(g^*) = \sigma^{(n_{\rm sp})}(g^*)/2=g^*/2$. Plugging this conjecture in \eqref{eq:ZtildeQAf3} the exponential growth on the right-hand side gets suppressed. Given that within the QA formalism the set ${\rm sp}(n_{\rm sp})$ is the one expected to dominate in \eqref{eq:ZtildeQAf}, the previous analytical argument provides a route to understand the absence of a sign problem for integrable initial states. 
    \item $\ket \psi$ is not an integrable initial state, but nonzero-overlap eigenstates possess instead constraints dictated by a \emph{multiparticle} structure similar to that investigated in Refs.~\cite{bertini_quantum_2017,bertini_entanglement_2018}. In such cases we expect $g^*/2 < \tilde \sigma^{(n_{\rm sp})}(g^*) < g^*=\sigma^{(n_{\rm sp})}(g^*)$. Using this in \eqref{eq:ZtildeQAf3} we obtain an exponentially growing lower bound, i.e.~a sign problem.
    \item $\ket{\psi}$ is a generic initial state with no constraints on the eigenstates it has a nonzero overlap with. Here the relevant entropic factor is identical to the one for equilibrium correlators, i.e.~$\tilde \sigma^{(n_{\rm sp})}(g^*) = \sigma^{(n_{\rm sp})}(g^*)=g^*$, leading to an even more pronounced sign problem.
\end{enumerate}

\begin{figure}[t!]
    \centering
    \begin{minipage}{0.35\textwidth}
        \centering
        \includegraphics[width=\textwidth]{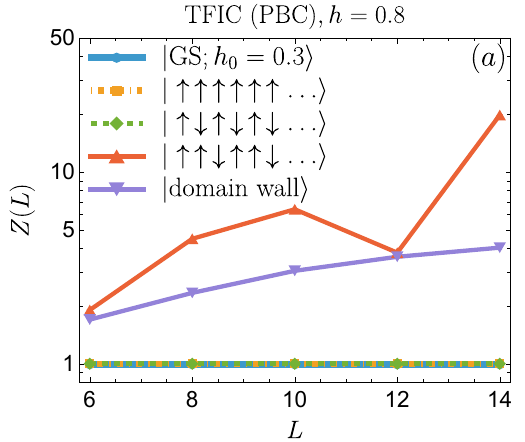}
    \end{minipage}
    \begin{minipage}{0.35\textwidth}
        \centering
        \includegraphics[width=\textwidth]{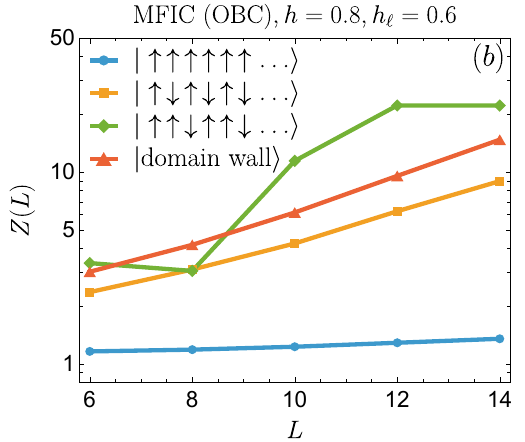}
    \end{minipage}

    \vspace{2ex} 

    \begin{minipage}{0.35\textwidth}
        \centering
        \includegraphics[width=\textwidth]{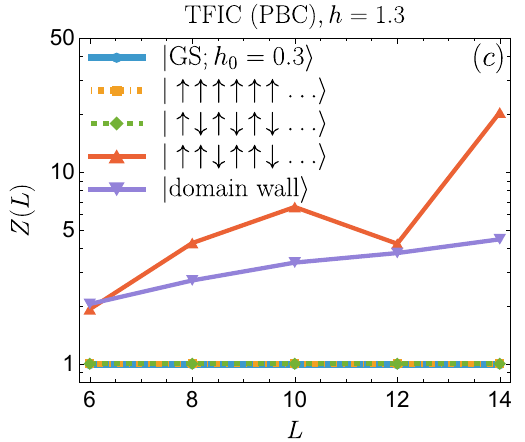}
    \end{minipage}
    \begin{minipage}{0.35\textwidth}
        \centering
        \includegraphics[width=\textwidth]{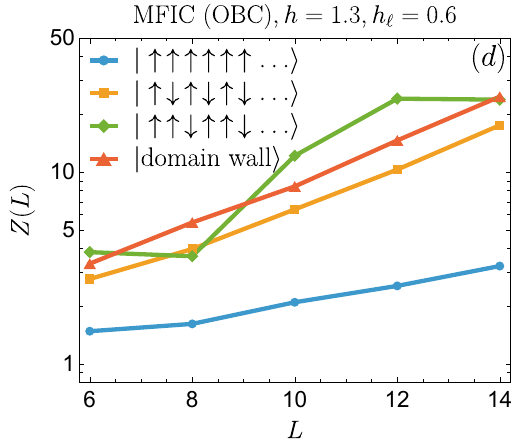}
    \end{minipage}
    
    \caption{Log-plots showing dependence of normalization $Z$ [see \cref{eq:ZforED}] on the system size $L$, as obtained by exact diagonalization in TFIC and MFIC [\cref{eq:MFIC}] for several initial states $\ket \psi$ (see text). (a) TFIC with transverse field $h = 0.8$. (b) MFIC with $h = 0.8$, $h_\ell = 0.6$. (c) TFIC with $h = 1.3$. (d) MFIC with $h = 1.3$, $h_\ell = 0.6$.}
    \label{fig:ED_TFIC}
\end{figure}
We verify these conclusions numerically by means of an exact diagonalization study in TFIC with periodic boundary conditions (PBC), and in the mixed-field Ising chain with open boundary conditions (OBC)
\begin{equation}
\label{eq:MFIC}
    H_{\rm MFIC} = -J \sum_{j=1}^L(\sigma^x_j \sigma_{j+1}^x + h\, \sigma_j^z + h_{\ell} \sigma_j^x) \ , 
\end{equation}
which differs from TFIC by the addition of an \emph{integrability-breaking} longitudinal field $h_{\ell}$ (and the boundary conditions). The reason for considering $H_{\rm MFIC}$ (which we know to possess a sign problem from the ETH discussion above) is that we will show it gives numerical results very similar to TFIC, substantiating the claim that also TFIC has a sign problem. For both $ H_{\rm TFIC}$ and $ H_{\rm MFIC}$ we numerically compute the full DS normalization
\begin{equation}
\label{eq:ZforED}
   Z = \sum_{m,n} \big|\braket{\psi|m}\braket{n|\psi}\braket{m|O|n}\big| \ ,
\end{equation}
for several initial states $\ket \psi$. In \eqref{eq:ZforED} we do not explicitly remove the diagonal terms just because they contribute at most $\mathcal{O}(L^0)$ to $Z$. We note that unlike the spectrum of TFIC (see previous section), the one of MFIC with OBC does not feature degeneracies. Therefore while the unphysical normalization $Z$ might depend slightly on the choice of the eigenbasis in TFIC (but this is not expected to affect the presence or absence of a sign problem), it is instead uniquely fixed in MFIC. The initial states on $L$ sites that we consider are
\begin{enumerate}[label=(\arabic*)]
    \item The $\mathbb Z_2$ symmetry-breaking ground state of $H_{\rm TFIC}(h_0)$, with $h_0 < h$ and $h_0<1$ (see \cref{sec:diagonTFIC}). 
    \item The fully polarized state along the $\hat x$ axis $\ket{\uparrow\, \uparrow\, \uparrow \, \uparrow\ldots}$. 
    \item The classical Néel state along $\hat x$ $\ket{\uparrow\,\downarrow\,\uparrow\,\downarrow\ldots}$. 
    \item The “$2+1$” alternating state along $\hat x$ $\ket{\uparrow\, \uparrow\,\downarrow\,\uparrow\,\uparrow\,\downarrow\ldots}$. 
    \item The domain wall along $\hat{x}$, with up (down) spins in the first (second) half of the chain, e.g.~$\ket{\uparrow\,\uparrow\,\uparrow\,\downarrow\,\downarrow\,\downarrow}$ for $L = 6$. 
\end{enumerate}
We choose $O=\sigma^x_j$ \footnote{For MFIC with OBC, this is chosen to be on the site just to the left of the center of the chain.}. All initial states $\ket \psi$ above have $\braket{\psi|\sigma_j^x|\psi}\neq 0$, and therefore exhibit interesting dynamics for $\braket{\psi(t)|\sigma_j^x|\psi(t)}$ under time evolution with $H_{\rm TFIC}$ or $ H_{\rm MFIC}$ (cf.~main text). In the case of evolution with $H_{\rm TFIC}$, the states (1), (2) and (3) represent integrable initial states.
This explains the behaviour seen in \cref{fig:ED_TFIC}(a) where $Z(L)$ is observed to stay constant as a function of $L$ (as expected for $h_0<h=0.8$, it coincides with the value $\braket{\psi|\sigma_j^x|\psi}$ at all $L$, see \cref{sec:QAandSPinTFIC}). We show in \cref{sec:QAandSPinTFIC} that for integrable initial states in the TFIC $Z(L)={\rm const}$, so there is no sign problem.
The states (4) and (5) do not have the pair structure characteristic of integrable initial states and as seen in
the log-plot of \cref{fig:ED_TFIC}(a) $Z(L)$ grows with $L$. While this growth is less pronounced than in  \cref{fig:signproblem} for the small system sizes available, the curves are still compatible with an exponential scaling with $L$. In fact, the growth with $L$ for TFIC is very similar to the one in MFIC of \cref{fig:ED_TFIC}(b), which we know is exponential for all initial states from the ETH arguments above. In \cref{fig:ED_TFIC}(c) and (d) we show analogous plots for $h = 1.3$ (paramagnetic regime of $H_{\rm TFIC}$ at $T=0$) both for TFIC and MFIC, finding very similar results. Overall, these results provide strong evidence for the existence of a sign problem for quenches from generic initial states in $H_{\rm TFIC}$, compatible with our analytic arguments.

\section{Details on the TFIC quench}

\subsection{Diagonalization, overlaps and form factors}
\label{sec:diagonTFIC}
Here we collect results regarding the diagonalization of the TFIC (\cref{eq:TFICh} with periodic boundary conditions in the ferromagnetic phase ($h<1$). For further details see, e.g., \cite{calabrese_quantum_2012, mbeng_quantum_2024} and references therein. 
Given that $[H_{\rm TFIC},e^{i \pi N}]=0$, where $N =\sum_{j=1}^L c_j^\dag c_j$ is the total Jordan-Wigner fermion number \cite{lieb_two_1961}, the Hamiltonian is block diagonal in the subspaces of the fermionic Fock space with even (e) and odd (o) number of fermions, i.e.~$H_{\rm TFIC}(h) = H_{\rm e}(h) \oplus H_{\rm o}(h)$. By going to Fourier space and performing a Bogoliubov transformation (which involves also a particle-hole transformation \cite{calabrese_quantum_2012}) one arrives at
\begin{equation}
    H_{\rm e}(h) = \sum_{k \in {\rm NS}}\varepsilon_h(k)\, \alpha_k^\dag \alpha_k + E_{\rm NS}(h) \ , \qquad  \qquad 
    H_{\rm o}(h) = \sum_{p \in {\rm R}}\varepsilon_h(p)\, \alpha_p^\dag \alpha_p + E_{\rm R}(h) \ ,
\end{equation}
\begin{equation}
    \varepsilon_h(k) = 2J \sqrt{1+h^2 - 2 h \cos (k)} \ .
\end{equation}
Here $\alpha_k$ ($\alpha_k^\dag$) indicate fermionic Bogoliubov annihilation (creation) operators labelled by free momenta in the so called Neveu–Schwarz (NS) and Ramond (R) sectors ($L$ even)
\begin{equation}
\label{eq:NSandRdef}
k_n = \frac{2\pi(n+\delta_{{\rm X},{\rm NS}}/2)}{L} \quad \qquad n = -\frac{L}{2}, \dots, \frac{L}{2} - 1 \in \mathbb{Z}, \ \ \  {\rm X} = {\rm R}, \, {\rm NS}
\end{equation}
The dependence of the operators $\alpha_k$ on $h$ is left implicit. 
The Fock space of fermions is spanned by the eigenstates
\begin{equation}
\label{eq:eigenstatesTFIC}
    \ket{k_1,\ldots,k_{2m};h}_{\rm X} \equiv \alpha_{k_1}^\dag \ldots \alpha_{k_{2m}}^\dag \ket{0;h}_{\rm X} \quad \qquad k_j \in {\rm X}, \ \ {\rm X} = {\rm NS}, \, {\rm R} \ \ , 
\end{equation}
where $\ket{0,h}_{\rm NS}$ ($\ket{0,h}_{\rm R}$) denotes the 0-fermion (1-fermion) state annihilated by all $\alpha_{k_j}$ with $k_j \in {\rm NS}$ (${\rm R}$). For any finite $L$ the ground state of $H_{\rm TFIC}(h)$ is $\ket{0;h}_{\rm NS}$. 
However, by spontaneous symmetry breaking for $L \to \infty$ the two possible physical ground states become
\begin{equation}
   \ket{\pm; h}\equiv \frac{1}{\sqrt{2}}(\ket{0;h}_{\rm NS}\pm \ket{0;h}_{\rm R}) \ .
\end{equation}
These possess nonvanishing expectation values for the order parameter $\sigma^x_j$. To capture this physical behaviour in the TFIC quench from the main text, we set as our initial state $\ket{+;h_0}$, and quench with $H_{\rm TFIC}(h)$ where $h \neq h_0$ and $h_0, h < 1$. 
The state $\ket{0;h_0}_{\rm X}$ with X = NS, R has non-zero overlap only with parity-invariant eigenstates \eqref{eq:eigenstatesTFIC} $\ket{k_1,-k_1,\ldots, k_m,-k_m;h}$ and has the same structure as integrable initial states \cite{ghoshal_boundary_1994, fioretto_quantum_2010, sotiriadis_zamolodchikov_2012, bertini_quantum_2014, piroli_what_2017}. The overlaps for these eigenstates are 
\begin{equation}
\label{eq:overlapsTFIC}
    _{\rm X}\braket{k_1,-k_1,\ldots, k_m,-k_m;h|0;h_0}_{\rm X} = (-i)^m\Bigg(\prod_{\substack{k \in {\rm X}\\k>0}}\frac{1}{\sqrt{1+K_k^2}}\Bigg) \prod_{j=1}^m K_{k_j} \quad \qquad k_j > 0 \ \ \forall \ j=1,\ldots, m\ ,
\end{equation}
\begin{equation}
\label{eq:KkTFIC}
    K_{k} = \frac{\sin(k) (h -h_0)}{\varepsilon_{h}(k)\varepsilon_{h_0}(k)/(2J)^{2}+1+h h_0 - (h+h_0)\cos(k)} \ .
\end{equation}
The convention for the bra in \eqref{eq:overlapsTFIC} is the standard one following from \eqref{eq:eigenstatesTFIC}
(and differs from the one used in \cite{calabrese_quantum_2012})
\begin{equation}
\label{eq:brafermordconv}
    _{\rm X}\langle k_1, \dots, k_{2m};h | \equiv \ _{\rm X}\bra{0;h} \alpha_{k_{2m}}\ldots \alpha_{k_1} \ .
\end{equation}
\eqref{eq:overlapsTFIC} provides the overlaps with the initial state $\ket{+,h_0}$ and shows that there are an exponentially large number of non-zero overlaps between the eigenstates of $H_{\rm TFIC}(h)$ and $\ket{+,h_0}$.

The order parameter $\sigma^x_j$ can only connect eigenstates in NS with those in R, and vice versa. The expressions for their form factors are known \cite{gehlen_formfactors_2008, iorgov_spin_2011, calabrese_quantum_2012, caux_time_2013}
{\allowdisplaybreaks
\begin{align}
\label{eq:ffTFIC}
& _{\rm NS}\langle k_1, \dots, k_{2n} ;h| \sigma^x_\ell | p_1, \dots, p_{2m} ;h\rangle_{\rm R} = (-1)^n e^{-i\ell \left[ \sum_{j=1}^{2n} k_j - \sum_{r=1}^{2m} p_r \right]} i^{(n+m )} (4J^2 h)^{(m-n)^2}  \nonumber \\
& \qquad \qquad \qquad \times \prod_{j<j'}^{2n} \left[ \frac{\sin[(k_j - k_{j'})/2]}{\epsilon_{k_j, k_{j'}}} \right] \prod_{r<r'}^{2m} \Bigg[ \frac{\sin[(p_r - p_{r'})/2]}{\epsilon_{p_r, p_{r'}}} \Bigg] \prod_{j=1}^{2n} \prod_{r=1}^{2m} \left[ \frac{\epsilon_{k_j, p_r}}{\sin[(k_j - p_r)/2]} \right] \nonumber \\
& \qquad \qquad \qquad \qquad \qquad \qquad \qquad \qquad \times \sqrt{\xi \xi_T} \prod_{j=1}^{2n} \left[ \frac{e^{\eta_{k_j}}}{L \varepsilon_h(k_j)} \right]^{1/2} \prod_{r=1}^{2m} \left[ \frac{e^{-\eta_{p_r}}}{L \varepsilon_h(p_r)} \right]^{1/2} \ .
\end{align}
}
Here $\epsilon_{k,k'}=[\varepsilon_h(k)+\varepsilon_h(k')]/2$, $\xi = |1-h^2|^{1/4}$ and 
\begin{equation}
\begin{aligned}
    \xi_T = \prod_{\substack{k \in {\rm NS} \\ p \in {\rm R}}} \epsilon_{k,p}^{1/2} \prod_{k,k' \in {\rm NS}} \epsilon_{k,k'}^{-1/4} \prod_{p,p' \in {\rm R}} \epsilon_{p,p'}^{-1/4} \ , \qquad \qquad \quad e^{\eta_q} = \frac{\prod_{k \in {\rm NS}} \epsilon_{q,k}}{\prod_{p \in {\rm R}} \epsilon_{q,p}} \ .
\end{aligned}
\end{equation}
Thanks to the different quantization conditions \eqref{eq:NSandRdef} for NS and R, the form factors \eqref{eq:ffTFIC} do \emph{not} feature singularities for any finite $L$. \cref{eq:ffTFIC} proves that there is an exponentially large (in system size $L$) number of nonzero form factors of $\sigma_\ell^x$, despite the fact that we are dealing with a free theory. This is related to the fact that $\sigma_\ell^x$, when expanded in Jordan-Wigner fermions, involves a Jordan-Wigner string.

\subsection{QA approach and absence of sign problem}
\label{sec:QAandSPinTFIC}

Given the initial state $\ket{+,h_0}$, the QA saddle-point eigenstate $\ket{\bs k_{\rm sp}}$ from \cref{eq:quenchQA} (main text) can be chosen to belong to NS, yielding the following QA representation of the time evolution of $\braket{\sigma_j^x(t)}$ in the limit of large $L$ \cite{caux_time_2013, granet_finite_2020}
\begin{equation}
\label{eq:QAsigmaxell}
\braket{+,h_0 |\sigma_\ell^x(t)|+,h_0} \quad \longrightarrow \quad  {\rm Re} \left[\frac{\braket{+,h_0|\sigma_\ell^x(t)|\bs k_{\rm sp}}_{\rm NS}}{\braket{+,h_0|\bs k_{\rm sp}}_{\rm NS}}\right] =  {\rm Re}\left[\frac{_{\rm R}\braket{0,h_0|\sigma_\ell^x(t)|\bs k_{\rm sp}}_{\rm NS}}{_{\rm NS}\braket{0,h_0|\bs k_{\rm sp}}_{\rm NS}}\right] \ .
\end{equation}
In the last equality we used the fact that $H_{\rm TFIC}(h)$ does not connect the NS and R sectors. Given that the initial state has nonzero overlap only with parity-invariant eigenstates, see \eqref{eq:overlapsTFIC}, $\ket{\bs k_{\rm sp}}_{\rm NS}$ must be parity invariant. The root density associated with $\ket{\bs k_{\rm sp}}_{\rm NS}$ is \cite{caux_time_2013}
\begin{equation}
    \rho_{\rm sp}(k) = \frac{1-\cos(\Delta_k)}{4\pi} \qquad \qquad  \cos(\Delta_k) = \frac{h h_0-(h+h_0)\cos(k)+1}{\varepsilon_h(k) \varepsilon_{h_0}(k)/(2J)^2} \ .
\end{equation}
In practice, at any large but finite $L$ the eigenstate $\ket{\bs k_{\rm sp}}_{\rm NS}=\ket{k_1,-k_1,\ldots,k_N,-k_N;h}_{\rm NS}$ is chosen such that its finite distribution of pairs of momenta $(k_j,-k_j)$ matches as closely as possible (see, e.g., \cite{essler_statistics_2024}) the thermodynamic root density $\rho_{\rm sp}(k)$, where $N = \big\lfloor L\int_{0}^\pi dk \, \rho_{\rm sp}(k)\big\rfloor$ indicates the number of positive-momentum Bogoliubov fermions present in $\ket{\bs k_{\rm sp}}_{\rm NS}$. Expanding \eqref{eq:QAsigmaxell} in a Lehmann sum as in \eqref{eq:quenchQA} (main text) we arrive at
\begin{align}
\label{eq:finalTFICQA}
\braket{+,h_0 |\sigma_\ell^x(t)|+,h_0} \ \  \longrightarrow \ \ &
{\rm Re}\Bigg\{\frac{\prod_{\substack{k \in {\rm NS}\\k>0}}\sqrt{1+K_k^2}}{\prod_{\substack{p \in {\rm R}\\p>0}}\sqrt{1+K_p^2}} \frac{e^{i E_{\rm R}(h)t}}{e^{i E_{\rm NS}(h)t}}\ \ \sum_{M = 0}^{L/2-1}\sum_{\substack{0<p_1<\ldots<p_M\\p_j \in R \ \ \forall j}} \Bigg[ i^{M-N}\frac{\prod_{r=1}^M K_{p_r}e^{2i \varepsilon_h(p_r)t}}{\prod_{j=1}^N K_{k_j}e^{2i \varepsilon_h(k_j)t}} \nonumber\\
& \qquad \times \,_{\rm R}\braket{p_1,-p_1,\ldots,p_M,-p_M;h|\sigma_\ell^x|k_1,-k_1,\ldots,k_N,-k_N;h}_{\rm NS}\Bigg] \Bigg\}  \ ,
\end{align}
In the case of TFIC, the correctness of the QA approach \eqref{eq:finalTFICQA} has been explicitly verified \cite{caux_time_2013}. The representation \eqref{eq:finalTFICQA} does not suffer from a sign problem as we will now demonstrate. Apart from the time-dependent phases (which do not give rise to a sign problem, see main text) in the QA sum we have
\begin{enumerate}[label=(\roman*)]
    \item The phase $i^{M-N}$ appearing explicitly in \eqref{eq:finalTFICQA}.
    \item The phase $(-1)^N (-i)^{M+N}$ from the complex conjugate of the form factor \eqref{eq:ffTFIC} (note that the phase involving the total momenta is absent because parity invariant states have zero total momentum). 
    \item The sign of the terms $\prod_{r=1}^M K_{p_r}$ and $\prod_{j=1}^K K_{k_j}$ appearing explicitly in \eqref{eq:finalTFICQA}. 
    \item The sign of the terms $\sin[(k_j - k_{j'})/2]$, $\sin[(p_r - p_{r'})/2]$ and $\sin[(k_j - p_r)/2]$ in the form factor \eqref{eq:ffTFIC}. 
\end{enumerate}
The phases in (i) and (ii) cancel, and as a result of the parity-invariant structure $(k,-k)$ of all eigenstates involved, the signs in (iv) also cancel. This can be easily seen in the case of X-sector states with 2 pairs $(k_1,-k_1,k_2,-k_2)$, where X = NS, R and $k_1,k_2>0$. From \eqref{eq:ffTFIC}, the terms within the same sector yield
\begin{equation}
   \left[ \sin\left(\frac{k_1 + k_1}{2}\right)\sin\left(\frac{k_2+k_2}{2}\right) \right] \left[\sin\left(\frac{k_1 - k_2}{2}\right) \sin\left(\frac{k_1 + k_2}{2}\right) \sin\left(\frac{-k_1-k_2}{2}\right)\sin\left(\frac{-k_1+k_2}{2}\right)\right] > 0 \ .
\end{equation}
The cancellation of all negative signs arises because 2 pairs of terms with opposite signs appear in the second square bracket. By induction the same mechanism applies to any number of pairs within the same sector X. Consider now the cross terms for single-pair states $(k,-k) \in {\rm NS}$ and $(p,-p) \in {\rm R}$, where $k,p>0$. From \eqref{eq:ffTFIC} the cross terms yield
\begin{equation}
    \sin\left(\frac{k-p}{2}\right)\sin\left(\frac{k+p}{2}\right)\sin\left(\frac{-k-p}{2}\right)\sin\left(\frac{-k+p}{2}\right) > 0 \ .
\end{equation}
Again, the presence of 2 pairs of terms with opposite signs yields a positive result. By induction, the cross terms for any parity-invariant states yield a positive contribution. The remaining phases to consider are those from (iii), which depend only on the sign of the terms $K_k$ from \eqref{eq:KkTFIC}. For $h_0,h \in (0,1)$ we have
\begin{equation}
    \text{Denominator}[K_k] \ge (1-h)(1-h_0)+1+hh_0-(h+h_0) = 2(1-h)(1-h_0)>0 \ .
\end{equation}
Therefore, the only signs come from the numerator of \eqref{eq:KkTFIC}. Recalling \eqref{eq:overlapsTFIC} that $k>0$ in $K_k$, we have
\begin{itemize}
    \item For $h > h_0$: \ $\text{Numerator}[K_k]> 0$. This proves the absence of a sign problem, as at $t=0$ each term in the sum \eqref{eq:finalTFICQA} is strictly positive, hence there are no strongly oscillating phases.
    \item For $h<h_0$: $\text{Numerator}[K_k]< 0$. In this case the terms in the sum \eqref{eq:finalTFICQA} (at $t=0$) are positive if $M+N$ is even, and negative otherwise. In the MC sampling, the vast majority of the sampled R eigenstates have $M = N$, as these are associated with dominant spectral weights. This implies that the vast majority of phases sampled are coherent, and hence MC faces no sign problem ($Z$ is not exponentially large). 
    We stress that the minority of samples with $M \neq N$, and in particular $M+N$ odd, play a crucial role in recovering the exact results of \cref{fig:TFIC} (main text). Indeed, only by the presence of some negative contributions in the Lehmann sum at $t=0$ it is possible for the curve with $h<h_0$ to grow at short times $t$. 
\end{itemize}

\subsection{MC sampling}
\label{sec:MCMCsamplingTFIC}
We now describe the algorithm used to sample $\ket{p_1,-p_1,\ldots,p_M,-p_M;h}_{\rm R}$ in \eqref{eq:finalTFICQA} in order to estimate $\braket{+,h_0 |\sigma_\ell^x(t)|+,h_0}$ as in \cref{eq:sampscheme} (main text). 
We sample sets of integers \eqref{eq:NSandRdef} in the R sector \cite{senese_finite_2026}. Given the parity invariance of the eigenstates involved, we can restrict our attention to sets of positive integers. Let $\{I_j\}_\ell$ be the set of $M_\ell$ positive, distinct integers at MC step $\ell$. The scheme for the update $\{I_j\}_\ell\rightarrow \{I_j\}_{\ell +1}$ is as follows:
\begin{itemize}
    \item If $0<M_\ell <L/2 - 1$: we randomly pick one of the following 3 possibilities (with equal probabilities). (i) we propose to add an unoccupied integer, which is chosen uniformly at random (we denote this move as “add”); (ii) we propose to remove an occupied integer, which is chosen uniformly at random (we denote this move as “rem”); (iii) we propose to perform a particle-hole excitation, i.e.~to swap an occupied integer chosen uniformly at random with an unoccupied one (we denote this move as “ph”). The unoccupied integer is chosen uniformly at random among the $Q/2$ unoccupied integers to the left, and $Q/2$ to the right, of the chosen occupied integer (if we hit the boundaries of the allowed positive integers, we wrap around). Here $Q$ is an arbitrary parameter fixed in each instance of the MC algorithm \cite{senese_finite_2026}. 
    \item If $M_\ell = 0$: we propose to add an unoccupied integer, i.e.~“add”.
    \item If $M_\ell = L/2 -1$: we propose to remove an occupied integer, i.e.~“rem”.
\end{itemize}
Let $\{I_j\}^*$ denote the set of integers obtained from  $\{I_j\}_\ell$ by accepting a proposed update. We then require the probabilities of proposing the moves $\{I_j\}_\ell\rightarrow \{I_j\}^*$ and $\{I_j\}^*\rightarrow \{I_j\}_{\ell}$ to implement a Metropolis-Hastings step \cite{metropolis_equation_1953, hastings_monte_1970}, cf.~Appendix A of main text. We have the following possibilities for the 3 types of proposal (i), (ii) and (iii) defined above. 
\begin{itemize}
    \item If $0<M_\ell <L/2 - 1$:
        \begin{enumerate}[label=(\roman*)]
        \item The probability of proposing “add” or its inverse (which is of type “rem”) are
        \begin{equation}
            P_{\rm add} = \frac{1}{3}\frac{1}{L/2 - 1 -M_{\ell}} \ , \qquad \qquad 
            P_{\rm add}^{\rm (inv)} = \Bigg\{
            \begin{aligned}
                \frac{1}{L/2 - 1} \qquad &M_\ell = L/2 - 2 \ ,\\
                \frac{1}{3}\frac{1}{M_\ell +1}\qquad &M_\ell < L/2 - 2 \ .
            \end{aligned}
        \end{equation}
        \item The probability of proposing “rem” or its inverse (which is of type “add”) are
        \begin{equation}
            P_{\rm rem} = \frac{1}{3}\frac{1}{M_\ell} \ , \qquad \qquad 
            P_{\rm rem}^{\rm (inv)} = \Bigg\{
            \begin{aligned}
                \frac{1}{L/2 - 1} \qquad &M_\ell = 1 \ ,\\
                \frac{1}{3}\frac{1}{L/2 - M_{\ell}}\qquad &M_\ell > 1 \ .
            \end{aligned}
        \end{equation}
        \item The probability of proposing “ph” or its inverse (which is of type “ph”) are
        \begin{equation}
            P_{\rm ph} = P_{\rm ph}^{\rm (inv)} =  \frac{1}{3}\frac{1}{M_{\ell}}\frac{1}{Q} \ .
        \end{equation}
        \end{enumerate}
    \item If $M_\ell = 0$: the probabilities of “add” and its inverse are
    \begin{equation}
        P_{\rm add} = \frac{1}{L/2 - 1} \qquad \qquad P_{\rm add}^{\rm (inv)} = 1/3 \ .
    \end{equation}
    \item If $M_\ell = L/2 - 1$: the probabilities of “rem” and its inverse are
    \begin{equation}
        P_{\rm rem} = \frac{1}{L/2 - 1} \qquad \qquad P_{\rm add}^{\rm (inv)} = 1/3 \ .
    \end{equation}
\end{itemize}
Denoting the probability distribution in \cref{eq:Csampling} (main text) by $\mathcal{P}(\{I_j\})=|F(\{I_j\})|/Z$ (here derived from \cref{eq:finalTFICQA}), we then accept the proposed move (“add”, “rem” or “ph”) with probability \cite{metropolis_equation_1953, hastings_monte_1970}
\begin{equation}
    \alpha_{\rm MH}=\min \left[\frac{\mathcal{P}(\{I_j\}^*)}{\mathcal{P}(\{I_j\}_{\ell })}\frac{P_{\rm move}^{\rm(inv)}}{P_{\rm move}},1\right] 
\end{equation}
and reject it with probability $1-\alpha_{\rm MH}$. We set $\{I_j\}_{\ell+1}=\{I_j\}^*$ in case of acceptance, and $\{I_j\}_{\ell+1}=\{I_j\}_{\ell}$ otherwise.  Iterating this procedure many times yields a collection of sets $\{I_j\}_\ell$ (approximately) sampled according to $\mathcal{P}(\{I_j\})$ \cite{rubinstein_simulation_2016, gilks_markov_1995}, which allows us to determine the expectation value of interest by means of \cref{eq:sampscheme} (main text). 
We have conducted a number of checks to verify that the TFIC MC algorithm appropriately converges (irrespective of the initial set $\{I_j\}_1$ chosen), and that the sampling is efficient \footnote{We stress that the ordering $p_1<p_2<\ldots<p_M$ chosen in \eqref{eq:finalTFICQA} is arbitrary. Indeed, swapping two pairs $(p_i,-p_i)$ and $(p_j,-p_j)$ for $i\neq j$ in the representation \eqref{eq:eigenstatesTFIC} of  $\ket{\bs p,h}_{\rm R}$ leads to no sign changes, and therefore the QA sum \eqref{eq:finalTFICQA} is invariant under permutations in the ordering of $p_1,\ldots,p_M$. This implies that in the MC sampling no time should be wasted in sorting the integers $\{I_j\}_\ell$.}. See \cref{sec:MCMCdetailsLL} or Appendix B of \cite{senese_finite_2026} for details. 
\section{Details on the LL quench}
\subsection{Initial state, overlaps and QA saddle-point}
\label{sec:inistateLL}
The LL Hamiltonian at $c = 0$ coincides with the free boson theory
\begin{equation}
    H = - \int_0^L dx \, \phi^\dag(x) \partial_x^2 \phi(x) = \sum_k k^2 \tilde\phi_k^\dag \tilde \phi_k \ .
\end{equation}
The Bose-Einstein condensate (BEC) ground state in the $N$-particle sector is
\begin{equation}
\label{eq:GSNLL}
    \ket{{\rm GS};N} = \frac{1}{\sqrt{N!}}(\tilde \phi_0^\dag)^N \ket 0 \ .
\end{equation}
This state does not respect the cluster decomposition principle for connected correlation functions of $\phi(x)$, given that $\braket{{\rm GS},N|\phi^\dag(x)\phi(0)|{\rm GS},N}_{\rm c} = N/L$ for any $x$. As the ground states $\ket{{\rm GS};N}$ for any $N$ are degenerate, we turn to the symmetry broken description of the ground state in terms of a coherent state of $k=0$ bosons
\begin{equation}
\label{eq:coherstateeta}
    \ket{{\rm GS}}_\eta = e^{-1/2 |\eta|^2} e^{\eta \tilde \phi_0^\dag }\ket 0  = e^{-1/2 |\eta|^2} \sum_{N = 0}^\infty \frac{\eta^N}{\sqrt{N!}}\ket{{\rm GS};N}\ .
\end{equation}
This restores clustering of $\phi(x)$ connected correlators. We stress that the fluctuations in particle number $\hat N$ of $\ket{\rm GS}_\eta$ become asymptotically negligible
\begin{equation}
\begin{aligned}
 \overline N  \equiv  \ _\eta\braket{{\rm GS}|\hat{N}|{\rm GS}}_\eta = |\eta|^2 & \ , \qquad \qquad {\rm sd}^2(N) = \ _\eta\braket{{\rm GS}|\hat{N}^2|{\rm GS}}_\eta - \overline N^2 = \overline{N} \ , \\
 & \frac{{\rm sd}(N)}{\overline N} = \frac{1}{(\overline N)^{1/2}} \ \ \xrightarrow[\overline N \to \infty]{} \ \ 0 \ .
 \end{aligned}
\end{equation}
The average number of particles $\overline N$ is chosen to yield a finite density in the thermodynamic limit, i.e. $\lim_{L \to \infty}\overline N/L = n > 0$. Importantly, the symmetry broken ground state features a non-vanishing order parameter (known as “macroscopic wavefunction” in the context of BEC physics)
\begin{equation}
    _\eta\braket{{\rm GS}|\phi(x)|{\rm GS}}_\eta = \frac{\eta}{\sqrt{L}} \neq 0 \ .
\end{equation}
Our goal is to compute its evolution $_\eta\braket{{\rm GS}|\phi(x,t)|{\rm GS}}_\eta$ after a quantum quench with $H_{\rm LL}(c)$, with $c>0$. 

To apply the MC method we need explicit expressions for the overlaps $\braket{\bs \lambda|{\rm GS}}_\eta$. Given \eqref{eq:coherstateeta}, these can be obtained straightforwardly from those of $\braket{\bs \lambda|{\rm GS};N}$. 
$\ket{{\rm GS};N}$ is an integrable initial state, therefore to have $\braket{\bs \lambda|{\rm GS};N}\neq 0$, $\ket{\bs \lambda}$ must be of the form
\begin{equation}
\begin{aligned}
    \ket{\bs \lambda}&=\ket{\lambda_1, -\lambda_1,\ldots,\lambda_{N/2},-\lambda_{N/2}} \qquad N \text{ even} \ , \\
    \ket{\bs \lambda}&=\ket{0,\lambda_1, -\lambda_1,\ldots,\lambda_{(N-1)/2},-\lambda_{(N-1)/2}} \qquad  N \text{ odd}\ ,
\end{aligned}
\end{equation}
where $\lambda_j >0 \ \ \ \forall \ j$. Explicit expressions for the overlaps for $N$ \emph{even} were obtained in Refs.~\cite{denardis_solution_2014, brockmann_overlaps_2014}
\begin{equation}
\label{eq:Nevenoverlap}
    \braket{\lambda_1, -\lambda_1,\ldots,\lambda_{N/2},-\lambda_{N/2}|{\rm GS};N} = \sqrt{\frac{(cL)^{-N}N!}{{\rm det}_{j,k=1}^NG_{jk}}}\frac{{\rm det}_{j,k=1}^{N/2}G^{(\rm even)}_{jk}}{\prod_{j=1}^{N/2}\frac{\lambda_j}{c}\sqrt{\frac{\lambda_j^2}{c^2}+\frac{1}{4}}} \ .
\end{equation}
Here $G_{jk}$ and $G_{jk}^{(\rm even)}$ denote the standard and generalized Gaudin matrices (we define $\{\tilde{\lambda}_1,\ldots,\tilde{\lambda}_N\} \equiv  \{\lambda_1,-\lambda_1,\ldots, \lambda_{N/2},-\lambda_{N/2}\}$)
\begin{equation}
    G_{jk}=\delta_{jk}\Big[ L + \sum_{\ell=1}^N K(\tilde\lambda_j - \tilde\lambda_\ell) \Big] - K(\tilde \lambda_j - \tilde \lambda_k) \qquad \qquad K(\lambda) = \frac{2\lambda}{\lambda^2 + c^2}  \ ,
\end{equation}
\begin{equation}
    G^{(\rm even)}_{jk}=\delta_{jk}\Big[ L + \sum_{\ell=1}^{N/2} K^{(\rm even)}(\lambda_j, \lambda_\ell) \Big] - K^{(\rm even)}(\lambda_j, \lambda_k) \qquad \qquad K^{(\rm even)}(\lambda,\mu) = K(\lambda - \mu)+K(\lambda + \mu) \ .
\end{equation}
Expression for $N$ \emph{odd} were obtained in Ref.~\cite{brockmann_neelxxz_2014}
\begin{equation}
\label{eq:Noddoverlap}
    \braket{0,\lambda_1, -\lambda_1,\ldots,\lambda_{(N-1)/2},-\lambda_{(N-1)/2}|{\rm GS};N} = 2 \sqrt{c} \, \sqrt{\frac{(cL)^{-N}N!}{{\rm det}_{j,k=1}^NG_{jk}}}\frac{{\rm det}_{j,k=0}^{(N-1)/2}G^{(\rm odd)}_{jk}}{\prod_{j=1}^{(N-1)/2}\frac{\lambda_j}{c}\sqrt{\frac{\lambda_j^2}{c^2}+\frac{1}{4}}}  \ .
\end{equation}
Here $G^{(\rm odd)}$ is defined as (we define $\{\tilde{\lambda}_0,\tilde \lambda_1,\tilde \lambda_2\ldots,\tilde{\lambda}_{(N-1)/2}\} \equiv  \{0,\lambda_1,\lambda_2,\ldots, \lambda_{(N-1)/2}\}$)
\begin{equation}
\begin{aligned}
    G^{(\rm odd)}&=\delta_{jk}\Big[ \frac{L}{1+\delta_{j,0}} + \sum_{\ell=0}^{(N-1)/2} K^{(\rm odd)}(\tilde \lambda_j, \tilde \lambda_\ell) \Big] - K^{(\rm odd)}(\tilde \lambda_j, \tilde \lambda_k) \qquad \quad j,k = 0, \ldots, (N-1)/2 \ ,\\
    K^{(\rm odd)}(\tilde \lambda_j,\tilde \lambda_k) &=\Bigg\{ 
    \begin{aligned}
        K(\tilde \lambda_j - \tilde \lambda_k) + K(\tilde \lambda_j + \tilde \lambda_k) \qquad &1\le j \le (N-1)/2 \ , \ \text{same for } k \\
        \frac{1}{2}\left[K(\tilde \lambda_j - \tilde \lambda_k) + K(\tilde \lambda_j + \tilde \lambda_k)\right] \qquad &\text{otherwise} \ .
    \end{aligned}
\end{aligned}
\end{equation}
The matrix $G$ appearing in \eqref{eq:Noddoverlap} is again the standard $N \times N$ Gaudin matrix (for this $N$-odd case it includes the $0$ rapidity).  In the $c = \infty$ limit (non-interacting fermions) \cref{eq:Nevenoverlap,eq:Noddoverlap} significantly simplify
\begin{equation}
\label{eq:overlapNoddcinf}
    \lim_{c \to \infty}\braket{\bs \lambda|{\rm GS};N} =\sqrt{N!}\bigg(\frac{2}{L}\bigg)^{(N-\delta)/2}\frac{1}{\prod_{j=1}^{(N-\delta)/2}\lambda_j} \qquad \qquad \delta = 1 \, (0) \text{ for } N \ {\rm odd}\  (\rm even) \ .
\end{equation}
These $c =\infty$ expressions will be useful to verify the absence of a sign problem in the MC sampling for the LL quench. The overlaps with the symmetry-broken ground state $\ket{\rm GS}_\eta$ can be obtained from \cref{eq:Nevenoverlap,eq:Noddoverlap,eq:overlapNoddcinf} by multiplying these expressions by $e^{-1/2 |\eta|^2}\eta^N/\sqrt{N!}$, cf.~\cref{eq:coherstateeta}. The results for overlaps prove that (given any UV cutoff in LL) there is an exponentially large in $L$ number of nonzero overlaps with $\ket{{\rm GS}}_\eta$. \\
From the overlap formulas, and the expression for the restricted Yang-Yang entropy density $s^{(\rm res)}_{\text{\tiny YY}}[\rho]$ (\cref{sec:briefreview}), it is possible to work out analytically the saddle-point macrostate $\rho_{\rm sp}(\lambda)$ needed in the QA approach. We note that $s^{(\rm res)}_{\text{\tiny YY}}[\rho]$ in LL differs from the standard Yang-Yang entropy density \eqref{eq:yangyangLL} only by restricting the $\lambda$ integral from the full real axis to $(0,\infty)$, due to the parity invariance of the relevant eigenstates for the quench from $\ket{{\rm GS}}_\eta$. 

In the $c = \infty$ case $\rho_{\rm sp}(\lambda)$ is obtained easily due to the simplicity of \cref{eq:overlapNoddcinf}.
Noting that the leading $\mathcal{O}(L^0)$ term in $-\ln |\braket{\bs \lambda|{\rm GS}}_\eta|/L$ depends only on macrostate information (cf.~\cref{sec:briefreview}), one obtains
\begin{equation}
   \lim_{c \to \infty} \rho_{\rm sp}(\lambda) = \frac{1}{2\pi}\frac{1}{\frac{L^2}{4\overline N^2}\lambda^2+1} \ .
\end{equation}
Due to the polynomial decay in $\lambda$, this is a very peculiar root density compared to the ones that define, for example, thermal equilibrium (exponentially decaying with $\lambda$). 
In the interacting case $0<c<\infty$ the saddle-point root density is significantly less straightforward to obtain.  It was shown in Ref.~\cite{denardis_solution_2014} that the leading $\mathcal{O}(L^0)$ term in $-\ln |\braket{\bs \lambda|{\rm GS}}_\eta|/L$ also in this case depends only on macrostate information $\rho(\lambda)$ and
\begin{align}
\label{eq:spcint}
    \rho_{\rm sp}(\lambda) &= \frac{1}{2\pi}\frac{a(\lambda/c)}{1+a(\lambda/c)}\left[ 1+\sqrt{\tau}\frac{I_{2-2i\lambda/c}(4\sqrt{\tau})}{I_{1-2i\lambda/c}(4\sqrt{\tau})}+\sqrt{\tau}\frac{I_{2+2i\lambda/c}(4\sqrt{\tau})}{I_{1+2i\lambda/c}(4\sqrt{\tau})}\right] \ , \\
    a(z) &= \frac{2\pi \tau}{z \sinh(2\pi z)}I_{1-2iz}(4\sqrt{\tau})I_{1+2iz}(4\sqrt{\tau}) \ .
\end{align}
Here $I_p(x)$ is the modified Bessel function of the first kind of order $p$, and $\tau=e^{h/2}$ (where $h$ represents a chemical potential) enforces the correct normalization of $\rho_{\rm sp}(\lambda)$ via
\begin{equation}
    n = \lim_{L \to \infty}\frac{\overline N}{L} = \int_{-\infty}^\infty d \lambda \, \rho_{\rm sp}(\lambda) = \tau \, c \qquad \qquad e = \int_{-\infty}^\infty d \lambda \, \rho_{\rm sp}(\lambda) \lambda^2 = \tau^2 c^3 \ .
\end{equation}
An important relation, useful to determine the saddle-point Yang-Yang entropy density \eqref{eq:yangyangLL}, is $\rho_{\rm sp}(\lambda)/\rho_{h,{\rm sp}}(\lambda) = a(\lambda/c)$. 

As in the case of TFIC, at large but finite $L$ the QA saddle-point eigenstate $\ket{\bs \lambda_{\rm sp}}=\ket{\lambda_1, -\lambda_1,\ldots,\lambda_{\overline N /2},-\lambda_{\overline N /2}}$ can be chosen to have an {even} number of particles $\overline N$ in such a way that the finite-$L$ distribution of rapidities matches as closely as possible $\rho_{\rm sp}(\lambda)$ (cf.~Ref.~\cite{essler_statistics_2024}). 
The QA time evolution will therefore take the form (we can set $\phi(x)\to\phi(0)$ due to translation invariance)
\begin{equation}
\label{eq:QALLexpr}
    _\eta\braket{{\rm GS}|\phi(x,t)|{\rm GS}}_\eta \quad \longrightarrow \quad \frac{1}{2}\sum_{\bs \mu}\left[ e^{i(E_{\bs \mu}-E_{\bs \lambda_{\rm sp}})t}\frac{_\eta\braket{{\rm GS}|\bs \mu}\braket{\bs \mu |\phi(0)|\bs \lambda_{\rm sp}}}{\,_\eta\braket{{\rm GS}|\bs \lambda_{\rm sp}}} + e^{i(E_{\bs \lambda_{\rm sp}}-E_{\bs \mu})t}\frac{\braket{\bs \lambda_{\rm sp}|\phi(0)|\bs \mu}\braket{\bs \mu |{\rm GS}}_\eta}{\braket{\bs \lambda_{\rm sp}|{\rm GS}}_\eta} \right] \ ,
\end{equation}
where the eigenstates $\bs \mu$ that yield non-zero contributions have {odd} numbers of particles, in particular $\overline N - 1$ and $\overline N +1$ respectively for the first and second terms in the brackets. The form factors that appear in \eqref{eq:QALLexpr} are reported in the next section. An important question is what is the relative contribution of the two terms within the brackets in \eqref{eq:QALLexpr} to the final expectation value. As the overlaps with the ground state $\ket{{\rm GS}}_\eta$ are all real for $\eta \in \mathbb R$, this amounts to understanding the relative contribution of $_\eta\braket{{\rm GS}|\phi(0,t)|\bs \lambda_{\rm sp}}$ and $\braket{\bs \lambda_{\rm sp} |\phi(0,t)|{\rm GS}}_\eta$.
Due to the fact that $\phi(0)$ is not Hermitian, these two terms are not related by any obvious conjugation or symmetry.  At $t = 0$ it is elementary to prove, using \cref{eq:coherstateeta}, that $_\eta\braket{{\rm GS}|\phi(0,0)|\bs \lambda_{\rm sp}} = \braket{\bs \lambda_{\rm sp} |\phi(0,0)|{\rm GS}}_\eta$ as a consequence of the fact that $\ket{\bs\lambda _{\rm sp}}$ has $\overline N = \ _\eta \braket{{\rm GS}|\hat N |{\rm GS}}_\eta$ particles. This implies that in the MC sampling scheme (cf.~\cref{eq:sampscheme} of main text) both terms are multiplied by the same normalization $Z$.  In fact, thanks to the MC sampling we find that for large $L$ both terms in \eqref{eq:QALLexpr} yield exactly the same result at all times $t$, and hence one can retain only one of the two (dropping the factor of $1/2$). In the QA method these two terms arise from an ad hoc symmetrization of the QA saddle-point expression \cite{caux_time_2013, caux_quench_2016} (cf.~\cref{eq:quenchQA} from main text), which for example automatically enforces that the time-evolving expectation values of Hermitian operators are real. However, when one follows the saddle-point reasoning leading to the simplified Lehmann representation of the QA method, the symmetrization should not be required, because whenever the QA assumptions are valid both terms yield the same result for $L$ large.

\subsection{Form factors and absence of a sign problem}
\label{eq:FFLL}

Expressions for the form factors $f_{\bmu,\blambda}=\braket{\bmu|\phi(0)|\blambda}$ of the Bose field $\phi(x)$ for generic interaction couplings $c$ can be obtained from the algebraic Bethe ansatz \cite{kojima_determinant_1997,korepin_quantum_1993}. As before $\ket{\blambda}$ and $\ket{\bmu}$ are normalized to 1. The expressions we use in this work have been derived in Refs.~\cite{caux_oneparticle_2007,piroli_exact_2015}.  
Consider a set of $N$ rapidities $\lambda_j$ and a set of $N-1$ rapidities $\mu_k$, which satisfy the Bethe equations (\cref{sec:briefreview}) and such that $\lambda_j \neq \mu_k$  $\forall \ j , k$. In general the previous condition is automatically satisfied due to the nontrivial correlations between rapidities induced by the Bethe equations. Denoting $\lambda_j - \lambda_k$ simply as $\lambda_{jk}$ (same for $\mu$), the logarithm of the form factor $f_{\blambda,\bmu}$ is given by
{\allowdisplaybreaks
\begin{align}
\label{eq:formfactorBose}
\ln (f_{\bmu,\blambda}) =& \ i\frac{\pi}{2}N(N+1) - \frac{1}{2}\ln\left[\text{det}_N G_\blambda\right]- \frac{1}{2}\ln\left[\text{det}_{N-1} G_\bmu\right] \nonumber \\
& - \sum_{j=1}^N \sum_{k=1}^{N-1}\ln\left[ \lambda_j - \mu_k \right] +\frac{1}{2}\sum_{1\le j<k\le N}\ln\left[\lambda_{jk}^2(\lambda_{jk}^2+c^2)\right] 
 + \frac{1}{2}\sum_{1\le j<k\le N-1}\ln\left[\frac{\mu_{jk}^2}{\mu_{jk}^2+c^2}\right] \nonumber \\
&+ \sum_{j=1}^N \ln\Big[ 2 \operatorname{Im}(V_{\lambda_j}) \Big]-\ln\Big[2 \operatorname{Im}(V_{\lambda_p}) \Big]-\ln\Big[ 2 \operatorname{Im}(V_{\lambda_s}) \Big] +\ln\Big[ \text{det}_{N}(\mathds{1}+B_{\lambda_s, \lambda_p})\Big] \ .
\end{align}
}
Here $\lambda_p$ and $\lambda_s$ are two arbitrary complex numbers, $\mathds{1}$ is the $N\times N$ identity matrix and
\begin{equation}
   (G_\blambda)_{jk} = \delta_{j,k}\left[ L + \sum_{\ell=1}^N K(\lambda_{j\ell}) \right] - K(\lambda_{jk}) \ , \qquad \qquad  V_{\lambda}=\frac{\prod_{m=1}^{N-1}(\mu_m-\lambda+ic)}{\prod_{m=1}^{N}(\lambda_m-\lambda+ic)} \ ,
\end{equation}
\begin{equation}
\begin{aligned}
    (B_{\lambda_s,\lambda_p})_{jk} &= \frac{1}{2 \operatorname{Im}(V_{\lambda_j})}\frac{\prod_{m=1}^{N-1}\ (\mu_m-\lambda_j)}{\prod_{\substack{m=1 \\ (m \neq j)}}^{N}(\lambda_m-\lambda_j)}\\
   & \ \ \  \times \big[K(\lambda_{jk})-K(\lambda_p - \lambda_j)K(\lambda_s - \lambda_k)\big] \quad \qquad j, k = 1, \ldots, N \ .
\end{aligned}
\end{equation}

The function $K(x)$ is given as before by $K(x)=2c/(x^2+c^2)$. The Gaudin matrix $G_\bmu$ is identical to $G_\blambda$ up to the substitution $\blambda \to \bmu$ and restriction of the $j,k$ indices from $1$ to $N-1$. We note that the expression \eqref{eq:formfactorBose} is obtained by multiplying the one reported in \cite{piroli_exact_2015} by a factor of $i$. This follows from the convention for the arbitrary phase of each eigenstate implicitly chosen in \cref{eq:Nevenoverlap,eq:Noddoverlap}. An important property of \eqref{eq:formfactorBose} is that the final value of $f_{\bmu,\blambda}$ does not depend on the choice of $\lambda_p$ and $\lambda_s$ \cite{piroli_exact_2015}. \cref{eq:formfactorBose} proves that, given an $N$-particle eigenstate $\ket \blambda$, the off-diagonal form factors $f_{\bmu,\blambda}=\braket{\bmu |\phi(x)|\blambda}$ are nonzero for \emph{any} $(N-1)$-particle eigenstate $\ket \bmu$. Therefore, given a UV cutoff in LL, there is an exponential number in $L$ of nonzero form factors that appear in Lehmann representations, including the ones for DS or QA. 

The expression for $\lim_{c \to \infty}f_{\bmu,\blambda}$ in the impenetrable case ($c=\infty$) is significantly simpler than \eqref{eq:formfactorBose}. By performing the $c\to \infty$ limit of the form factor expression from \cite{caux_oneparticle_2007} (which is slightly simpler than starting from \eqref{eq:formfactorBose}, but the result is the same) it is easy to see that the divergent terms cancel out and one gets
\begin{equation}
\label{eq:formfactoBoseImpenetrable}
\begin{aligned}
    \ln(f_{\bmu,\blambda})\bigg|_{c=\infty} &= i \frac{\pi}{2}N(N-1) +(N-1)\ln 2 - (N-1/2)\ln L 
    \\ & \quad + \frac{1}{2}\sum_{1\le j<k\le N}\ln\left[\lambda_{jk}^2\right]+\frac{1}{2} \sum_{1\le j<k\le N-1}\ln\left[\mu_{jk}^2\right] - \sum_{j=1}^N \sum_{k=1}^{N-1}\ln\left[ \lambda_j - \mu_k \right]   \ .
    \end{aligned}
\end{equation}
The previous expression proves that also in the impenetrable case $c = \infty$ (free fermions), there is an exponential number in $L$ of nonvanishing form factors of $\phi(x)$. The reason, similar to what seen for $\sigma_j^x$ in TFIC, is that $\phi(x)$ represents a nonlocal Jordan-Wigner string when expressed in terms of the free fermionic operators that diagonalize the impenetrable LL Hamiltonian \cite{creamer_quantum_1980}. \\

For $c = \infty$, we can easily convince ourselves that there is no “sign problem” in the DS and QA expressions for the time-evolving expectation value of the Bose field. It follows from \cref{eq:overlapNoddcinf} that the $c = \infty$ overlaps between eigenstates and $\ket{\rm{GS}}_\eta$ (with $\eta$ chosen real and positive) are positive. Hence the only potentially rapidly varying phases in the DS and QA expressions arise from the form factors \eqref{eq:formfactoBoseImpenetrable}. The phases in $f_{\bmu,\blambda}$ are
\begin{enumerate}[label=(\roman*)]
    \item The factor $(-1)^{N(N-1)/2}$.
    \item The signs arising from $\prod_{j=1}^N \prod_{k=1}^{N-1}(\lambda_j - \mu_k)$.
\end{enumerate}
Given the parity invariance of the relevant eigenstates, the phases (ii) can be easily be worked out considering $\bs \lambda = (\lambda_1, -\lambda_1,\ldots,\lambda_m,-\lambda_m)$ and $\bs \mu=(0,\mu_1,-\mu_1,\ldots,\mu_n,-\mu_n)$, where $m = N/2$ and $n=(N/2 -1)$ (we are assuming $N$ even). We note that the multiplication between any pair $(\lambda_j,-\lambda_j)$ and the rapidity $0$ in $\bs \mu$ yields
\begin{equation}
    (\lambda_j - 0)(-\lambda_j -0) = - \lambda_j^2 < 0 \ .
\end{equation}
On the other hand, the multiplication between a pair $(\lambda_j,-\lambda_j)$ and a pair $(\mu_k,-\mu_k)$ gives
\begin{equation}
    (\lambda_j - \mu_k)(\lambda_j +\mu_k)(-\lambda_j - \mu_k)(-\lambda_j +\mu_k) > 0 \ ,
\end{equation}
similarly to what seen in the case of TFIC. Therefore the overall phase from (ii) is $(-1)^{N/2}$. When multiplied by the phase (i) we get $(-1)^{N^2/2}=1$, which establishes the absence of a sign problem. The same conclusions are reached for $N$ odd. The emergence of a sign problem is not expected to be related to taking the limit $c \to \infty$. Indeed, we have checked numerically using \cref{eq:Nevenoverlap,eq:Noddoverlap,eq:formfactorBose}  that the overall phase of the DS or QA spectral weight $F_{\bs \lambda,\bs\mu}$ is always $1$, and hence that there is no sign problem for $0<c < \infty$.

\subsection{MC sampling}
\label{sec:MCMCdetailsLL}
Here we report details of the MC sampling for DS (\cref{eq:quench} of main text) and QA (\cref{eq:quenchQA} of main text). 
As discussed in Appendix A (main text) (cf.~also \cref{sec:briefreview}), sampling can be performed directly over sets of (half-odd) integers for $N$ (even) odd. 
The sampled eigenstates are then used to estimate $\braket{\psi(t)|\phi(x)|\psi(t)}$ via \cref{eq:sampscheme} (main text). 
Given the results on overlaps of the previous sections, the sampling space can be restricted to sets $\{I_j\}$ that include only positive (half-odd) integers, because of the parity invariance of the relevant eigenstates.\\
 
\textbf{\emph{QA sampling. }}\\
The QA representation \eqref{eq:QALLexpr} involves a single sum over eigenstates $\ket{\bs \mu}$. As already discussed, the second term appearing on the right-hand side of \eqref{eq:QALLexpr} gives MC results identical to the first term. Therefore we restrict the following discussion to the first term. Given the choice of the saddle-point eigenstate $\ket{\bs \lambda_{\rm sp}}$ as an $\overline N$-particle state with $\overline N$ even, the sets $\{I_j\}$ associated with the $(\overline N-1)$-particle eigenstates $\ket{\bs \mu}$ include $\overline{N}/2 -1$ integers (the integer $0$ is excluded from $\{I_j\}$ because the rapidity $0$ is always present in $\ket{\bs \mu}$).  
We want to sample the eigenstates $\ket{\bs \mu}$ according to $\mathcal{P}(\{I_j\})=|F(\{I_j\})|/Z$ (cf.~\cref{eq:Csampling} of main text), where 
\begin{equation}
    |F(\{I_j\})| = \frac{_\eta\braket{{\rm GS}|\bs \mu}\braket{\bs \mu |\phi(0)|\bs \lambda_{\rm sp}}}{\,_\eta\braket{{\rm GS}|\bs \lambda_{\rm sp}}} > 0 \ .
\end{equation}
At each MC step $\ell$ we {propose} a “particle-hole” (ph) move within $\{I_j\}$, identical to the one described for TFIC in \cref{sec:MCMCsamplingTFIC} (cf.~also Appendix B of \cite{senese_finite_2026}). In particular, the unoccupied integer is chosen uniformly at random among the $Q/2$ unoccupied integers to the left and to the right of the chosen occupied integer. We introduce a large cutoff $I_{\rm max}$ and if in the proposal step we hit the boundaries of the allowed positive integers, we wrap around. We {accept} or {reject} the proposed “ph” move according to the “ph” Metropolis scheme (see \cref{sec:MCMCsamplingTFIC} or Appendix B of Ref.~\cite{senese_finite_2026}). \\

\textbf{\emph{DS sampling. }}\\
We now turn to the DS representation \eqref{eq:quench} (main text) for $\braket{\psi(t)|\phi(0)|\psi(t)}$.  
Given the expansion \eqref{eq:coherstateeta} of $\ket{\psi}=\ket{{\rm GS}}_\eta$ in terms of the states $\ket{{\rm GS};N}$ of definite particle number, we can write $\ket{\psi} = \ket{\psi_{\rm even}} + \ket{\psi_{\rm odd}}$, where $\ket{\psi_{\rm even}}$ ($\ket{\psi_{\rm odd}}$) includes all the $\ket{{\rm GS};N}$ with $N$ even (odd). Given the action of $\phi(0)$, one has
\begin{equation}
    \braket{\psi(t)|\phi(0)|\psi(t)} = \braket{\psi_{\rm odd}|\phi(0,t)|\psi_{\rm even}} + \braket{\psi_{\rm even}|\phi(0,t)|\psi_{\rm odd}} \ .
\end{equation}
Both terms on the right-hand side give identical results in the limit of large $L$, so that
we can restrict our attention to 
\begin{equation}
\label{eq:actualDSsum}
    \braket{\psi_{\rm odd}|\phi(0,t)|\psi_{\rm even}} = \sum_{\substack{N=0\\(N = \text{even})}}^\infty \sum_{\substack{\blambda \, \rightarrow \, N\\ \bmu \,\rightarrow\, N-1}} e^{i(E_{\bmu}-E_{\blambda})t} \braket{\psi_{\rm odd}|\bmu}\braket{\bmu|\phi(0)|\blambda}\braket{\blambda|\psi_{\rm even}} \ .
\end{equation}
We denote by $\{J_j\}$ ($\{I_j\}$ ) the set of positive half-odd integers associated with $\ket{\blambda}$ ($\ket \bmu$). Given \eqref{eq:actualDSsum}, we want to sample pairs of eigenstates  $(\ket{\bs \lambda},\ket{\bs \mu})$ (where $\ket \blambda$ has $N$ particles and $\ket \bmu$ has $N-1$, with $N$ varying among even integers) according to $\mathcal{P}(\{J_j\},\{I_j\})=|F(\{J_j\},\{I_j\})|/Z$ (cf.~\cref{eq:Csampling} of main text), where 
\begin{equation}
    |F(\{J_j\},\{I_j\})| = \braket{\psi_{\rm odd}|\bmu}\braket{\bmu|\phi(0)|\blambda}\braket{\blambda|\psi_{\rm even}} > 0 \ .
\end{equation}
At each MC step $\ell$ we choose a proposal among: a particle-hole (“ph”) move with probability $p_{\rm ph}$; a creation (“add”) or destruction (“rem”) move with probabilities $p_{\rm add} = p_{\rm rem}$. By construction we have $p_{\rm ph}+p_{\rm add} +p_{\rm rem} = 1$. Then:
\begin{enumerate}[label=(\roman*)]
\item If the move is “ph”, we select either $\ket{\bs \lambda}_{\ell}$ or $\ket{\bs \mu}_{\ell}$ with equal probability and update the corresponding set $\{J_j\}_\ell$ or $\{{I}_j\}_\ell$ with the same particle-hole single-integer Metropolis scheme of QA above (see also \cref{sec:MCMCsamplingTFIC} or Appendix B of Ref.~\cite{senese_finite_2026}). This leads to $\ket{\bs \lambda}_{\ell+1}$ and $\ket{\bs \mu}_{\ell+1}$, where one of the two has necessarily remained invariant given that the move has been (if accepted) performed only on the other.
\item  If the move is “add”, we pick uniformly at random an {unoccupied}, with respect to $\{J_j\}_\ell$, positive half-odd integer $J^*$, up to a fixed cutoff $J_{\rm max}$. We denote by $w_{\bs \lambda}$ the number of possible choices. We then consider the {unoccupied}, with respect to $\{I_j\}_\ell$, integers at a distance from $J^*$ less or equal than a constant $\Delta$, and select one uniformly at random, call it $I^*$. We denote by $w_{\bs \mu}$ the number of possible choices. We then propose the move to the new sets $\{J_j\}^*$ and $\{I_j\}^*$, in which $J^*$ is added to $\{J_j\}_{\ell}$ and $I^*$ is added to $\{I_j\}_\ell$. Note that by the parity-invariance of the eigenstates, this move always involves an increase of 2 in the number of particles in $\ket \blambda$ and $\ket \bmu$, therefore leaving $N$ even as required by \eqref{eq:actualDSsum}.  We denote by $w^*\ge 1$ the number of pairs $(J_j, I_k)$ of (half-odd) integers in $\{J_j\}^*$ and $\{I_j\}^*$ such that
\begin{equation}
\label{eq:DeltaineqMC}
|J_j - I_k| \le \Delta \ .
\end{equation}
Here $1/w^*$ is needed to define the probability of proposing the inverse move (a “rem” move, see below). The probabilities of proposing “add” ($\{J_j\}_\ell \to \{J_j\}^*$ and $\{I_j\}_\ell \to \{I_j\}^*$) or its inverse ($\{J_j\}^* \to \{J_j\}_\ell$ and $\{I_j\}^* \to \{I_j\}_\ell$) are
\begin{equation}
    P_{\rm add} = p_{\rm add} \frac{1}{w_\blambda w_\bmu}\ , \qquad \qquad P_{\rm add}^{(\rm inv)} = p_{\rm rem} \frac{1}{w^*} \ .
\end{equation}
Clearly, if $w_\bmu=0$ the “add” move is automatically rejected and one sets $\{J_j\}_{\ell+1}=\{J_j\}_{\ell}$, $\{I_j\}_{\ell+1}=\{I_j\}_{\ell}$.
\item If the move is “rem”, we denote by $w$ the number of pairs $(J_j, I_k)$ of (half-odd) integers in $\{J_j\}_   \ell$ and $\{I_j\}_\ell$ that respect the inequality \eqref{eq:DeltaineqMC} (if $w = 0$ the “rem” move is automatically rejected and we set $\{J_j\}_{\ell+1}=\{J_j\}_{\ell}$, $\{I_j\}_{\ell+1}=\{I_j\}_{\ell}$). We pick at random one pair $(J^*,I^*)$ among the $w$ possible ones. We then propose the move to new sets $\{J_j\}^*$ and $\{I_j\}^*$ defined by removing $J^*$ ($I^*$) from $\{J_j\}_\ell$ ($\{I_j\}_\ell$). We denote by $w^*_\blambda>1$ and $w^*_\bmu>1$ the analogues of $w_\blambda$ and $w_\bmu$ from point (ii), in this case associated with the inverse “add” move $\{J_j\}^* \to \{J_j\}_\ell$ and $\{I_j\}^* \to \{I_j\}_\ell$.  The probabilities of proposing “rem” and its inverse are
\begin{equation}
    P_{\rm rem} = p_{\rm rem} \frac{1}{w} \ , \qquad \qquad
    P_{\rm add}^{(\rm inv)} = p_{\rm add} \frac{1}{w^*_\blambda w^*_\bmu} \ .
\end{equation}
\end{enumerate}
Given a proposed move, we accept it with Metropolis-Hastings probability \cite{metropolis_equation_1953, hastings_monte_1970}
\begin{equation}
    \alpha_{\rm MH}=\min \left[\frac{\mathcal{P}(\{J_j\}^*,\{I_j\}^*)}{\mathcal{P}(\{J_j\}_{\ell },\{I_j\}_{\ell })}\frac{P_{\rm move}^{\rm(inv)}}{P_{\rm move}},1\right] 
\end{equation}
and reject it with probability $1-\alpha_{\rm MH}$. We set $\{J_j\}_{\ell+1}=\{J_j\}^*$ in case of acceptance and $\{J_j\}_{\ell+1}=\{J_j\}_{\ell}$ otherwise (same for $\{I_j\}$).  

We note that unlike in \cref{sec:MCMCsamplingTFIC}, we have not taken into account the possibility for the sets $\{J_j\}_\ell$ and $\{I_j\}_\ell$ to become empty or fill all the available (half-odd) integers up to the cutoff $J_{\rm max}$. Taking into account these scenarios would complicate the algorithm above, as one needs to consider also other modified proposals (see \cref{sec:MCMCsamplingTFIC}). However, in practice, in all the samplings performed the number of (half-odd) integers in $\{J_j\}_\ell$ and $\{I_j\}_\ell$ remained always far from both extrema (the maximum relative change in the value of $N$ during a single Markov chain run usually being at most $50$-$70\%$ of the value of $\overline N=|\eta|^2$, and never close to $N = 0$). Therefore the Metropolis-Hastings scheme defined above remains exact for the purposes of unbiased sampling based on detailed balance \cite{rubinstein_simulation_2016, gilks_markov_1995}.\\

\textbf{\emph{Convergence checks, statistical errors and other technical details.}}\\
In all the MC algorithms we implemented for the LL quench we have:
\begin{enumerate}
    \item Tuned the values of the free parameters in the MC schemes such to always have an average acceptance rate close to or higher than $10\%$. In particular, setting $Q \in [2,10]$ (relevant for both QA and DS), $\Delta \in [2,6]$ (DS) and $p_{\rm ph} = [0.7,0.9]$ (DS) always leads to accurate sampling of the stationary distribution $\mathcal P(\curlb j) = |F_{\curlb j}|/Z$. 
    \item Verified convergence with the choices of UV cutoffs (needed in LL because it is a field theory) on the maximal allowed (half-odd) integers in the sampling. For example, we verified that the cutoffs chosen were sufficiently high that integers in the sampling never reach them.
    \item Verified that irrespective of the initial sets $\curlb j_1$ chosen, the Markov chain always converges towards the same distribution after a sufficiently large number of steps $\ell$. Once reached, the distribution appears to be stationary. These checks were achieved by monitoring the sampled values of $g_{\{\blambda^{(j)}\}}$ like in Fig.~4 (main text) and \cref{fig:weights_c_8}, or those for the energies $E_{\blambda}$, momenta $P_\blambda$ and number of particles. We have also verified convergence of the results with respect to the choice of $\ell_{\rm max} \in [10^5, 10^7]$ in all regimes of physical interest in which the time-evolving expectation value is not vanishingly small. 
    \item Always discarded a “burn-in” number of initial MC steps $\ell$, to allow time for the Markov chain to reach the stationary distribution. We set $\ell_\text{burn-in}$ to be 1/11 of the total number of MC steps $\ell_\text{burn-in} + \ell_{\rm max}$. 
    \item Extracted the statistical uncertainty on the MC results by computing the standard error on the mean (at each time $t$) associated with the output of $C$ Markov chains run in parallel (we verified that the statistical fluctuations are well fitted by normal distributions). Usually we set $C \in [50,200]$. The uncertainties obtained were negligible in all regimes of physical interest. They are shown in \cref{fig:n_c_plots} of main text. 
\end{enumerate}

An important practical aspect is that the MC sampling must be as efficient as possible in the numerical implementation, given the large number of steps $\ell_{\rm max}$ to be performed. The main bottleneck is the numerical computation of the determinants ($\mathcal{O}(N^3)$ complexity) appearing in the expressions for the overlaps and form factors, see \cref{sec:inistateLL,eq:FFLL}. However, the matrices $G$, $G^{(\rm even)}$, $G^{(\rm odd)}$ and $\mathds{1}+B_{\lambda_s,\lambda_p}$ from \cref{eq:Nevenoverlap,eq:Noddoverlap,eq:formfactorBose} are all centrosymmetric \cite{weaver_centrosymmetric_1985} thanks to the parity-invariance of the eigenstates involved. $N \times N$ centrosymmetric matrices can be block-diagonalized into two $m \times m$ blocks, where $m =N/2$ or $N/2 +1$. The product of the determinants of the two blocks can therefore be computed with overall complexity reduced by a factor $4$ compared to the original cost. This also improves the stability of the determinant evaluation by reducing the condition number of the matrices.

\section{Truncation of the BBGKY hierarchy for the LL quench}

Here we report the coupled sets of differential equations corresponding to the 1$^{\rm st}$ (Gross-Pitaevskii), 2$^{\rm nd}$ (SCTDMFT) and 3$^{\rm rd}$ order truncations of the BBGKY hierarchy for the time evolution of $\braket{\psi(t)|\phi(x)|\psi(t)}$ in the LL quench (see Appendix B of main text). 
We note that truncations beyond the $2^{\rm nd}$ order seem to have received little attention in the literature \cite{kronke_bornbogoliubovgreenkirkwoodyvon_2018}, and therefore the truncation at 3$^{\rm rd}$ order is an interesting result on its own. 
\subsection{First order truncation: Gross-Pitaevskii}
The first order truncation can be worked out directly in real space. Using the exact Heisenberg equations of motion (EOM), and denoting as before $\ket \psi = \ket{{\rm GS}}_\eta$ and $\ket{\psi(t)}=e^{-iHt}\ket \psi$ (where $H = H_{\rm LL}(c)$), we get
\begin{equation}
   \frac{d}{dt} \braket{\psi(t)|\phi(x)|\psi(t)} = i\braket{\psi(t)|[H,\phi(x)]|\psi(t)} = i\braket{\psi(t)|\partial_x^2 \phi(x) - 2c\phi^\dag(x)\phi^2(x) |\psi(t)} \ .
\end{equation}
At the first order of the BBGKY truncation we neglect all $n$-point cumulants with $n\ge2$, i.e.
\begin{equation}
\langle \psi(t) | \phi^\dagger(x) \phi^2(x) | \psi(t) \rangle \quad \to \quad \langle \psi(t) | \phi^\dagger(x) | \psi(t) \rangle \langle \psi(t) | \phi(x) | \psi(t) \rangle^2 = |\varphi(x,t)|^2 \varphi(x,t) \ ,
\end{equation}
where $\varphi(x,t)=\braket{\psi(t)|\phi(x)|\psi(t)}$. Inserting this approximation in the exact EOM we obtain the time-dependent Gross-Pitaevskii equation
\begin{equation}
\frac{d}{dt} \varphi(x,t) = i \partial_x^2 \varphi(x,t) - 2ic |\varphi(x,t)|^2 \varphi(x,t) \ .
\end{equation}
The first term on the right-hand side vanishes by translation invariance and writing $\varphi(t) = \sqrt{\varrho(t)} e^{i\theta(t)}$ (we write $\varphi(t)=\varphi(0,t)$ to ease notations) we obtain
\begin{equation}
\underbrace{\frac{1}{2} \frac{\dot{\varrho}(t)}{\varrho(t)}}_{\text{purely real}} + \underbrace{i\dot{\theta}(t) + 2ic \varrho(t)}_{\text{purely imaginary}} = 0 \ .
\end{equation}
This implies that
\begin{equation}
\dot{\varrho}(t) = 0\ , \qquad \qquad  \dot{\theta}(t) = -2c \varrho(t) = -2c \varrho(0) = -2c \frac{\overline{N}}{L} \ .
\end{equation}
We conclude that in the first-order approximation the order parameter exhibits persistent oscillations
\begin{equation}
\varphi(x,t) = \sqrt{\frac{\overline{N}}{L}} e^{i\omega t} \  \qquad \qquad \omega = - 2c \frac{\overline{N}}{L}\ .
\end{equation}
This behaviour is unphysical as it does not exhibit a restoration of the $U(1)$ symmetry at late times. 

\subsection{Second order truncation: SCTDMFT}
\label{sec:secondorderBBGKY}
The truncations at order $r\ge2$ are best performed in Fourier space
\begin{equation}
    \phi_k = \frac{1}{\sqrt{L}}\int_0^L dx \, e^{ikx}\phi(x)\qquad \qquad k = \frac{2\pi}{L}n, \ \  n \in \mathbb Z .
\end{equation}
The LL Hamiltonian in Fourier space reads
\begin{equation}
H = \sum_{k} k^2 \phi_k^\dagger \phi_k + \frac{c}{L} \sum_{k_1, k_2, k_3} \phi_{k_1+k_3}^\dagger \phi_{k_2-k_3}^\dagger \phi_{k_2} \phi_{k_1} \ .
\end{equation}
We start with the exact Heisenberg EOM (we drop the explicit time dependence in $\phi_k(t)=e^{iHt}\phi_k e^{-iHt}$)
{\allowdisplaybreaks
\begin{align}
\label{eq:2pointeoms}
\frac{d}{dt}\phi_k &= -i k^2 \phi_k - \frac{2ic}{L}\sum_{k_1, k_2} \phi_{k_1+k_2-k}^\dagger \phi_{k_2}\phi_{k_1} \ ,  \nonumber \\
\frac{d}{dt}(\phi_k^\dagger\phi_k) &=  -\frac{2ic}{L}\sum_{k_1, k_2} \phi_k^\dagger\phi_{k_1+k_2-k}^\dagger\phi_{k_2}\phi_{k_1}  + {\rm H.c.} \ , \nonumber \\
\frac{d}{dt}(\phi_k\phi_{-k}) &=  -2ik^2 \phi_k\phi_{-k} - \frac{2ic}{L}\sum_{k_1}\phi_{k_1}\phi_{-k_1} - \left[ \frac{2ic}{L}\sum_{k_1, k_2}\phi_{k_1+k_2-k}^\dagger\phi_{k_2}\phi_{k_1}\phi_{-k} + (k \rightarrow -k) \right] \ .
\end{align}
}
In the 2$^{\rm nd}$ order truncation, i.e.~SCTFMFT, one neglects all $n$-point {cumulants} with $n\ge 3$. Schematically, this leads to the truncations (we denote $\braket{\psi(t)|\ldots|\psi(t)}\equiv\braket{\ldots}$) 
{\allowdisplaybreaks
\begin{align}
    \braket{ABC} &= \braket{AB}\braket C + \braket{AC}\braket B+ \braket{BC} \braket A - 2\braket A \braket B \braket C \ , \nonumber\\
\braket{ABCD} &= \braket{AB}\braket{CD} + \braket{AC}\braket{BD} + \braket{AD}\braket{BC} - 2\braket{A}\braket{B}\braket{C}\braket{D} \ .
\end{align}
}
By translational invariance the only non-vanishing correlators $\braket{\psi(t)|\phi^\dag_{p_1}\ldots\phi^\dag_{p_n}\phi_{k_1}\ldots\phi_{k_m}|\psi(t)}$ are those in which $\sum_{j=1}^n p_j - \sum_{j=1}^m k_j = 0$. Defining
\begin{equation}
\label{eq:deftandd}
    s = \sum_k \braket{\phi_k^\dagger \phi_k} \ , \qquad \quad \qquad d = \sum_k \braket{\phi_{k} \phi_{-k}} \ .
\end{equation}
we can cast the SCTDMFT in the form
\begin{align}
\label{eq:EOMSCTDMFT}
\frac{d}{dt}\braket{\phi_0} &= -\frac{2ic}{L} \left( 2\braket{\phi_0}s + \braket{\phi_0^\dagger}d - 2\braket{\phi_0^\dagger}\braket{\phi_0}^2 \right) \ , \\
\frac{d}{dt}\braket{\phi_k^\dagger \phi_k} &= -\frac{2ic}{L} \braket{\phi_k^\dagger \phi_{-k}^\dagger} d + \text{H.c.} \ , \\
\frac{d}{dt}\braket{\phi_k\phi_{-k}} &=  \left(-2ik^2 - 8ic\frac{s}{L}\right) \braket{\phi_k\phi_{-k}} - \frac{2ic}{L}\left( \braket{\phi_k^\dagger \phi_k} + \braket{\phi_{-k}^\dagger \phi_{-k}} + 1 \right) d  + \frac{8ic}{L}\delta_{k,0}\braket{\phi_0^\dagger}\braket{\phi_0}^3 \ .
\end{align}
The initial conditions are obtained from \cref{eq:coherstateeta} by noting that $\phi_0 \ket{\rm GS}_\eta = \sqrt{\overline N} \ket{\rm GS}_\eta $ (we choose $\eta$ always real and positive) and $\phi_k \ket{\rm GS}_\eta = 0$ for $k \neq 0$. To obtain the dynamics of $\varphi(x,t)$ presented in the main text, this closed systems of ODEs is integrated numerically by a $4^{\rm th}$ order Runge-Kutta (RK4) method.

We have verified convergence with respect to the RK4 time step $dt$, the cutoff $k_{\rm max}$ (for each $k_{\rm max}$ chosen, in \eqref{eq:EOMSCTDMFT} we retain only the equations involving momenta $|k|<k_{\rm max}$) and the system size $L$. Another useful check of the correctness of the integration is that $s(t)$ remains pinned to $\braket{\psi|\hat N |\psi}=\overline N$ at all times. At later times than the ones shown in \cref{fig:BBGKY_c_p2} of main text, the approximation becomes as poor as the $1^{\rm st}$ order result because it again displays persistent oscillations with
an amplitude which is only slightly smaller than the one of the $1^{\rm st}$ order truncation (cf.~also \cref{fig:BBGKY_c_p5}, which shows same plots as \cref{fig:BBGKY_c_p2} of main text but for $c = 0.5$).

\subsection{Third order truncation}
\begin{figure}[!t]
     \centering
     \begin{minipage}[b]{0.35\textwidth}
         \centering
         \includegraphics[width=\textwidth]{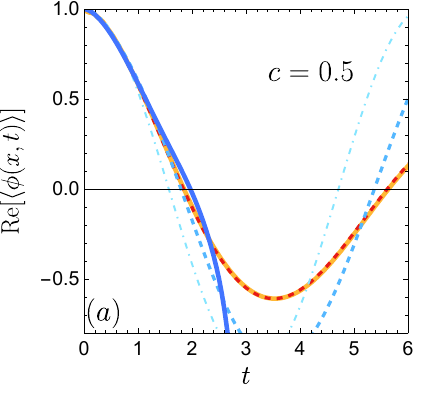}
     \end{minipage}
     \hspace{0pt}
     \begin{minipage}[b]{0.35\textwidth}
         \centering
         \includegraphics[width=\textwidth]{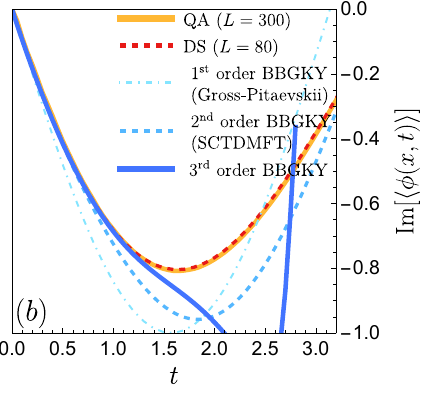}
     \end{minipage}
     \vspace{-10pt}
     \caption{MC results from QA \eqref{eq:quenchQA} and the full double sum (DS) \eqref{eq:quench} for $\braket{\phi(x,t)}$ in the shallow quench $c = 0 \to c =0.5$ (with density $n=\overline{N}/L = 1$). The MC results are benchmarked against the prediction of the truncated BBGKY hierarchy. The MC data is obtained by running 100 Markov chains in parallel with $\ell_{\rm max}=10^6$-$10^7$. (a) Real part. (b) Imaginary part.}
     \label{fig:BBGKY_c_p5}
\end{figure}

To go one order beyond SCTDMFT we need, in addition to \eqref{eq:2pointeoms}, also the following Heisenberg EOMs
{\allowdisplaybreaks
\begin{align}
\frac{d}{dt} (\phi_{k+p} \phi_{-k} \phi_{-p}) &= -i ((k+p)^2 + k^2 + p^2) \phi_{k+p} \phi_{-k} \phi_{-p} \nonumber \\
&\quad - \frac{2ic}{L} \Bigg\{ \sum_{k_1} \Big[ (\phi_{k_1+p} \phi_{-k_1} \phi_{-p} + k \leftrightarrow p) + \phi_{k_1-k-p} \phi_{-k_1} \phi_{k+p} \Big] \nonumber \\
&\quad + \sum_{k_1, k_2} \Big[ (\phi_{k_1+k_2+p}^\dagger \phi_{k_1} \phi_{k_2}\phi_{-k} \phi_{k+p} + k \leftrightarrow p) + \phi_{k_1+k_2-k-p}^\dagger \phi_{k_1} \phi_{k_2} \phi_{-k} \phi_{-p} \Big] \Bigg\} \ ,
\end{align}
\begin{align}
\frac{d}{dt} (\phi_{k+p}^\dagger \phi_k \phi_p) &= 2ikp \,\phi_{k+p}^\dagger \phi_k \phi_p - \frac{2ic}{L} \Bigg\{ \sum_{k_1} \phi_{k+p}^\dagger \phi_{k_1} \phi_{k+p-k_1} \nonumber \\
&\quad + \sum_{k_1, k_2} \Big[ (\phi_{k_1+k_2-p}^\dagger \phi_{k+p}^\dagger \phi_{k_1} \phi_{k_2} \phi_k + k \leftrightarrow p) - \phi_{-k_2+k+p}^\dagger \phi_{k_1+k_2}^\dagger \phi_{k_1} \phi_k \phi_p \Big] \Bigg\} \ .
\end{align}
}
In the 3$^{\rm rd}$ order truncation one neglects all $n$-point cumulants with $n\ge 4$. Schematically, this leads to the truncations (the expansions are first performed in term of cumulants $\braket{\ldots}_{\rm c}$ with $n \le 3$ and then simplified to write them in terms of the non-connected parts $\braket{\ldots}$) 
{\allowdisplaybreaks
\begin{align}
    \braket{ABCD} &=  \ \big( \braket{ABC} \braket{D} + 3 \text{\ cyclic terms} \big) + \big( \braket{AB}\braket{CD} + 2 \text{\ cyclic terms} \big) \nonumber \\
    & \quad \quad -2 \big( \braket{AB}\braket C \braket D + 5 \text{\ cyclic terms}  \big) + 6 \braket A\braket B\braket C\braket D  \ , \\
    \braket{ABCDE} &= \ \big( \braket{ABC} \braket{DE} + 9 \text{\ cyclic terms}  \big) \nonumber \\
   & \quad \quad- \big(\braket{AB}\braket{CD}\braket{E} + 14 \text{\ cyclic terms}  \big) 
    + 6 \braket A\braket B\braket C\braket D \braket E \ ,
\end{align}
}
where in the cyclic permutations the alphabetic order inside each $\braket{\ldots}$ must be respected. 
In addition to \eqref{eq:deftandd}, we define
\begin{align}
    &s_k = \braket{\phi_k^\dag \phi_k} = s_{-k} \ , \qquad \qquad d_k = \braket{\phi_k \phi_{-k}}=d_{-k}  \ ,\nonumber\\
    & S_{k,p} = S_{p,k} = \braket{\phi^\dag_{k+p}\phi_k \phi_p} = S_{-k,-p} = S_{-p,-k} \ , \nonumber\\
    & D_{k,p} = D_{p,k} = \braket{\phi_{k+p}\phi_{-k} \phi_{-p}} = D_{-k,-p} = D_{-p,-k} \ ,
\end{align}
where we have used the parity-invariance $\mathbb P H \mathbb P^{-1}=H$ of the LL Hamiltonian (and $\mathbb P \phi_k \mathbb P= \phi_{-k} $). 
Then the $3^{\rm rd}$ order BBGKY hierarchy reads
{\allowdisplaybreaks
\begin{align}
\frac{d}{dt}\braket{\phi_0} &= -\frac{2ic}{L}\sum_{k_1,k_2} S_{k_1,k_2} \ ,
\end{align}

\begin{align}
\frac{d}{dt} s_k &= \frac{8c}{L} \text{Im} \left( \braket{\phi_0^\dagger}^2 d_k \right) 
 + \frac{4c}{L} \text{Im} \bigg[ \sum_{k_1} \Big( 2\braket{\phi_0} S_{k,k_1}^* - \braket{\phi_0} S_{k-k_1,k_1}^* \Big) + d_k^* d - 2\delta_{k,0}\braket{\phi_0^\dagger}^2 d \bigg] \nonumber \\
&\quad + \frac{4c}{L} \text{Im} \bigg[ \delta_{k,0}\braket{\phi_0^\dagger} \sum_{k_1,k_2} S_{k_1,k_2} \bigg] \ ,
\end{align}

\begin{align}
\frac{d}{dt} d_k &= -2ik^2 d_k - \frac{4ic}{L} \left[ d_k \left(2s - 4|\braket{\phi_0}|^2\right) - 2\braket{\phi_0}^2 s_k + \delta_{k,0} \left(6|\braket{\phi_0}|^2\braket{\phi_0}^2 - 4\braket{\phi_0}^2 s\right) \right] \nonumber \\
&\quad - \frac{2ic}{L} \bigg[ d \left(1 + 2s_k - 4\delta_{k,0}|\braket{\phi_0}|^2\right) + \sum_{k_1} \left( 4\braket{\phi_0} S_{k,k_1} + 2\braket{\phi_0^\dagger} D_{k,k_1} \right) \bigg] \nonumber \\
&\quad- \frac{4ic}{L} \delta_{k,0}\braket{\phi_0} \sum_{k_1,k_2} S_{k_1,k_2} \ ,
\end{align}

\begin{align}
\frac{d}{dt} D_{k,p} &= -i\left[(k+p)^2 + k^2 + p^2\right]D_{k,p} - \frac{2ic}{L} \bigg\{ -2d_{k+p} \left( \braket{\phi_0}(s_k + s_p) + \braket{\phi_0^\dagger}(d_k + d_p) \right) \nonumber \\
&\quad\quad - 2\braket{\phi_0} s_{k+p} (d_k + d_p) - 2\braket{\phi_0} (s_k d_p + s_p d_k) - 2\braket{\phi_0^\dagger} d_k d_p + 18\delta_{k,0}\delta_{p,0}|\braket{\phi_0}|^2\braket{\phi_0}^3 \nonumber \\
&\quad\quad + 6s \Big[ D_{k,p} - \delta_{p,-k}\braket{\phi_0} d_k - \delta_{k,0}\braket{\phi_0} d_p - \delta_{p,0}\braket{\phi_0} d_k \Big] \nonumber \\
& \quad\quad + d \Big[ S_{k+p,-k} + S_{k+p,-p} + S_{k,p} - \delta_{p,-k}\braket{\phi_0}(s_k + s_p) - \delta_{p,-k}\braket{\phi_0^\dagger} d_k \nonumber \\
&\quad\quad - \delta_{k,0}\left(2\braket{\phi_0} s_p + \braket{\phi_0^\dagger} d_p\right) - \delta_{p,0}\left(2\braket{\phi_0} s_k + \braket{\phi_0^\dagger} d_k\right) \Big] \nonumber \bigg\} \nonumber \\
&\quad - \frac{2ic}{L} \sum_{k_1} \Big[ D_{k,k_1}(1 + s_{k+p} + s_p) + D_{p,k_1}(1 + s_{k+p} + s_k) + D_{k+p,k_1}(1 + s_k + s_p) \nonumber \\
&\quad\quad + 2S_{k,k_1}(d_{k+p} + d_p) + 2S_{p,k_1}(d_{k+p} + d_k) + 2S_{k+p,k_1}(d_k + d_p) \Big] \nonumber \\
&\quad - \frac{2ic}{L} \sum_{k_1,k_2} \Big[ S_{k_1,k_2}(\delta_{p,0} d_k + \delta_{k,0} d_p) + \delta_{p,-k} S_{k_1,k_2} d_k \Big] \ ,
\end{align}

\begin{align}
\frac{d}{dt} S_{k,p} &= 2ikp S_{k,p} - \frac{2ic}{L} \bigg\{ -2s_{k+p} \left( \braket{\phi_0}(s_k + s_p) + \braket{\phi_0^\dagger}(d_k + d_p) \right) \nonumber \\
&\quad\quad - 2\braket{\phi_0} d_{k+p}^* (d_k + d_p) + 2\braket{\phi_0} s_k s_p + 2\braket{\phi_0^\dagger}(d_p s_k + d_k s_p) + 6\delta_{k,0}\delta_{p,0}|\braket{\phi_0}|^4\braket{\phi_0} \nonumber \\
&\quad\quad + 2s \left[ S_{k,p} - \delta_{k,0}\braket{\phi_0} s_p - \delta_{p,0}\braket{\phi_0} s_k - \delta_{p,-k}\braket{\phi_0^\dagger} d_k \right] \nonumber \\
&\quad\quad + d \Big[ S_{k+p,-k}^* + S_{k+p,-p}^* - \delta_{k,0}\left(\braket{\phi_0} d_p^* + \braket{\phi_0^\dagger} s_p\right) - \delta_{p,0}\left(\braket{\phi_0} d_k^* + \braket{\phi_0^\dagger} s_k\right) - \delta_{p,-k}\braket{\phi_0^\dagger}(s_k + s_p) \Big] \nonumber \\
&\quad\quad + d^* \Big[ -D_{k,p} + \delta_{k,0}\braket{\phi_0} d_p + \delta_{p,0}\braket{\phi_0} d_k + \delta_{p,-k}\braket{\phi_0} d_k \Big] \bigg\} \nonumber \\
&\quad - \frac{2ic}{L} \sum_{k_1} \Big[ d_{k+p}^*(D_{k,k_1} + D_{p,k_1}) + S_{k_1,k+p-k_1}(1 + s_k + s_p) - S_{k_1,k-k_1}^* d_p - S_{k_1,p-k_1}^* d_k \nonumber \\
&\quad\quad + 2S_{k+p,k_1}^*(d_k + d_p) + 2S_{k,k_1}(s_{k+p} - s_p) + 2S_{p,k_1}(s_{k+p} - s_k) \Big] \nonumber \\
&\quad - \frac{2ic}{L} \sum_{k_1,k_2} \Big[ S_{k_1,k_2}(\delta_{p,0} s_{k} + \delta_{k,0} s_p) - \delta_{p,-k} S_{k_1,k_2}^* d_k \Big] \ .
\end{align}
}
The considerations regarding the initial conditions and the numerical integration are the same as for the SCTDMFT. 
Numerical integration of the $3^{\rm rd}$ order truncation exhibits a runaway behaviour of the order paramter $\varphi(x,t)$. This occurs at times slightly larger than the ones shown in \cref{fig:BBGKY_c_p2} of main text (see \cref{fig:BBGKY_c_p5} for $c = 0.5$, where this is more visible). This behaviour persists if we decrease the RK4 time step $dt$, and increase the cutoff $k_{\rm max}$ and system size $L$, suggesting that this is an intrinsic property of the set of ODEs (cf.~Ref.~\cite{kronke_bornbogoliubovgreenkirkwoodyvon_2018}). However, at sufficiently short times the truncation gives accurate results for the dynamics. In particular we have verified that $s(t)$ remains pinned at  $\overline N$ before the runaway behaviour sets in. In \cref{fig:BBGKY_c_p5} we show results analogous to the ones of \cref{fig:BBGKY_c_p2} from main text, but this time for the larger interaction strength $c = 0.5$. The accuracy of the MC results for QA and DS at short times is apparent by comparison with the increasing order in the truncations of the BBGKY hierarchy. 

\section{A benchmark in the impenetrable LL}
Using the QA method, Ref.~\cite{denardis_analytical_2014} obtained an exact expression in terms of Fredholm determinants for the time evolution of the one-body density matrix after a quench with $H_{\rm LL}(c = \infty)$
\begin{equation}
\label{eq:1BDM}
    \lim_{L \to \infty} \braket{\psi(t)|\phi^\dag(x)\phi(0)|\psi(t)} \ ,
\end{equation}
where as before the initial state coincides with the BEC ground state of the $c = 0$ theory (in this case one can equally well consider $\ket \psi = \ket{\rm GS}_\eta$ or $\ket \psi = \ket{{\rm GS};N}$). This provides us with another benchmark of the MC sampling for QA in a non-interacting theory (LL at $c = \infty$). We stress that the operator for which we track the dynamics in \eqref{eq:1BDM}, i.e.~$O=\phi^\dag(x)\phi(0)$, is different from $\phi(x)$ considered in the main text. The overlaps with the initial state are again the ones reported in \ref{sec:inistateLL}, but in this case the $c = \infty$ parity-invariant eigenstates $\ket{\blambda_{\rm sp}}$ and $\ket \bmu$ have the same number of particles $N$. Explicit expressions for the relevant form factors
\begin{equation}
    f_{\bmu,\blambda}^{(c=\infty)}(x) = \braket{\bmu | \phi^\dag(x)\phi(0)|\blambda} \ 
\end{equation}
can be found in \cite{denardis_analytical_2014}. We observe that in the limit of small $x$ the form factors $f_{\bmu,\blambda}^{(c=\infty)}(x)$ develop extreme \emph{lumpiness} (see Appendix H of \cite{senese_finite_2026}) as a consequence of the fact that for $x \to 0$ the only non-zero off-diagonal matrix elements are those corresponding to a \emph{single} particle-hole excitation distinguishing $\ket{\bmu}$ from $\ket \blambda$. With “lumpiness” we refer to the tendency to observe a very large change in the value of $-\ln|f_{\bmu,\blambda}^{(c=\infty)}(x)|$ as one varies the number of particle-hole excitations distinguishing $\ket{\bmu}$ from $\ket \blambda$. We verified that the degree of lumpiness increases with increasing $L$, and it intrinsically reduces the acceptance rate associated with the type of MC algorithms implemented in this work. Lumpiness is associated with observables for which, at small and intermediate sizes $L$, the number of \emph{relevant} terms to include in the Lehmann sum to reconstruct dynamical correlators is not too large \cite{senese_finite_2026}, and for which therefore MC sampling might not even be necessary \cite{caux_dynamical_2006, caux2009correlation, panfil_finitetemperature_2014}. For $x \gg 1$, the degree of lumpiness is reduced and at large $L$ the MC becomes again necessary to numerically evaluate the Lehmann sum. However, there remain drastic differences with respect to form factors of generic operators in interacting theories $(c<\infty)$: the form factors $f_{\bmu,\blambda}^{(c=\infty)}(x)$ for finite $x$ decay merely as power-laws with $L$ for any $\ket \blambda$ and $\ket \bmu$ that differ by an $\mathcal{O}(L^0)$ number of particle-hole excitations, even if these carry an $\mathcal{O}(1)$ momentum transfer. In interacting theories, form factors associated with such particle-hole excitations decay exponentially with $L$ \cite{essler_statistics_2024, senese_finite_2026}. A consequence of this structural difference in the matrix elements is that in the QA Lehmann sum associated with \eqref{eq:1BDM} only eigenstates $\ket \bmu$ differing from $\ket{\blambda_{\rm sp}}$ by a {sub-extensive} (i.e.~$o(L)$) number of particle-hole excitations meaningfully contribute in the $L \to \infty$ limit \cite{denardis_analytical_2014}. This must be contrasted with interacting theories, where relevant eigenstates in Lehmann sums at large $L$ feature {extensive} numbers of particle-hole excitations compared to the reference state \cite{essler_statistics_2024, senese_finite_2026} (see also \cite{granet_finite_2020}). 

We have implemented a MC sampling of the QA representation for \eqref{eq:1BDM} similar to the one discussed in \cref{sec:MCMCdetailsLL}. In particular, also in this case we make use of single-integer particle-hole moves and the Metropolis acceptance step. The main differences with respect to the interacting case $c < \infty$ for $\phi(x)$ are:
\begin{enumerate}
    \item We need to separate the contribution of diagonal matrix elements ($\braket{\blambda_{\rm sp} | \phi^\dag(x)\phi(0)|\blambda_{\rm sp}}$) from that of off-diagonal ones ($\braket{\bmu | \phi^\dag(x)\phi(0)|\blambda_{\rm sp}}$ for $\ket{\bmu} \neq \ket{\blambda_{\rm sp}}$). This is achieved by setting $\mathcal{P}(\bmu = \blambda_{\rm sp})=0$ in the Metropolis acceptance step, and estimating the correlator as
    \begin{equation}
        \braket{\psi(t)|\phi^\dag(x)\phi(0)|\psi(t)} \quad \longrightarrow \quad e^{-2|x|}+\bigg(\frac{\overline N}{L}-e^{-2|x|}\, \text{Y}_{\rm MC}(t)\bigg) \ ,
    \end{equation}
    where $Y_{\rm MC}(t)$ indicates the output of the MC (such that $Y_{\rm MC}(0)=1$). The terms $\overline N / L$ and $e^{-2|x|}$ derive respectively from $\braket{\psi|\phi^\dag(x)\phi(0)|\psi}=\overline N / L$ and (see \cite{denardis_analytical_2014, caux_time_2013})
    \begin{equation}
        \lim_{t \to \infty}\lim_{L \to \infty} \braket{\psi(t)|\phi^\dag(x)\phi(0)|\psi(t)} = \lim_{L \to \infty} \braket{\blambda_{\rm sp} | \phi^\dag(x)\phi(0)|\blambda_{\rm sp}} = e^{-2|x|} \ .
    \end{equation}
    \item We observe that even though $f_{\bmu,\blambda}(x)$ can be either positive or negative, it is positive for the vast majority of the eigenstates sampled (no sign problem). This is very similar to the case $h_0>h$ of TFIC, see \cref{sec:QAandSPinTFIC}. 
    \item The saddle-point eigenstate $\ket{\blambda_{\rm sp}}$ is constructed as in \cref{sec:inistateLL}. However, due to the very slow $\sim \lambda^{-2}$ decay of the QA saddle-point root density at $c = \infty$ (see \cref{sec:inistateLL}), which must be contrasted with the $\sim \lambda^{-4}$ decay of the interacting case $0<c<\infty$, the set of positive integers $\{J_j\}$ defining $\ket{\blambda_{\rm sp}}$ for large $L$ includes very large integers. We find that this enhances the statistical uncertainty of the MC estimate for the given MC algorithm we are implementing. The performance is also affected by lumpiness (which is suppressed at larger $x$, but not removed, and increases with $L$). Developing an algorithm more suitable for sampling in this landscape of eigenstates is beyond the scope of the present work, and will be addressed in future publications. With the present algorithm, fixing $x = 5$, we obtain good convergence of the QA MC approach for $L \le 100$. 
    \item The eact result for \eqref{eq:1BDM} of Ref.~\cite{denardis_analytical_2014} in terms of a difference of Fredholm determinants can be evaluated by standard numerical methods based on quadrature rules \cite{bornemann_numerical_2009}. Fixing $x = 5$, we employ a Gauss-Legendre quadrature grid with several thousands of point to reach convergence (up to $n = 8000$). The cutoff is varied in the range $\lambda_{\rm max} \in [200,600]$ to check convergence.
    \end{enumerate}

\begin{figure}[ht]
    \centering
    \includegraphics[width=0.35\linewidth]{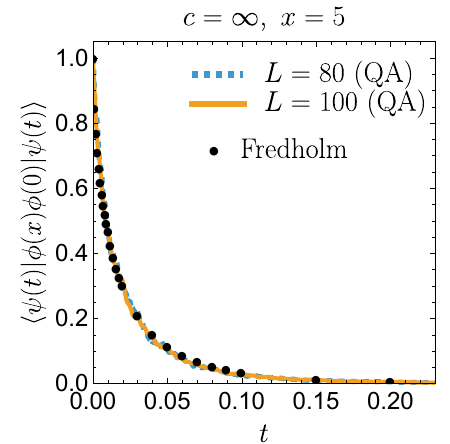}
    \caption{Time evolution of the one-body density matrix $\braket{\psi(t)|\phi^\dag(x)\phi(0)|\psi(t)}$ for $x = 5$ in a quench with Hamiltonian $H_{\rm LL}(c = \infty)$ from the ground state of $H_{\rm LL}(c=0)$. The curves show MC results for the QA representation with $L = 80,100$, obtained by running 100 parallel Markov chains with $\ell_{\rm max}= 4\times 10^7$ MC steps each. The dots indicate exact thermodynamic-limit results obtained from the analytic Fredholm-determinant expression from Ref.~\cite{denardis_analytical_2014}, evaluated with a Gauss-Legendre quadrature rule at $n = 8000$ points and cutoff $\lambda_{\rm max}=600$ \cite{bornemann_numerical_2009}.}
    \label{fig:c_inf_benchmark}
\end{figure}

In \cref{fig:c_inf_benchmark} we present a comparison of the MC output for QA and the Fredholm results at $x = 5$. We see that in spite of the technical difficulties discussed above, the agreement is very good. This provides an independent check of the ability of MC to correctly reconstruct QA Lehmann representations. 

\begin{figure}[ht]
     \centering
     \begin{minipage}[b]{0.3\textwidth}
         \centering
         \includegraphics[width=\textwidth]{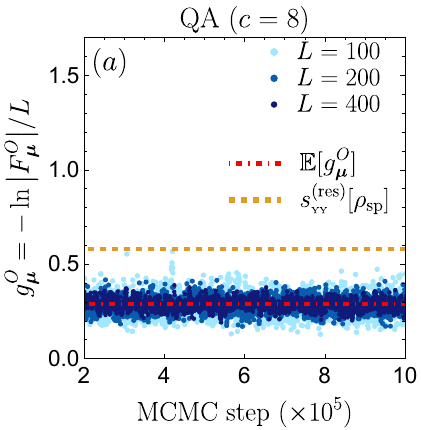}
     \end{minipage}
     \hspace{-7pt}
     \begin{minipage}[b]{0.3\textwidth}
         \centering
         \includegraphics[width=\textwidth]{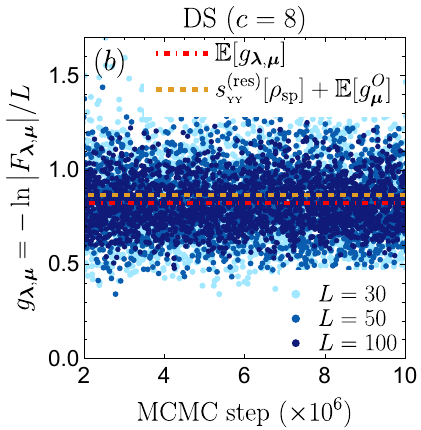}
     \end{minipage}
     \vspace{-8pt}
     \caption{Analogue of Fig.~4 from main text, but for $c = 8$.}
     \label{fig:weights_c_8}
\end{figure}

\section{Further results and plots}

In \cref{fig:weights_c_8} we plot the analogue of \cref{fig:weights_c_2} ($c = 2$) from main text but now at the higher interaction strength $c = 8$. The prediction $g^*_{\rm DS} = g^*_{\rm QA}+s_\text{\tiny YY}^{(\rm res)}[\rho_{\rm sp}] $ appears to again hold (within uncertainties on the values of $g^*_{\rm DS}$ and $g^*_{\rm QA}$).

\begin{figure}[ht]
    \centering
    
    \begin{minipage}{0.8\linewidth}
        \centering
        \includegraphics[width=0.9\linewidth]{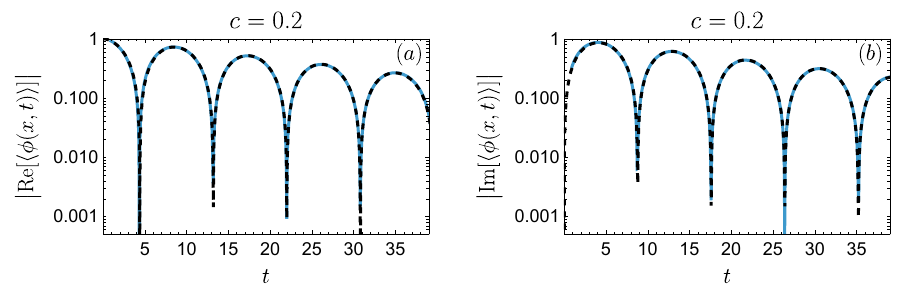}
    \end{minipage}
    \vspace{-.5cm} 

    \begin{minipage}{0.8\linewidth}
        \centering
        \includegraphics[width=0.9\linewidth]{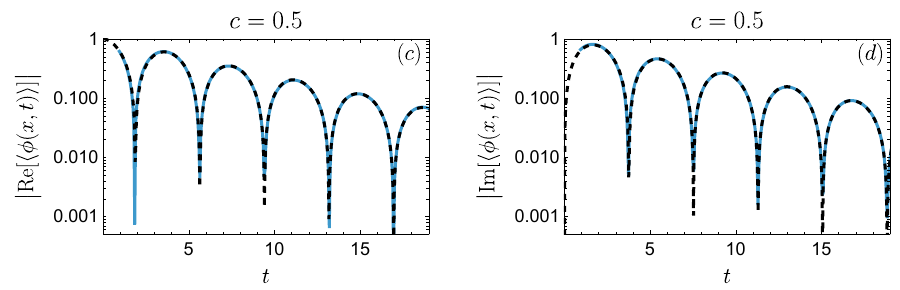}
    \end{minipage}
    \vspace{-.5cm}

    \begin{minipage}{0.8\linewidth}
        \centering
        \includegraphics[width=0.9\linewidth]{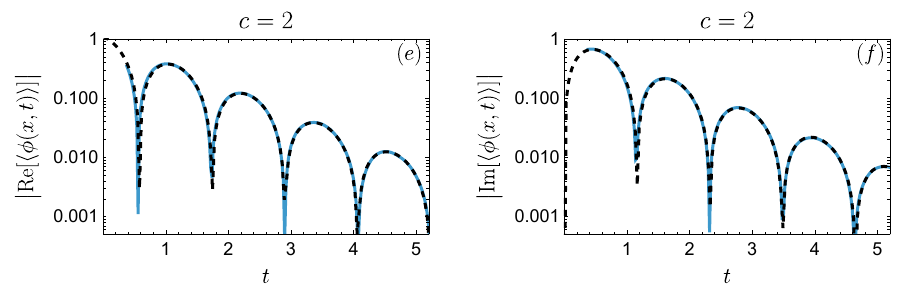}
    \end{minipage}
    \vspace{-.5cm}

    \begin{minipage}{0.8\linewidth}
        \centering
        \includegraphics[width=0.9\linewidth]{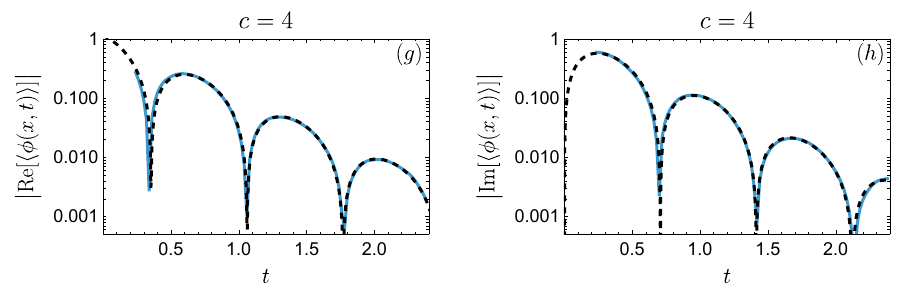}
    \end{minipage}

    \caption{Log-plots for the absolute value of the real and imaginary parts of $\braket{\phi(x,t)}$ in LL quench and associated fits, for several values of interaction strength $c$. Continuous blue curves: QA MC data for $L = 300$ (see Fig.~3 of main text). Black dashed curves: $\braket{\phi(x,t)} \rightarrow  a \, e^{-i \omega t}e^{-t/\tau}$ fits (see text for details on the fitting procedure). The fitted frequencies are $\omega \approx $ 0.357 (a)-(b), 0.834 (c)-(d), 2.70 (e)-(f), 4.44 (g)-(h). The decay times for the same subplots are $\tau \approx $  26.3, 6.96, 1.02, 0.43.}
    \label{fig:fits}
\end{figure}

In \cref{fig:fits} we show log-plots for $\big|\text{Re}[\braket{\phi(x,t)}]\big|$ and $\big|\text{Im}[\braket{\phi(x,t)}]\big|$ in LL quench obtained by QA MC at $L = 300$ (see \cref{fig:n_c_plots} of main text), and compare them with fitting functions 
\begin{equation}
    \braket{\phi(x,t)} \quad \longrightarrow \quad y_{\rm fit}(t) = a \, e^{-i \omega t}e^{-t/\tau} \ ,
\end{equation}
with 3 fitting parameters $a, \omega, \tau$. The fitting parameters are obtained by first considering $|\braket{\phi(x,t)}|$ and performing an exponential 2-parameter fit to extract $a$ and $\tau$, followed by 1-parameters fits $\cos(\omega t)$ and $\sin(-\omega t)$ respectively on $\text{Re}[\braket{\phi(x,t)}]e^{t/\tau}/a$ and $\text{Im}[\braket{\phi(x,t)}]e^{t/\tau}/a$, to extract $\omega$. The agreement of the fits with the MC data is excellent.

\clearpage

\begingroup
\renewcommand{\addcontentsline}[3]{}
\putbib[references,outputNotes]
\endgroup

\end{bibunit}

\end{document}